\documentclass[12pt]{article}
\usepackage{amssymb}
\usepackage{graphicx}
\usepackage{amsmath}
\usepackage{makeidx}
\usepackage{indentfirst}
\usepackage[T1]{fontenc}
\usepackage[utf8]{inputenc}
\usepackage{authblk}

\newcounter{resultnum}[section]
\newcounter{conclusionnum}[section]
\newcounter{conditionnum}[section]
\newcounter{conjecturenum}[section]
\newcounter{examplenum}[section]
\newcounter{exercisenum}[section]
\newcounter{lemmanum}[section]
\newcounter{notationnum}[section]
\newcounter{theoremnum}[section]
\newcounter{definitionnum}[section]
\newcounter{corollarynum}[section]
\newcounter{remarknum}[section]
\newcounter{propositionnum}[section]
\newcounter{acknowledgementnum}[section]
\newcounter{algorithmnum}[section]
\newcounter{axiomnum}[section]
\newcounter{casenum}[section]
\newcounter{claimnum}[section]
\newcounter{summarynum}[section]
\newcounter{problemnum}[section]
\begin{document}

\title{Geometric Flows and Perelman's Thermodynamics for Black Ellipsoids in $R^2$ and Einstein Gravity Theories}

\date{January 3, 2016}

\author{% ${}$ \\
%EndAName
{\bf Tamara Gheorghiu}\\
%\vspace{.1 in}
 {\small \textit{University "Al.  I. Cuza" Ia\c si, Project IDEI,}}\\
 {\small \textit{14 Alex. Lapu\c sneanu str., Corpus R, UAIC, office 323, Ia\c si, Romania, 700057 }}\\
{\small and } \\
{\small \textit{Univeristy of Medicine and Pharmacy "Gr. T. Popa" Ia\c si, Faculty of Medicine,}} \\
{\small \textit{16 University str., Ia\c si, Romania, 700114}} \\
{\small \textit{email: tamara.gheorghiu@yahoo.com  ${}$ }}\\
\vspace{.1 in}{\bf Vyacheslav Ruchin}\\
\vspace{.1 in}
{\small \textit{Heinrich-Wieland-Str. 182,\  81735 M\"{u}nchen, Germany }}\\
{\small \textit{email: v.ruchin-software@freenet.de  ${}$ }}\\
\vspace{.1 in}{\bf Olivia Vacaru}\\
\vspace{.1 in} {\small \textit{National College of Ia\c si; 4 Arcu street, Ia\c si, Romania, 700125 }}\\
{\small \textit{email: olivia.vacaru@yahoo.com  ${}$ }}\\
\vspace{.1 in} {\bf Sergiu I. Vacaru}\\
\vspace{.1 in} {\small \textit{University "Al. I. Cuza" Ia\c si, Rector's
Department }}\\
{\small \textit{14 Alexandru Lapu\c sneanu street, Corpus R, UAIC, office
323, Ia\c si, Romania 700057 }}\\
{\small and\thanks{two DAAD fellowship affiliations} }\\
{\small \textit{Max-Planck-Institute for Physics,
Werner-Heisenberg-Institute, }}\\
{\small Foehringer Ring 6, M\"{u}nchen, Germany D-80805 ;}\\
{\small and }\\
{\small \textit{    Leibniz University of Hannover, Institute for Theoretical Physics, }}\\
{\small Appelstrasse 2, Hannover, Germany 30167}\\
{\small \textit{emails: sergiu.vacaru@uaic.ro; sergiu.vacaru@gmail.com}}}

%%%%%
\maketitle

\begin{abstract}
We study geometric relativistic flow and Ricci soliton equations which (for
respective nonholonomic constraints and self-similarity conditions) are
equivalent to the gravitational field equations of $R^2$ gravity and/or to
the Einstein equations with scalar field in general relativity, GR.
Perelman's functionals are generalized for modified gravity theories, MGTs,
which allows to formulate an analogous statistical thermodynamics for
geometric flows and Ricci solitons. There are constructed and analyzed
generic off--diagonal black ellipsoid, black hole and solitonic exact
solutions in MGTs and GR encoding geometric flow evolution scenarios and
nonlinear parametric interactions. Such new classes of solutions in MGTs can
be with polarized and/or running constants, nonholonomically deformed
horizons and/or imbedded self-consistently into solitonic backgrounds. They
exist also in GR as generic off--diagonal  vacuum configurations
with effective cosmological constant and/or mimicking effective scalar field
interactions. Finally, we compute Perelman's energy and entropy for black
ellipsoids and evolution solitons in $R^2$ gravity.

\vskip5pt

\end{abstract}

\tableofcontents

\section{Introduction}

The Ricci flow theory was formulated mathematically in R. Hamilton's works
beginning 1982 \cite{hamilt1,hamilt2,hamilt3}. It became famous after an
enormous success related to Grisha Perelman's proof in 2002-2003 of the
Poincar\'{e} and Thurston conjectures \cite{perelman1,perelman2,perelman3}.
There were proposed various constructions and applications of geometric flow
methods in theoretical and mathematical particle physics and gravity before
and after 1980, see reviews and original results in \cite%
{friedan1,friedan2,friedan3,vnhrf,vnrflnc,vvrfthbh,bakaslust,carfora}. A
series of problems for developing such directions in physics is related to
technical difficulties in constructing exact solutions of relativistic
geometric flow equations and understanding their physical meaning and
properties in modified gravity theories, MGTs, and general relativity, GR.
Rigorous geometric analysis methods elaborated in mathematical works do not
provide an effective  tool for investigating physical problems
 on possible geometric flow evolution scenarios related to modern
gravity and cosmology.

The main purpose of this work is to elaborate on the anholonomic frame
deformation method, AFDM, (see reviews and original results in refs. \cite%
{sv2001,sv2014,svvvey,tgovsv}), as a geometric method for constructing exact
solutions of geometric flow equations and gravitational field equations in $%
R^{2}$--gravity. As explicit examples, we shall generalize the black hole
solutions considered by A. Kehagias, C. Kounnas, D. L\"{u}st and A. Riotto
in \cite{kehagias} and study the conditions when corresponding modifications
encode geometric flow evolution scenarios. Such new classes of generic
off--diagonal solutions depend, in general, on all spacetime coordinates and
on a geometric evolution parameter. They can be with prescribed ellipsoid
and/or solitonic symmetries, nonholonomically deformed horizons and physical
constants with locally anisotropic polarizations and running on flow
parameter. For certain conditions, there are generated solutions of Ricci
soliton equations modelling MGT effects and/or off--diagonal interactions in
GR.

\subsection{Geometric flows in string theory and modified gravity}

Geometric flows appear naturally in off--critical string theory via the
renormalization--group equations and nonlinear sigma model, $\sigma $%
--models. Physicists knew independently (perhaps, some years before
mathematicians) about certain models with geometric flow equations. Here we
cite D. Friedan's works \cite{friedan1,friedan2,friedan3} on nonlinear
models in two + epsilon dimensions, $2+\varepsilon $, published in 1980.
Those papers were related to developments of Polyakov's research \cite%
{polyakov} on renormalization of the $\mathit{O}(N)$--invariant nonlinear $%
\sigma $--model, which in a low--temperature regime is dominated by small
fluctuations around ordered states.

The standard nonlinear $\sigma $--models consist some special cases when $M$
is a homogeneous space (the quotient $G/H$ of a Lie group $G$ by a compact
subgroup $H$) and $g_{ij}$ is some $G$--invariant Riemannian metric on $M.$
The renormalization of such models considers a techniques used for the
standard power counting arguments combined with generalizations of the BRST
transformation and the method of quadratic identities. In result, the
renormalization group equation for the metric coupling is
\begin{equation*}
\ell ^{-1}\frac{\partial g_{ij}}{\partial \ell ^{-1}}=-\beta _{ij}(g),
\end{equation*}%
where the $\beta $-function
\begin{equation*}
\beta _{ij}(\wp ^{-1}g)=-\varepsilon (\wp ^{-1}g)_{ij}+\ _{\shortmid }R_{ij}+%
\frac{1}{2}\wp \ _{\shortmid }R_{iklm}\ _{\shortmid }R_{jklm}+O(\wp ^{2})
\end{equation*}%
is a vector field on the infinite dimensional space of Riemannian metrics on
$M.$ The value $(\wp ^{-1}g)_{ij}$ is a (positive definite) Riemannian
metric on $M,$ called the metric coupling and $\ell $ is the short distance
cutoff (in certain models $\ell ^{-1}$ is treated as a temperature like
parameter).

Two important results on global properties of above type $\beta $-functions
were obtained. When a manifold $M$ is a homogeneous space $G/H,$ the $\beta $%
--function is shown to be a gradient type function for a finite dimensional
space of $G$--invariant metric couplings on $M.$ If $M$ is a two dimensional
compact manifold, the $\beta $--function is shown to be a gradient on the
infinite dimensional space of metrics of $M.$

A series of mathematical and physical results were obtained for self-similar
configurations defined by equations which are similar to field equations in
modern gravity. Such fixed points of Ricci flows are described by (latter
called Ricci solitons) equations
\begin{equation}
R_{ij}-\lambda g_{ij}=\nabla _{i}v_{j}+\nabla _{j}v_{i},
\label{friedsoliton}
\end{equation}%
for $\lambda =\pm 1,0.$ In above formulas, $g_{ij}(x^{k}),\nabla _{i}$ and $%
\ R_{ij}$ are respectively the symmetric metric field, Levi--Civita
connection, LC--connection, and the Ricci tensor. For instance, on a two
dimensional, 2-d, Riemannian manifold $M,$ the local coordinates can be
written $x=\{x^{i}\},$ for indices $i,j,k...=1,2,..n$ and $v_{j}(x)\subset
TM $ is a vector field defined by sections of tangent bundle $TM.$ Equations
of type (\ref{friedsoliton}) are considered in various MGTs as gravitational
field equations for pseudo--Riemannian metrics and generalized connections
or for LC--connections.

\subsection{Ricci flows of Riemannian metrics on 3--d manifolds}

Ricci flows (as a particular but very important example of a geometric
evolution theory) are known in mathematics due to Hamilton's programme on
geometric analysis and attempts to prove the Poincar\'{e} conjecture. In the
"standard" Ricci flow theory on three dimensional, 3-d, Riemannian
manifolds, one considers the evolution of a family of metrics $g_{\grave{%
\imath}\grave{j}}(\chi )=g_{\grave{\imath}\grave{j}}(\chi ,x^{\grave{k}})$
of signature $(+++)$ parameterized by real parameter $\chi ,$ with respect
to the coordinate base $\partial _{\grave{\imath}}:=\partial /\partial x^{%
\grave{\imath}}$ and $\partial _{\chi }:=\partial /\partial \chi ,$ for $%
\grave{\imath},\grave{j},\grave{k}=1,2,3.$ The R. Hamilton equations where
postulated in the form
\begin{equation}
\partial _{\chi }g_{\grave{\imath}\grave{j}}=-2\ _{\shortmid }R_{\grave{%
\imath}\grave{j}}+\frac{2}{5}\ _{\shortmid }\rho \ g_{\grave{\imath}\grave{j}%
},  \label{heq}
\end{equation}%
where the normalizing factor$\ _{\shortmid }\rho =\int\nolimits_{Vol}\sqrt{%
|\ _{\shortmid }g|}d^{3}x\ _{\shortmid }R/\int\nolimits_{Vol}\sqrt{|\
_{\shortmid }g|}d^{3}x,\ $ (for $_{\shortmid }R:=g^{\grave{\imath}\grave{j}%
}\ _{\shortmid }R_{\grave{\imath}\grave{j}}$ and $\ _{\shortmid }g:=\det |g_{%
\grave{\imath}\grave{j}}|$), is introduced in order to preserve a 3-d
compact volume $Vol.$\footnote{%
Our notations are different from those in mathematical books because we
follow a system of notations which is useful for constructing generic
off--diagonal exact solutions of such equations. For instance, we use
"primed" indices and a left "vertical line" label like$\ _{\shortmid }R$ in
order to emphasize that such values are used for 3--d Riemannian spaces/
hypersurfaces.} We can take $\ _{\shortmid }\rho =0$ and consider a zero
effective cosmological constant $\lambda =\frac{2}{5}\ _{\shortmid }\rho =0$
for non--renormalized Ricci flows. In certain sense, such equations consist
a generalized nonlinear diffusion equation for a tensor filed $g_{\grave{%
\imath}\grave{j}}$ because $\ _{\shortmid }R_{\grave{\imath}\grave{j}}\simeq
\Delta ,$ where $\Delta $ is the Laplace operator if $g_{\grave{\imath}%
\grave{j}}$ $\simeq 1+\delta _{\grave{\imath}\grave{j}}$ for small
fluctuations $\delta _{\grave{\imath}\grave{j}}$ of the Euclidean metric and
$\chi $ treated as a temperature type parameter. \ If $\partial _{\chi }g_{%
\grave{\imath}\grave{j}}=0,$ we obtain the equations for 3--d Einstein
spaces with metrics of positive definite signature.

One of the most important results due to G. Perelman is that the equations (%
\ref{heq}) can be derived as gradient flows \cite{perelman1} from certain
Lyapunov type functionals for dynamical systems,
\begin{eqnarray}
\ _{\shortmid }\mathcal{F}(g_{\grave{\imath}\grave{j}},\nabla ,f)
&=&\int\nolimits_{Vol}\sqrt{|\ _{\shortmid }g|}d^{3}x\left( \ _{\shortmid
}R+|\nabla f|^{2}\right) e^{-f},  \label{fpfunct} \\
\ _{\shortmid }\mathcal{W}(g_{\grave{\imath}\grave{j}},\nabla ,f,\tau )
&=&\int\nolimits_{Vol}\sqrt{|\ _{\shortmid }g|}d^{3}x\left[ \tau \left( \
_{\shortmid }R+|\nabla f|\right) ^{2}+f-3\right] \mu ,  \label{wpentr}
\end{eqnarray}%
where the function $\tau =\tau (\chi )>0,\mu :=(4\pi \tau )^{-3/2}e^{-f}$
for $\int\nolimits_{Vol}\mu \sqrt{|\ _{\shortmid }g|}d^{3}x=1.$ Such
functionals are called Perelman's F--functional and W--entropy.

It should be noted that the W--entropy was used by G. Perelman \cite%
{perelman1} in order to elaborate a statistical thermodynamics approach to
the theory of Ricci flows. There were considered generalizations of such
functionals in refs. \cite{vnhrf,vnrflnc,vvrfthbh} related to nonholonomic
and noncommutative geometric flows and black hole entropy and geometric
flows. Certain classes of generic off--diagonal solutions of 4-d Ricci
soliton equations were constructed and studied for nonholonomic dynamical
systems. We also emphasize that 3-d extensions of the Ricci soliton
equations (\ref{friedsoliton}), written in the form
\begin{equation}
R_{\grave{\imath}\grave{j}}-\lambda g_{\grave{\imath}\grave{j}}=\nabla _{%
\grave{\imath}}v_{\grave{j}}+\nabla _{\grave{j}}v_{\grave{\imath}},
\label{riccisol3d}
\end{equation}%
can be derived as self--similar fixed configurations of functionals (\ref%
{fpfunct}), or (\ref{wpentr}), and Ricci flow evolution equations (\ref{heq}%
).

\subsection{$R^{2}$ gravity and Ricci solitons}

Ricci soliton type equations exist naturally in $R^{2}$ gravity. Let us
consider the equations (61) and (62) from \cite{kehagias} (for well defined
conditions, such equations are equivalent to the gravitational field
equations in MGT),%
\begin{eqnarray}
\overline{R}_{\mu \nu }-\overline{\nabla }_{\mu }\phi \ \overline{\nabla }%
_{\nu }\phi -2\varsigma ^{2}\overline{g}_{\mu \nu } &=&0,  \label{riccisol4d}
\\
\overline{\nabla }^{2}\phi &=&0,  \label{aux2}
\end{eqnarray}%
where the non-scale mode $\Phi =\frac{1}{2}e^{\sqrt{2/3}\phi }$ plays the
role of a Lagrange multiplier used in conformal (Jordan frames ) frames
without spacetime derivatives and $\varsigma ^{2}$ is a non-zero
cosmological constant. There are considered conformal transforms of the
metric
\begin{equation}
g_{\mu \nu }\rightarrow \overline{g}_{\mu \nu }=e^{\sqrt{1/3}\phi }g_{\mu
\nu },  \label{overlmetric}
\end{equation}%
with
\begin{equation*}
e^{\sqrt{1/3}\phi }=\frac{1}{8\varsigma ^{2}}R\text{\mbox{ and }}\overline{g}%
_{\mu \nu }=e^{\sqrt{1/3}\phi }g_{\mu \nu }=\frac{R}{8\varsigma ^{2}}g_{\mu
\nu },R\neq 0,
\end{equation*}%
where $R$ is the Ricci scalar determined by the metric $g_{\mu \nu }$ and
corresponding Levi--Civita, LC, connection. Over-lined values like $%
\overline{R}_{\mu \nu }$ are determined by $\overline{g}_{\mu \nu }$ and $%
\overline{\nabla }_{\mu },$ where local coordinates are labels $u^{\mu }$
for indices with conventional 3+1 splitting $\alpha ,\beta ,\mu
,...=1,2,3,4, $ when $\alpha =(\grave{\imath},4).$

The equations (\ref{riccisol4d}) can be considered as gravitational field
equations for certain MGTs with $R^2$ terms in Lagrangians and Einstein
gravity models with additional massless scalar propagating field $\phi $ and
nontrivial cosmological constant $\varsigma ^{2}.$ Using respective
conformal transforms, such theories can be derived equivalently from the
action%
\begin{equation}
S=\int \sqrt{|g|}d^{4}u\left( \frac{1}{16\varsigma ^{2}}R^{2}\right) ,
\label{r2action}
\end{equation}%
and/or
\begin{equation}
S=\int \sqrt{|g|}d^{4}u\left( \Phi R-4\varsigma ^{2}\Phi ^{2}\right) ,
\label{tmtact}
\end{equation}%
and/or
\begin{equation}
S=\int \sqrt{|g|}d^{4}u\left( \frac{1}{2}\overline{R}-\frac{1}{2}g^{\mu \nu
}\partial _{\mu }\phi \partial _{\nu }\phi -\varsigma ^{2}\right) ,
\label{einstgrscact}
\end{equation}%
see respective formulas (40)-(43) in \cite{kehagias}.

Theories of type (\ref{r2action}) are of great interest in cosmology \cite%
{starob,much} and inflationary scenarios in the early universe \cite%
{guth,linde,albrecht}. We note that in string theory such higher curvatures
corrections appear naturally but as infinite series which result in the
appearance of ghost-like modes originating both from the square oft eh
Riemann and Weyl tensors. Nevertheless, such a pathology is absent for the $%
R+R^{2}$ theory, which is equivalent to standard Einstein gravity with an
additional scalar field $\phi $ as in (\ref{einstgrscact}). Here it should
be emphasized that the action of type (\ref{tmtact}) is an example of action
for two measure theories, TMT, see refs. \cite{eghnsr,srsv2}.

An alternative interpretation of equations (\ref{riccisol4d}) is to consider
them as a 4--d generalization of the Ricci solitonic equations (\ref%
{friedsoliton}) and (\ref{riccisol3d}). Such a modification is not trivial
even we can find solutions of a system
\begin{equation*}
\overline{\nabla }_{\grave{\imath}}v_{\grave{j}}+\overline{\nabla }_{\grave{j%
}}v_{\grave{\imath}}=\overline{\nabla }_{\grave{\imath}}\phi \ \overline{%
\nabla }_{\grave{j}}\phi ,
\end{equation*}%
where $\overline{\nabla }_{\mu }=(\overline{\nabla }_{\grave{\imath}},%
\overline{\nabla }_{4}),$ for a 3-d $v_{\grave{j}}$ extended to 4-d in order
to define a scalar field $\phi (u^{\nu }).$ For instance, we can take a
gradient vector field $v_{\grave{\imath}}=\frac{1}{2}$ $\phi \overline{%
\nabla }_{\grave{\imath}}\phi $

We have to perform a relativistic 4-d generalization for metrics of
signature $(+++-)$ of functionals (\ref{fpfunct}), or (\ref{wpentr}) which
result in geometric evolution equations which are not of parabolic type but
correspond to a new class of physically important equations. For stationary
configurations (we can consider over-lined or other type values),
\begin{eqnarray}
R_{\alpha \beta }-\lambda g_{\alpha \beta } &=&\nabla _{\alpha }v_{\beta
}+\nabla _{\beta }v_{\alpha }  \label{4dprflows} \\
&=&\overline{\nabla }_{\alpha }\phi \ \overline{\nabla }_{\beta }\phi +\frac{%
1}{2}(\overline{\nabla }_{\alpha }\ \overline{\nabla }_{\beta }+\ \overline{%
\nabla }_{\beta }\overline{\nabla }_{\alpha })\phi .  \notag
\end{eqnarray}%
For gradient Ricci solitons, the term $\frac{1}{2}g^{\alpha \beta }(%
\overline{\nabla }_{\alpha }\ \overline{\nabla }_{\beta }+\ \overline{\nabla
}_{\beta }\overline{\nabla }_{\alpha })\phi $ does not contribute in the
action (\ref{einstgrscact}) if there are satisfied the equations (\ref{aux2}%
). In general, terms of type $\nabla _{\alpha }v_{\beta }+\nabla _{\beta
}v_{\alpha }$ are contained in some classes of MGTs, for instance, in Ho\v{r}%
ava--Lifshitz gravity \cite{hl1,hl2} and corresponding Ricci flow
anisotropic cosmological models \cite{bakaslust}.

We conclude this subsection with the remark that the gravitational field
equations in the $R^{2},$ TMT, and Einstein gravity theories determine
self--similar fixed point configurations (with corresponding Ricci soliton
equations) of a relativistic 4-d generalization of standard Ricci flow
models of 3-d Riemannian metrics. In certain sense, a large class of MGTs
can be reproduced via (nonholonomic) geometric flow evolution scenarios,
when modified Perelman's functionals include (as Ricci solitons) various
types of actions for MGTs. Such constructions have a rigorous mathematical
and physical motivation if we are able to construct in explicit form certain
classes of exact solutions for geometric flow scenarios which model
important black hole, cosmological and other type solutions in $R^{2}$ and
Einstein gravity.

\vskip5pt

The paper is structured as follows. Section \ref{s2} is devoted to
generalizations and formulation of Perelman's functionals including $R^2$
gravity in nonholonomic variables. We provide an introduction into the
geometry of Lorentz manifolds with nonholonomic 2+2 splitting and adapted
physical/ geometric objects. There are derived the equations for
relativistic geometric flows and generic off--diagonal Ricci solitons. The
concepts of W--entropy and statistical thermodynamics are revised in the
context of generalizations for nonholonomic relativistic geometric flows and
MGTs.

In section \ref{s3}, we provide an introduction to the AFDM as a geometric
method for constructing exact solutions with coefficients of metrics and
generalized connections depending on all spacetime coordinates. We develop
this method for modelling relativistic flows, modified Ricci solitons and
generic off--diagonal configurations in $R^{2}$ gravity. It is proved the
decoupling property of such systems of nonlinear partial differential
equations, PDEs, which allows us to integrate such systems in very general
forms. There are considered four classes of such solutions for 1) geometric
flows of metric coefficients with non-evolution of nonlinear connection
coefficients; 2) nonholonomic Ricci soliton equations; 3) geometric
evolution models with running physical constants and/or deformed horizons;
geometric flows with nonholonomic vacuum. There are studied the equations
for geometric evolution and generating Ricci solitons for Levi-Civita
configurations.

In section \ref{s4}, we construct exact solutions for geometric evolution of
black ellipsoids as Ricci solitons and/or solutions in $R^2$ gravity. There
are analyzed two classes of generic off--diagonal metrics describing 1)
solitonic black ellipsoids and limits to $R^2$ and Einstein gravity theories
and 2) geometric flows and solitons for asymptotically de Sitter solutions.

Section \ref{s5} is devoted to formulation of W--thermodynamics for black
ellipsoids and solitonic flows in $R^{2}$ gravity. We show how such values
can defined for 3--d hypersurface configurations and computed in general
form for generic off--diagonal solutions describing geometric flows and
Ricci solitons, or $R^{2}$ gravity and (modified) Einstein equations.

Conclusions are formulated in section \ref{s6}. We provide some necessary
coefficient formulas in Appendix.

\section{Perelman's Functionals \& MGTs in Nonholonomic Variables}

\label{s2} For elaborating a geometric method of constructing exact
solutions for geometric flow, Ricci soliton and gravitational field
equations, it is important to formulate such theories in nonholonomic
variables with nonlinear connection splitting. A nonholonomic 3+1 splitting
is convenient for relativistic generalizations of the Hamilton-Pereman
theory, see details in \cite{vvrfthbh}. In another turn, nonholonomic 2+2
splitting is important for decoupling (modified) geometric evolution and
gravity equations \cite{sv2001,sv2014,svvvey,tgovsv}. In general, we can
work with models of geometric evolution of certain classes of exact
solutions in gravity theories by considering double 3+1 and 2+2 splitting.
In this section, we generalize/ modify Perelman's functionals for
pseudo--Riemannian signatures and in nonholonomic variables and prove the
main evolution and gravitational field equations for nonholonomic 2+2
splitting and canonical distortions of linear connection structures.

\subsection{Nonholonomic $2+2+...$ splitting and adapted geometric objects}

We consider a (pseudo) Riemannian manifold $\mathbf{V}$ enabled with a
conventional 2+2 splitting into horizontal (h) and vertical (v) components
defined by a Whitney sum
\begin{equation}
\mathbf{N}:\ T\mathbf{V}=h\mathbf{\mathbf{V\oplus }}v\mathbf{V},
\label{ncon}
\end{equation}%
where $T\mathbf{V}$ is the tangent bundle. \ A N--connection structure (\ref%
{ncon}) is determined locally by a corresponding set of coefficients $%
N_{i}^{a},$ when $\mathbf{N}=N_{i}^{a}(u)dx^{i}\otimes \partial _{a}.$%
\footnote{%
The local coordinates are labelled $u^{\mu }=(x^{i},y^{a}),$ (in brief, we $%
u=(x,y)$), where indices run respectively values of type $i,j,...=1,2$ and $%
a,b,...=3,4.$ The cumulative small Greek indices run values $\alpha ,\beta
,...=1,2,3,4,$ where $u^{4}=y^{4}=t$ is a time like coordinate. An arbitrary
local basis is denoted $e^{\alpha }=(e^{i},e^{a})$ and the corresponding
dual one, co-basis, is $e_{\beta }=(e_{j},e_{b}).$ We consider that there
are always nontrivial frame transforms to corresponding coordinate bases, $%
\partial _{\alpha ^{\prime }}=(\partial _{i^{\prime }},\partial _{a^{\prime
}})$ [for instance, $\partial _{i^{\prime }}=\partial /\partial x^{i^{\prime
}}],$ and cobasis. The values $e_{\beta }=A_{\beta }^{\ \beta ^{\prime
}}(u)\partial _{\beta ^{\prime }}$ and $e^{\alpha }=A_{\ \alpha ^{\prime
}}^{\alpha }(u)du^{\alpha ^{\prime }},$ for $du^{\alpha ^{\prime
}}=(dx^{i^{\prime }},dy^{a^{\prime }}),$ can be treated as frame (vierbein)
transforms. On convenience, we shall use primed, underlined indices etc. The
Einstein summation rule on repeating up--low indices will be applied if the
contrary will be not stated.}

For any h-v--splitting, there are structures of N--adapted local bases, $%
\mathbf{e}_{\nu }=(\mathbf{e}_{i},e_{a}),$ and cobases, $\mathbf{e}^{\mu
}=(e^{i},\mathbf{e}^{a}),$ when
\begin{eqnarray}
\mathbf{e}_{\nu } &=&(\mathbf{e}_{i}=\partial /\partial x^{i}-\
N_{i}^{a}\partial /\partial y^{a},\ e_{a}=\partial _{a}=\partial /\partial
y^{a}),  \label{nader} \\
\mathbf{e}^{\mu } &=&(e^{i}=dx^{i},\mathbf{e}^{a}=dy^{a}+\ N_{i}^{a}dx^{i}).
\label{nadif}
\end{eqnarray}%
Such N--adapted bases are nonholonomic because, in general, there are
satisfied relations of type
\begin{equation}
\lbrack \mathbf{e}_{\alpha },\mathbf{e}_{\beta }]=\mathbf{e}_{\alpha }%
\mathbf{e}_{\beta }-\mathbf{e}_{\beta }\mathbf{e}_{\alpha }=W_{\alpha \beta
}^{\gamma }\mathbf{e}_{\gamma },  \label{anhcoef}
\end{equation}%
with nontrivial anholonomy coefficients $W_{ia}^{b}=\partial
_{a}N_{i}^{b},W_{ji}^{a}=\Omega _{ij}^{a}=\mathbf{e}_{j}\left(
N_{i}^{a}\right) -\mathbf{e}_{i}(N_{j}^{a}).$ We obtain holonomic
(integrable) bases if and only if $W_{\alpha \beta }^{\gamma }=0.$\footnote{%
We can elaborate a N--adapted differential and integral calculus and a
corresponding variational formalism for (modified) gravity theories using
N--elongated operators (\ref{nader}) and (\ref{nadif}). The geometric
constructions are performed with distinguished objects, in brief, d--objects
with coefficients determined with respect to N--adapted (co) frames and
their tensor products. For instance, a vector $Y(u)\in T\mathbf{V}$ can be
parameterized as a d--vector, $\mathbf{Y}=$ $\mathbf{Y}^{\alpha }\mathbf{e}%
_{\alpha }=\mathbf{Y}^{i}\mathbf{e}_{i}+\mathbf{Y}^{a}e_{a},$ or $\mathbf{Y}%
=(hY,vY),$ with $hY=\{\mathbf{Y}^{i}\}$ and $vY=\{\mathbf{Y}^{a}\}.$
Similarly, we can determine and compute the coefficients of d--tensors,
N--adapted differential forms, d--connections, d--spinors etc.} A manifold $(%
\mathbf{V,N})$ endowed with a nontrivial structure $W_{\alpha \beta
}^{\gamma }$ is called nonholonomic.

For nonholonomic manifolds, we can consider a class of linear connections
which are adapted to the N--connection structure. A distinguished
connection, d--connection, $\mathbf{D}=(h\mathbf{D},v\mathbf{D})$ on $%
\mathbf{V}$ is such a linear connection which preserves under parallel
transport the N--connection splitting (\ref{ncon}). A general linear
connection $D$ is not adapted to a chosen $h$-$v$--decomposition, i.e. it is
not a d--connection (for instance, the Levi--Civita connection in GR is not
a d--connection). We do not use boldface symbols for not N--adapted
geometric objects. For any d--connection $\mathbf{D,}$ we can consider as an
operator of covariant derivative, $\mathbf{D}_{\mathbf{X}}\mathbf{Y}$, for a
d--vector $\mathbf{Y}$ in the direction of a d--vector $\mathbf{X}.$ With
respect to N--adapted frames (\ref{nader}) and (\ref{nadif}), we can compute
N--adapted coefficients for $\mathbf{D}=\{\mathbf{\Gamma }_{\ \alpha \beta
}^{\gamma }=(L_{jk}^{i},L_{bk}^{a},C_{jc}^{i},C_{bc}^{a})\},$ see details
and explicit formulas in Refs. \cite{sv2001,sv2014,svvvey,tgovsv}. The
coefficients $\mathbf{\Gamma }_{\ \alpha \beta }^{\gamma }$ are computed
geometrically for the h--v--components of $\mathbf{D}_{\mathbf{e}_{\alpha }}%
\mathbf{e}_{\beta }:=$ $\mathbf{D}_{\alpha }\mathbf{e}_{\beta }$ using $%
\mathbf{X}=\mathbf{e}_{\alpha }$ and $\mathbf{Y}=\mathbf{e}_{\beta }.$

The d--torsion, $\mathbf{T,}$ the nonmetricity, $\mathbf{Q},$ and the
d--curvature, $\mathbf{R},$ tensors (for N--adapted constructions, it is
used the term d--tensor) are defined in standard form. For any d--connection
$\mathbf{D}$ and d--vectors $\mathbf{X,Y\in }T\mathbf{V,}$
\begin{eqnarray*}
\mathbf{T}(\mathbf{X,Y}):= &&\mathbf{D}_{\mathbf{X}}\mathbf{Y}-\mathbf{D}_{%
\mathbf{Y}}\mathbf{X}-[\mathbf{X,Y}],\ \mathbf{Q}(\mathbf{X}):=\mathbf{D}_{%
\mathbf{X}}\mathbf{g,} \\
\mathbf{R}(\mathbf{X,Y}):= &&\mathbf{D}_{\mathbf{X}}\mathbf{D}_{\mathbf{Y}}-%
\mathbf{D}_{\mathbf{Y}}\mathbf{D}_{\mathbf{X}}-\mathbf{D}_{\mathbf{[X,Y]}}.
\end{eqnarray*}%
The N--adapted coefficients are correspondingly labeled using $h$- and $v$%
--indices,
\begin{eqnarray}
\mathbf{T} &=&\{\mathbf{T}_{\ \alpha \beta }^{\gamma }=\left( T_{\
jk}^{i},T_{\ ja}^{i},T_{\ ji}^{a},T_{\ bi}^{a},T_{\ bc}^{a}\right) \},%
\mathbf{Q}=\mathbf{\{Q}_{\ \alpha \beta }^{\gamma }\},  \label{rnmc} \\
\mathbf{R} &\mathbf{=}&\mathbf{\{R}_{\ \beta \gamma \delta }^{\alpha }%
\mathbf{=}\left( R_{\ hjk}^{i}\mathbf{,}R_{\ bjk}^{a}\mathbf{,}R_{\ hja}^{i}%
\mathbf{,}R_{\ bja}^{c},R_{\ hba}^{i},R_{\ bea}^{c}\right) \},  \notag
\end{eqnarray}%
see explicit formulas in \cite{sv2001,sv2014,svvvey,tgovsv}. In Appendix, we
provide such formulas for the case of the canonical distinguished connection
(defined in (\ref{lcconcdcon})), see (\ref{dtors}) and (\ref{dcurv}).

Any metric tensor $\mathbf{g}$ on $\left( \mathbf{V,N}\right) $ can be
parameterized in
\begin{equation}
\mathbf{g}=\underline{g}_{\alpha \beta }du^{\alpha }\otimes du^{\beta },%
\mbox{\ where \ }\underline{g}_{\alpha \beta }=\left[
\begin{array}{cc}
g_{ij}+N_{i}^{a}N_{j}^{b}g_{ab} & N_{j}^{e}g_{ae} \\
N_{i}^{e}g_{be} & g_{ab}%
\end{array}%
\right] ,  \label{ofdans}
\end{equation}%
with respect to a dual local coordinate basis $du^{\alpha }.$ Equivalently,
we can write a metric as a d--tensor (d--metric)
\begin{equation}
\mathbf{g}=g_{\alpha }(u)\mathbf{e}^{\alpha }\otimes \mathbf{e}^{\beta
}=g_{i}(x)dx^{i}\otimes dx^{i}+g_{a}(x,y)\mathbf{e}^{a}\otimes \mathbf{e}%
^{a},  \label{dm1}
\end{equation}%
in brief, $\mathbf{g}=(h\mathbf{g},v\mathbf{g}),$ with respect to a tensor
product of N--adapted dual frame (\ref{nadif}). A metric $\mathbf{g}$ (\ref%
{ofdans}) with N--coefficients $N_{j}^{e}$ is generic off--diagonal if the
anholonomy coefficients $W_{\alpha \beta }^{\gamma }$ (\ref{anhcoef}) are
not zero.

For any d--metric $\mathbf{g,}$ we can define two important linear
connection structures following such geometric conditions: {%
\begin{equation}
\mathbf{g}\rightarrow \left\{
\begin{array}{ccccc}
\nabla : &  & \nabla \mathbf{g}=0;\ ^{\nabla }\mathbf{T}=0, &  &
\mbox{ the
Levi--Civita connection;} \\
\widehat{\mathbf{D}}: &  & \widehat{\mathbf{D}}\ \mathbf{g}=0;\ h\widehat{%
\mathbf{T}}=0,\ v\widehat{\mathbf{T}}=0, &  &
\mbox{ the canonical
d--connection.}%
\end{array}%
\right.  \label{lcconcdcon}
\end{equation}%
The LC--connection }$\nabla $ can be introduced without any N--connection
structure but the canonical d--connection $\widehat{\mathbf{D}}$ depends
generically on a prescribed nonholonomic $h$- and $v$-splitting, see
formulas (\ref{candcon}). In above formulas, $h\widehat{\mathbf{T}}$ and $\ v%
\widehat{\mathbf{T}}$ are respective torsion components which vanish on
conventional h- and v--subspaces. Nevertheless, there are nonzero torsion
components, $hv\widehat{\mathbf{T}}$, (see coefficients $\widehat{T}_{\
ja}^{i},\widehat{T}_{aj}^{c}$ and $\widehat{T}_{\ ji}^{a}$ in (\ref{dtors}))
with nonzero mixed indices with respect to a N-adapted basis (\ref{nader})
and/or (\ref{nadif}).

On $\mathbf{V,}$ all geometric constructions can be performed equivalently
with $\nabla $ and/or $\widehat{\mathbf{D}}$ and related via a canonical
distorting relation
\begin{equation}
\widehat{\mathbf{D}}\mathbf{[g,N]}=\nabla \lbrack \mathbf{g}]+\widehat{%
\mathbf{Z}}\mathbf{[g,N]},  \label{distr}
\end{equation}%
when both linear connections and the distorting tensor $\widehat{\mathbf{Z}}$
are uniquely determined by data $(\mathbf{g,N)}$ as an algebraic combination
of coefficients of $\widehat{\mathbf{T}}_{\ \alpha \beta }^{\gamma }$. The
N--adapted coefficients for $\widehat{\mathbf{D}}$ and corresponding
torsion, $\widehat{\mathbf{T}}_{\ \alpha \beta }^{\gamma }$ (\ref{dtors}),
Ricci d--tensor, $\widehat{\mathbf{R}}_{\ \beta \gamma }$ (\ref{driccic}),
and Einstein d--tensor, $\widehat{\mathbf{E}}_{\ \beta \gamma }$ (\ref%
{enstdt}), can be computed in standard form \cite%
{sv2001,sv2014,svvvey,tgovsv}. The canonical distortion relation (\ref{distr}%
) defines respective distortion relations of the Riemiann, Ricci and
Einstein tensors and respective curvature scalars which are uniquely
determined by data $(\mathbf{g,N).}$ Any (pseudo) Riemannian geometry can be
equivalently formulated using $(\mathbf{g,\nabla )}$ or $(\mathbf{g},%
\widehat{\mathbf{D}}).$

The canonical d--connection $\widehat{\mathbf{D}}$ has a very important role
in elaborating the AFDM of constructing exact solutions in geometric flows
and MGTs. It allows to decouple the gravitational and matter field equations
with respect to N--adapted frames of reference. This is not possible if we
work only with $\nabla .$ Constructing certain general classes of solutions
for $\widehat{\mathbf{D}}$, we can impose at the end the condition $\widehat{%
\mathbf{T}}=0$ and extract LC--configurations $\widehat{\mathbf{D}}_{\mid
\widehat{\mathbf{T}}=0}=\nabla .$

\subsection{Relativistic nonholonomic geometric flows}

We can consider a family of 4--d Lorentz nonholonomic manifolds $\mathbf{V}%
(\tau )$ with metrics $\mathbf{g}(\tau )=\mathbf{g}(\tau ,u)$ of signature $%
(+++-)$ and N--connection $\mathbf{N}(\tau )$ parameterized by a positive
parameter $\tau ,0\leq \tau \leq \tau _{0}.$ Any $\mathbf{V}$ $\subset
\mathbf{V}(\tau )$ can be enabled with a double nonholonomic 2+2 and 3+1
splitting (see \cite{vvrfthbh} for the geometry of spacetimes enabled with
such double distributions). Conventionally, the local coordinates are
labeled $u^{\alpha }=(x^{i},y^{a})=($ $x^{\grave{\imath}},u^{4}=t)$ for $%
i,j,k,...=1,2;$ $a,b,c,...=3,4;$ and $\grave{\imath},\grave{j},\grave{k}%
=1,2,3.$ The 3+1 splitting can be chosen in such form that any open region $%
U\subset $ $\mathbf{V}$ is covered by a family of 3-d spacelike
hypersurfaces $\widehat{\Xi }_{t}$ parameterized by a time like parameter $%
t. $ There are two generic different types of geometric flow theories when
1) $\tau (\chi )$ is a re-parametrization of the temperature like parameter
used for labeling 4-d Lorentz spacetime configurations and 2) $\tau (t)$ is
a time like parameter when 3-d spacelike configurations evolve
relativistically on a "redefined" time like coordinate. For simplicity, we
shall study in this work only theories of type 1.

For arbitrary frame transforms on 4-d nonholonomic Lorentz manifolds, we can
generalize the Perelman's functionals (\ref{fpfunct}) and (\ref{wpentr}),
respectively, in terms of data $(\mathbf{g}(\tau ),\widehat{\mathbf{D}}(\tau
)).$ We postulate
\begin{eqnarray}
\widehat{\mathcal{F}} &=&\int_{t_{1}}^{t_{2}}\int_{\widehat{\Xi }_{t}}e^{-%
\widehat{f}}\sqrt{|\mathbf{g}_{\alpha \beta }|}d^{4}u(\widehat{R}+|\widehat{%
\mathbf{D}}\widehat{f}|^{2}),  \label{ffperelm4} \\
&&\mbox{ and }  \notag \\
\widehat{\mathcal{W}} &=&\int_{t_{1}}^{t_{2}}\int_{\widehat{\Xi }_{t}}%
\widehat{M}\sqrt{|\mathbf{g}_{\alpha \beta }|}d^{4}u[\tau (\widehat{R}+|\
^{h}\widehat{D}\widehat{f}|+|\ ^{v}\widehat{D}\widehat{f}|)^{2}+\widehat{f}%
-8],  \label{wfperelm4}
\end{eqnarray}%
where the normalizing function $\widehat{f}(\tau ,u)$ satisfies $%
\int_{t_{1}}^{t_{2}}\int_{\widehat{\Xi }_{t}}\widehat{\mu }\sqrt{|\mathbf{g}%
_{\alpha \beta }|}d^{4}u=1$ for $\widehat{\mu }=\left( 4\pi \tau \right)
^{-3}e^{-\widehat{f}},$ see formula (\ref{sdcurv}) for the "hat" scalar
curvature. It should be noted that $\widehat{\mathcal{W}}$ (\ref{wfperelm4})
do not have a character of entropy for pseudo--Riemannian metrics but can be
treated as a value characterizing relativistic geometric hydrodynamic flows
\cite{vvrfthbh}. We can compute entropy like values of type (\ref{wpentr})
for any 3+1 splitting with hypersurface fibrations $\widehat{\Xi }_{t}.$

For 4-d configurations with a corresponding re--definition of the scaling
function, $\widehat{f}\rightarrow f,$ we can construct models of geometric
evolution with $h$-- and $v$--splitting for $\mathbf{\widehat{\mathbf{D}},}$
\begin{eqnarray}
\partial _{\tau }g_{ij} &=&-2(\widehat{R}_{ij}-\widehat{\mathbf{D}}_{i}\phi
\widehat{\mathbf{D}}_{j}\phi -2\varsigma ^{2}g_{ij}),\   \label{rfcandch} \\
\partial _{\tau }g_{ab} &=&-2(\widehat{R}_{ab}-\widehat{\mathbf{D}}_{a}\phi
\widehat{\mathbf{D}}_{b}\phi -2\varsigma ^{2}g_{ab}),  \label{rfcandcv} \\
\widehat{R}_{ia} &=&\widehat{R}_{ai}=0;\widehat{R}_{ij}=\widehat{R}_{ji};%
\widehat{R}_{ab}=\widehat{R}_{ba};  \notag \\
\widehat{\mathbf{D}}^{2}\phi &=&0;  \notag \\
\partial _{\tau }f &=&-\widehat{\square }f+\left\vert \widehat{\mathbf{D}}%
f\right\vert ^{2}-\ ^{h}\widehat{R}-\ ^{v}\widehat{R}.  \notag
\end{eqnarray}%
These formulas can be derived from the functionals (\ref{ffperelm4}) and/or (%
\ref{wfperelm4}) following a calculus which is similar to that presented in
N--adapted form in Ref. \cite{vnhrf,vnrflnc}. In abstract geometric form, we
can apply the strategy elaborated originally for such proofs in \cite%
{perelman1} for metric compatible connections. The conditions $\widehat{R}%
_{ia}=0$ and $\widehat{R}_{ai}=0$ are necessary if we wont to keep the total
metric to be symmetric under Ricci flow evolution. The general relativistic
character of 4-d geometric flow evolution is encoded in operators like $%
\widehat{\square }=\widehat{\mathbf{D}}^{\alpha }\widehat{\mathbf{D}}%
_{\alpha },$ d-tensor components $\widehat{R}_{ij}$ and $\widehat{R}_{ab},$
theirs scalars $\ ^{h}\widehat{R}=g^{ij}\widehat{R}_{ij}$ and $\ ^{v}%
\widehat{R}=g^{ab}\widehat{R}_{ab}$ with data $(g_{ij},g_{ab},\widehat{%
\mathbf{D}}_{\alpha }).$

In order to study geometric flow evolution of solutions of $R^{2}$ and
equivalent Einstein -- scalar field theories, the $\ \widehat{\mathcal{F}}$
\ and $\widehat{\mathcal{W}}$ \ functionals can be written
\begin{eqnarray*}
\widehat{\mathcal{F}} &=&\int_{t_{1}}^{t_{2}}\int_{\widehat{\Xi }_{t}}e^{-%
\overline{f}}\sqrt{|\mathbf{g}_{\alpha \beta }|}d^{4}u(\widehat{R}-(\widehat{%
\mathbf{D}}\phi )^{2}-8\varsigma ^{2}+|\widehat{\mathbf{D}}\overline{f}%
|^{2}), \\
&&\mbox{ and } \\
\widehat{\mathcal{W}} &=&\int_{t_{1}}^{t_{2}}\int_{\widehat{\Xi }_{t}}%
\overline{\mu }\sqrt{|\mathbf{g}_{\alpha \beta }|}d^{4}u[\tau (\widehat{R}-(%
\widehat{\mathbf{D}}\phi )^{2}-8\varsigma ^{2}+|\ ^{h}\widehat{D}\overline{f}%
|+|\ ^{v}\widehat{D}\overline{f}|)^{2}+\overline{f}-8],
\end{eqnarray*}%
for $\widehat{f}\rightarrow \overline{f},$ where $\overline{f}(\tau ,u)$
satisfies $\int_{t_{1}}^{t_{2}}\int_{\widehat{\Xi }_{t}}\overline{\mu }\sqrt{%
|\overline{\mathbf{g}}_{\alpha \beta }|}d^{4}u=1$ for $\overline{\mu }%
=\left( 4\pi \tau \right) ^{-3}e^{-\overline{f}}.$ These formulas are
important for investigation of non--stationary configurations of
gravitational fields. We can consider $\widehat{\mathbf{D}}_{\mid \widehat{%
\mathbf{T}}=0}=\overline{\nabla },$ where the LC--connection is determined
by the over-lined metric $\overline{g}_{\mu \nu }$ (\ref{overlmetric}).

\subsection{Generic off--diagonal Ricci solitons}

For self--similar fixed point $\tau =\tau _{0}$ configurations, the
equations (\ref{rfcandch}) and (\ref{rfcandcv}) transform into relativistic
Ricci soliton equations, respectively, into a system of nonholonomically
deformed Einstein -- scalar field equations%
\begin{eqnarray}
\ _{\flat }\widehat{R}_{ij} &=&\ \ _{\flat }^{\phi }\Upsilon
_{ij}+2\varsigma ^{2}\ _{\flat }g_{ij}  \label{rsolit1} \\
\ _{\flat }\widehat{R}_{ab} &=&\ \ _{\flat }^{\phi }\Upsilon
_{ab}+2\varsigma ^{2}\ _{\flat }g_{ab},  \label{rsolit2} \\
\ _{\flat }\widehat{R}_{ia} &=&0;\ _{\flat }\widehat{R}_{aj}=0;  \notag \\
\ _{\flat }\widehat{\mathbf{D}}^{2}\phi &=&0,  \notag
\end{eqnarray}%
where the corresponding $h$-- and $v$--sources are $\ \ _{\flat }^{\phi
}\Upsilon _{ij}=\ _{\flat }\widehat{\mathbf{D}}_{i}\ _{\flat }\phi \ _{\flat
}\widehat{\mathbf{D}}_{j}\ _{\flat }\phi $ and $\ \ _{\flat }^{\phi
}\Upsilon _{ab}=\ _{\flat }\widehat{\mathbf{D}}_{a}\ _{\flat }\phi \ _{\flat
}\widehat{\mathbf{D}}_{b}\ _{\flat }\phi .$ In this work, we put the left
low label $\flat $ for necessary values in order to emphasize that such
geometric/ physical objects are computed for certain Ricci soliton
configurations with $\tau =\tau _{0}.$ Such labels will be omitted if that
will not result in ambiguties.

Using N--adapted 2+2 frame and coordinate transforms of the metric and
source $\ ^{\phi }\Upsilon _{\alpha \beta },$
\begin{eqnarray*}
\mathbf{g}_{\alpha \beta }(\tau ,x^{i},t) &=&e_{\ \alpha }^{\alpha ^{\prime
}}(\tau ,x^{i},y^{a})e_{\ \beta }^{\beta ^{\prime }}(\tau ,x^{i},y^{a})%
\widehat{\mathbf{g}}_{\alpha ^{\prime }\beta ^{\prime }}(\tau ,x^{i},y^{a})%
\mbox{ and } \\
\ ^{\phi }\Upsilon _{\alpha \beta }(\tau ,x^{i},t) &=&e_{\ \alpha }^{\alpha
^{\prime }}(\tau ,x^{i},y^{a})e_{\ \beta }^{\beta ^{\prime }}(\tau
,x^{i},y^{a})\widehat{\Upsilon }_{\alpha ^{\prime }\beta ^{\prime }}(\tau
,x^{i},y^{a}),
\end{eqnarray*}%
for a time like coordinate $y^{4}=t$ ($i^{\prime },i,k,k^{\prime },...=1,2$
and $a,a^{\prime },b,b^{\prime },...=3,4),$ we introduce certain canonical
parameterizations which will allow us to decouple and solve the system the
parameterize the metric and effective source in certain adapted forms. The
generic off--diagonal metric ansatz is taken in the form
\begin{eqnarray}
\mathbf{g} &=&\mathbf{g}_{\alpha ^{\prime }\beta ^{\prime }}\mathbf{e}%
^{\alpha ^{\prime }}\otimes \mathbf{e}^{\beta ^{\prime }}=g_{i}(\tau
,x^{k})dx^{i}\otimes dx^{j}+\omega ^{2}(\tau ,x^{k},y^{3},t)h_{a}(\tau
,x^{k},y^{3})\mathbf{e}^{a}\otimes \mathbf{e}^{a}  \label{offdans} \\
&=&q_{i}(\tau ,x^{k})dx^{i}\otimes dx^{i}+q_{3}(\tau ,x^{k},y^{3},t)\mathbf{e%
}^{3}\otimes \mathbf{e}^{3}-\breve{N}^{2}(\tau ,x^{k},y^{3},t)\mathbf{e}%
^{4}\otimes \mathbf{e}^{4},  \label{lapsnonh} \\
\mathbf{e}^{3} &=&dy^{3}+w_{i}(\tau ,x^{k},y^{3})dx^{i},\mathbf{e}%
^{4}=dt+n_{i}(\tau ,x^{k},y^{3})dx^{i}.  \notag
\end{eqnarray}%
This ansatz is a general one for a 4--d metric which can be written as an
extension of a 3--d metric $q_{ij}=diag(q_{\grave{\imath}})=(q_{i},q_{3})$
on a hypersurface $\widehat{\Xi }_{t}$ \ if
\begin{equation}
q_{3}=g_{3}=\omega ^{2}h_{3}\mbox{ and }\breve{N}^{2}(\tau ,u)=-\omega
^{2}h_{4}=-g_{4},  \label{shift1}
\end{equation}%
where $\breve{N}(\tau ,u)$ is the lapse function. It allows a
straightforward extension of 3--d ansatz to 4-d configurations by
introducing the values $\breve{N}^{2}(\tau ,x^{k},y^{3})$ and $w_{i}(\tau
,x^{k},y^{3})$ in order to generate exact solutions of the Ricci soliton
and/or (modified) Einstein equations.

The nontrivial N--connection and d-metric coefficients with running
parameter $\tau $ are denoted
\begin{eqnarray}
\ N_{i}^{3}(\tau ) &=&w_{i}(\tau ,x^{k},y^{3});N_{i}^{4}=n_{i}(\tau
,x^{k},y^{3})\mbox{ and }  \label{coefft} \\
g_{i^{\prime }j^{\prime }}(\tau )
&=&diag[g_{i}],g_{1}=g_{2}=q_{1}=q_{2}=e^{\psi (\tau ,x^{k})};\   \notag \\
g_{a^{\prime }b^{\prime }}(\tau ) &=&diag[\omega ^{2}h_{a}],h_{a}=h_{a}(\tau
,x^{k},y^{3}),q_{3}=\omega ^{2}h_{3},\breve{N}^{2}=\breve{N}^{2}(\tau
,x^{k},y^{3},t).  \notag
\end{eqnarray}%
We shall be able to construct exact solutions in explicit form if the source
will be parameterized with respect to N--adapted frames as
\begin{equation}
\widehat{\Upsilon }_{\alpha \beta }=diag[\Upsilon _{i}(\tau );\Upsilon
_{a}(\tau )],\mbox{ for }\Upsilon _{1}(\tau )=\Upsilon _{2}(\tau )=%
\widetilde{\Upsilon }(\tau ,x^{k}),\Upsilon _{3}(\tau )=\Upsilon _{4}(\tau
)=\Upsilon (\tau ,x^{k},y^{3}).  \label{sourc2}
\end{equation}%
The ansatz (\ref{offdans}) determines 4-d generalizations of 3-d
hypersurface metrics as it is given by (\ref{lapsnonh}) for a nontrivial
lapse function (\ref{shift1}). The N--adapted coefficients (\ref{sourc2})
can be very general ones. To be able to construct solutions in explicit
forms, we suppose that the scalar field can be described with respect to
N--adapted frames when the exact solutions for $\omega =1$ are with Killing
symmetry on $\partial /\partial t.$ For such configurations, there are
N--adapted bases when the geometric and physical values do not depend on
coordinate $y^{4}=t.$ We can work for simplicity with solutions with one
Killing symmetry. Technically, it is possible to construct very general
classes of generic off--diagonal solutions depending on all spacetime
variables, see details and examples in Refs. \cite%
{sv2001,sv2014,svvvey,tgovsv} for "non--Killing" configurations.

\subsection{W--entropy and statistical thermodynamics for geometric flows
and MGTs}

Let us denote by$\ _{\shortmid }\widehat{\mathbf{D}}=\widehat{\mathbf{D}}%
_{\mid \widehat{\Xi }_{t}}$ the canonical d--connection $\widehat{\mathbf{D}}
$ defined on a 3-d hypersurface $\widehat{\Xi }_{t},$ when all values depend
on temperature like parameter $\tau (\chi ).$ We define also $\ _{\shortmid }%
\widehat{R}:=$ $\widehat{R}_{\mid \widehat{\Xi }_{t}}.$ Using $(q_{\grave{%
\imath}})=(q_{i},q_{3}),$ the Perelman's functionals parameterized in
N--adapted form are constructed%
\begin{eqnarray}
\ _{\shortmid }\widehat{\mathcal{F}} &=&\int_{\widehat{\Xi }_{t}}e^{-f}\sqrt{%
|q_{\grave{\imath}\grave{j}}|}d\grave{x}^{3}(\ _{\shortmid }\widehat{R}-(\
_{\shortmid }\widehat{\mathbf{D}}\phi )^{2}-8\varsigma ^{2}+|\ _{\shortmid }%
\widehat{\mathbf{D}}f|^{2}),  \label{perelm3f} \\
&&\mbox{ and }  \notag \\
\ _{\shortmid }\widehat{\mathcal{W}} &=&\int_{\widehat{\Xi }_{t}}\mu \sqrt{%
|q_{\grave{\imath}\grave{j}}|}d\grave{x}^{3}[\tau (\ _{\shortmid }\widehat{R}%
-(\ _{\shortmid }\widehat{\mathbf{D}}\phi )^{2}-8\varsigma ^{2}+|\ \
_{\shortmid }^{h}\widehat{\mathbf{D}}f|+|\ \ _{\shortmid }^{v}\widehat{%
\mathbf{D}}f|)^{2}+f-6],  \label{perelm3w}
\end{eqnarray}%
where we chose a necessary type scaling function $f$ which satisfies $\int_{%
\widehat{\Xi }_{t}}\mu \sqrt{|q_{\grave{\imath}\grave{j}}|}d\grave{x}^{3}=1$
for $\mu =\left( 4\pi \tau \right) ^{-3}e^{-f}$ $.$ These functionals \
transform into standard Perelman functionals for 3-d Riemannian metrics on $%
\widehat{\Xi }_{t}$ if $\ _{\shortmid }\widehat{\mathbf{D}}\rightarrow \
_{\shortmid }\nabla .$

For any closed hypersurface $\widehat{\Xi }_{t},$ the W--entropy $\
_{\shortmid }\widehat{\mathcal{W}}$ \ is a Lyapunov type non--decreasing
functional which is analogous to minus entropy. We can formulate a
statistical thermodynamics model associated to (non) holonomic modified
Ricci flow evolution of metrics with local Euclidean structure and
nonholonomically modified connections. The constructions in statistical
thermodynamics begin with a partition function
\begin{equation*}
\breve{Z}=\exp \left\{ \int_{\widehat{\Xi }_{t}}\mu \sqrt{|q_{\grave{\imath}%
\grave{j}}|}d\grave{x}^{3}[-\breve{f}+n]~\right\}
\end{equation*}%
for the conditions stated for definition of (\ref{perelm3f}) \ and (\ref%
{perelm3w}). This allows us to compute main thermodynamical values for the
Levi--Civita connection $\ _{\shortmid }\nabla $ and $n=3.$\footnote{%
We remember that a statistical model can be elaborated for any prescribed
partition function $Z=\int \exp (-\ell ^{-1}E)d\omega (E)$ considering
canonical ensemble at temperature $\ell .$ The measure is taken to be the
density of states $\omega (E).$ We compute the standard thermodynamical
values for the average energy, $\mathcal{E}=\ \left\langle E\right\rangle
:=-\partial \log Z/\partial (\ell ^{-1}),$ the entropy $S:=\ell
^{-1}\left\langle E\right\rangle +\log Z$ and the fluctuation $\sigma
:=\left\langle \left( E-\left\langle E\right\rangle \right)
^{2}\right\rangle =\partial ^{2}\log Z/\partial (\ell ^{-1})^{2}.$}

To elaborate the analogous thermodynamics constructions in N--adapted form
we consider a family of $q_{\grave{\imath}\grave{j}}(\tau ),$ with $\partial
\tau /\partial \ell ^{-1}=-1,$ being a real re-parametrization of $\ell
^{-1}.$ We compute (see similar details in \cite{vvrfthbh,vnhrf})
\begin{eqnarray}
\ _{\shortmid }\widehat{\mathcal{E}}\ &=&-\tau ^{2}\int_{\widehat{\Xi }%
_{t}}\mu \sqrt{|q_{\grave{\imath}\grave{j}}|}d\grave{x}^{3}\left( \mathbf{\ }%
_{\shortmid }\widehat{R}-(\ _{\shortmid }\widehat{\mathbf{D}}\phi
)^{2}-8\varsigma ^{2}+|\ _{\shortmid }\widehat{\mathbf{D}}\tilde{f}|^{2}-%
\frac{3}{\tilde{\tau}}\right) ,  \label{3dthv} \\
\ _{\shortmid }\widehat{S} &=&-\int_{\widehat{\Xi }_{t}}\mu \sqrt{|q_{\grave{%
\imath}\grave{j}}|}d\grave{x}^{3}\left[ \tau \left( \ \mathbf{\ }_{\shortmid
}\widehat{R}-(\ _{\shortmid }\widehat{\mathbf{D}}\phi )^{2}-8\varsigma
^{2}+|\ _{\shortmid }\widehat{\mathbf{D}}\tilde{f}|^{2}\right) +\tilde{f}-6%
\right] ,  \notag \\
\ _{\shortmid }\widehat{\sigma } &=&2\ \tau ^{4}\int_{\widehat{\Xi }_{t}}\mu
\sqrt{|q_{\grave{\imath}\grave{j}}|}d\grave{x}^{3}[|\ _{\shortmid }\widehat{%
\mathbf{R}}_{\grave{\imath}\grave{j}}-\ _{\shortmid }\widehat{\mathbf{D}}_{%
\grave{\imath}}\phi \ _{\shortmid }\widehat{\mathbf{D}}_{\grave{j}}\phi +\
_{\shortmid }\widehat{\mathbf{D}}_{\grave{\imath}}\ _{\shortmid }\widehat{%
\mathbf{D}}_{\grave{j}}\tilde{f}-(\frac{1}{2\tau }+2\varsigma ^{2})q_{\grave{%
\imath}\grave{j}}|^{2}].  \notag
\end{eqnarray}%
These formulas can be considered for 4--d configurations taking the lapse
function $\breve{N}=1$ for N-adapted Gaussian coordinates. In such cases it
is more difficult to compute in explicit form the values corresponding to
4--d physically important solutions. They characterize stationary exact
solutions in 4-d MGTs and GR if the values $q_{\grave{\imath}\grave{j}},\
_{\shortmid }\widehat{\mathbf{D}}$ and $\phi $ are computed on a closed 3-d
hypersurface $\widehat{\Xi }_{t}.$

The formulas (\ref{3dthv}) provide a thermodynamic characterization of
nonholonomic geometric flows and (modified) gravitational field equations,
in particular, in $R^{2}$ gravity and its equivalent Einstein -- scalar
field formulation. We can re--define the normalization function $\tilde{f}$
\ and flow parameter in such a form with terms
\begin{equation*}
-(\ _{\shortmid }\widehat{\mathbf{D}}\phi )^{2}-8\varsigma ^{2}+|\
_{\shortmid }\widehat{\mathbf{D}}\tilde{f}|^{2}=0\mbox{ and }-\ _{\shortmid }%
\widehat{\mathbf{D}}_{\grave{\imath}}\phi \ _{\shortmid }\widehat{\mathbf{D}}%
_{\grave{j}}\phi +\ _{\shortmid }\widehat{\mathbf{D}}_{\grave{\imath}}\
_{\shortmid }\widehat{\mathbf{D}}_{\grave{j}}\tilde{f}=2\varsigma ^{2}q_{%
\grave{\imath}\grave{j}}.
\end{equation*}%
This is similar to re--definition of the "vacuum" thermodynamical state.

Considering N--connection adapted foliations $\widehat{\Xi }_{t}$
parameterized by a spacetime coordinate $t,$ we can generalize the
constructions for 4--d functionals
\begin{eqnarray}
&&\widehat{\mathcal{F}}(\mathbf{q},\ _{\shortmid }\widehat{\mathbf{D}}%
,f)=\int_{t_{1}}^{t_{2}}dt\ \breve{N}(\tau )\ _{\shortmid }\widehat{\mathcal{%
F}}(\mathbf{q},\ _{\shortmid }\widehat{\mathbf{D}},f),  \label{ffperel} \\
\mbox{ and } &&  \notag \\
&&\widehat{\mathcal{W}}(\mathbf{q},\ _{\shortmid }\widehat{\mathbf{D}}%
,f)=\int_{t_{1}}^{t_{2}}dt\ \breve{N}(\tau )\ \ _{\shortmid }\widehat{%
\mathcal{W}}(\mathbf{q},\ _{\shortmid }\widehat{\mathbf{D}},f(t)),\
\label{wfperel}
\end{eqnarray}%
where the lapse function is taken for an exact solution of certain
nonholonomic Ricci flow/soliton and/or (modified) gravitational field
equations. To elaborate a 4-d general relativistic thermodynamic formulation
is a more difficult task. For instance, we can consider relativistic
hydrodynamical type generalizations, see \cite{vvrfthbh}. Nevertheless, we
can compute the corresponding average energy, entropy and fluctuations for
evolution both on redefined parameter $\tau $ and on a time like coordinate $%
t$ for a time interval from $t_{1}$ to $t_{2}$ of any family of closed \
hypersurfaces all determined by $\ _{\shortmid }\widehat{\mathbf{D}},$
\begin{equation}
\widehat{\mathcal{E}}(\tau )=\int_{t_{1}}^{t_{2}}dt\breve{N}(\tau )\
_{\shortmid }\widehat{\mathcal{E}}(\tau ),\ \widehat{\mathcal{S}}(\tau
)=\int_{t_{1}}^{t_{2}}dt\breve{N}(\tau )\ _{\shortmid }\widehat{S}(\tau ),\
\widehat{\Sigma }(\tau )=\int_{t_{1}}^{t_{2}}dt\breve{N}(\tau )\ _{\shortmid
}\widehat{\sigma }(\tau ).  \label{thermodv}
\end{equation}%
Working with distortion formulas (\ref{distr}), we can compute similar
values in terms of $\ _{\shortmid }\nabla ,$
\begin{equation*}
\ ^{\nabla }\mathcal{E}(\tau )=\int_{t_{1}}^{t_{2}}dt\breve{N}(\tau )\
_{\shortmid }^{\nabla }\mathcal{E}(\tau ),\ ^{\nabla }\mathcal{S}(\tau
)=\int_{t_{1}}^{t_{2}}dt\breve{N}(\tau )\ _{\shortmid }^{\nabla }S(\tau ),\
^{\nabla }{\Sigma }(\tau )=\int_{t_{1}}^{t_{2}}dt\breve{N}(\tau )\
_{\shortmid }^{\nabla }\sigma (\tau ).
\end{equation*}%
We can provide a gravitational thermodynamics interpretation only on 3--d
closed hypersurfaces. For more special classes of solutions, we can model
the standard black hole thermodynamics by considering 2+1+1 splitting and
solutions with horizons and 2-d hypersurface.

\section{A Geometric Method for Generating Solutions for Ricci Flows \& $%
R^{2}$ Gravity}

\label{s3}

We develop a geometric method for integrating 4--d geometric flow and
gravitational field equations in MGTs, see reviews of former results and
examples in Refs. \cite{sv2001,sv2014,svvvey,tgovsv}. Working with
nonholonomic (equivalently, anholonomic, i.e. non-integrable) variables is
possible to decouple and integrate such systems of nonlinear partial
differential equations, PDEs, in certain general forms with generic
off--diagonal metrics $g_{\alpha \beta }(\tau ,u^{\gamma })$ depending on
all spacetime coordinates $u^{\gamma }$ and on flow parameter $\tau .$

\subsection{PDEs for off--diagonal geometric flows and Ricci solitons}

In this work, the effective scalar field $\phi $ is subjected to constraints
of type $\mathbf{e}_{\alpha }\phi =\ ^{0}\phi _{\alpha }=const,$ which
results in $\widehat{\mathbf{D}}^{2}\phi =0.$ We restrict our models to
configurations of $\phi ,$ which can be encoded into N--connection
coefficients\footnote{%
we shall use brief denotations for partial derivatives like $a^{\bullet
}=\partial _{1}a,b^{\prime }=\partial _{2}b,h^{\ast }=\partial _{3}h$ if it
will be necessary}
\begin{eqnarray*}
\mathbf{e}_{i}\phi &=&\partial _{i}\phi -w_{i}\phi ^{\ast }-n_{i}\partial
_{4}\phi =\ ^{0}\phi _{i};\partial _{3}\phi =\ ^{0}\phi _{3};\partial
_{4}\phi =\ ^{0}\phi _{4}; \\
\mbox{ for }\ ^{0}\phi _{1} &=&\ ^{0}\phi _{2}\mbox{ and }\ ^{0}\phi _{3}=\
^{0}\phi _{4}.
\end{eqnarray*}%
This results in an additional source $~\ ^{\phi }\widetilde{\Upsilon }=~\
^{\phi }\widetilde{\Lambda }_{0}=const$ \ and $\ ^{\phi }\Upsilon =\ ^{\phi
}\Lambda _{0}=const.$ Under geometric flows, it is possible running of such
configurations when
\begin{equation}
\mathbf{e}_{\alpha }\phi (\tau ,x^{k},y^{a})=\ ^{0}\phi _{\alpha }+\ \
_{\bot }^{0}\phi _{\alpha }(\tau )  \label{scfl}
\end{equation}%
modify the effective h- and v--sources%
\begin{equation}
~\ ^{\phi }\widetilde{\Upsilon }=~\ ^{\phi }\widetilde{\Lambda }_{0}+\
^{\phi }\widetilde{\check{\Lambda}}(\tau )\mbox{ and }\ ^{\phi }\Upsilon =\
^{\phi }\Lambda _{0}+\ ^{\phi }\check{\Lambda}(\tau ).  \label{scfsourc}
\end{equation}%
We shall use such effective sources as additional nonholonomic and geometric
flow deformations of the evolution and modified gravitational field
equations.

\subsubsection{Geometric flows of d--metric coefficients}

Let us consider a set of coefficients $\alpha _{\beta }=(\alpha _{i},\alpha
_{3}=0,\alpha _{4})$ determined by a generating function $\Psi $ when
\begin{eqnarray}
\alpha _{i} &=&h_{4}^{\ast }\partial _{i}\Psi /\Psi ,\alpha _{3}=h_{4}^{\ast
}\ \Psi ^{\ast }/\Psi ,\gamma =\left( \ln |h_{4}|^{3/2}/|h_{3}|\right)
^{\ast }  \label{coefgenf} \\
\mbox{ for }\Psi &:=& h_{4}^{\ast }/\sqrt{|h_{3}h_{4}|}.  \label{genfunct}
\end{eqnarray}

Using the ansatz for d--metric (\ref{offdans}) and sources (\ref{sourc2}),
with $\tau $--parameter dependencies of coefficients (\ref{coefft}), and
expressing the coefficients (\ref{coefgenf}) and related formulas in terms
of generating functions like (\ref{genfunct}), we transform\footnote{%
see details of such computation in \cite{sv2014,svvvey,tgovsv}} the system (%
\ref{rfcandch}) and (\ref{rfcandcv}) into a system of nonlinear PDEs
\begin{eqnarray}
\psi ^{\bullet \bullet }+\psi ^{\prime \prime } &=&2(\ \ ^{\thicksim }%
\overline{\Upsilon }-\frac{1}{2}\partial _{\tau }\psi ),  \label{rf1} \\
\ \Psi ^{\ast }h_{4}^{\ast } &=&2h_{3}h_{4}(\overline{\Upsilon }-\partial
_{\tau }\ln |\omega ^{2}h_{4}|)\Psi ,  \label{rf2} \\
\partial _{\tau }\ln |\omega ^{2}h_{3}| &=&\partial _{\tau }\ln |\omega
^{2}h_{4}|=\partial _{\tau }\ln |\breve{N}^{2}|\ ,  \label{rf2a} \\
\alpha _{3}w_{i}-\alpha _{i} &=&0,\   \label{rf3a} \\
n_{i}^{\ast \ast }+\gamma n_{i}^{\ast } &=&0,\   \label{rf4a} \\
\mathbf{e}_{k}\omega &=&\partial _{k}\omega +w_{k}\omega ^{\ast
}+n_{k}\partial _{4}\omega =0,  \label{rfc} \\
\mathbf{e}_{\alpha }\phi &=&\ ^{0}\phi _{\alpha }=const,  \label{rfsc}
\end{eqnarray}%
with effective h- and v--sources,
\begin{equation}
\ ^{\thicksim }\overline{\Upsilon }:=~\widetilde{\Upsilon }+~\ ^{\phi }%
\widetilde{\Lambda }_{0}+2\varsigma ^{2}\mbox{ and }\overline{\Upsilon }%
:=\Upsilon +\ ^{\phi }\Lambda +2\varsigma ^{2}.  \label{effs}
\end{equation}%
The un--known functions for this system are $\psi (\tau ,x^{i}),\omega (\tau
,x^{k},y^{3},t),h_{a}(\tau ,x^{k},y^{3}),w_{i}(\tau ,x^{k},y^{3})$ and $%
n_{i}(\tau ,x^{k},y^{3}).$ The first two equations contain possible
additional sources determined by other effective polarized cosmological
constants or matter fields written as $~\widetilde{\Upsilon }(\tau ,x^{k})$
and $~\Upsilon (\tau ,x^{k},y^{3}).$ We omitted the last equation for the
re-scaling function $f$ \ because it can be found at the end when other
values are determined by a class of solutions. As a matter of principle, we
can work with not normalized geometric flow solutions.

\subsubsection{Nonholonomic Ricci soliton equations}

For stationary configurations with $\partial _{\tau }g_{\alpha \beta }=0$
and $\tau =\tau _{0},$ the first three equations in the system of nonlinear
PDEs (\ref{rf1})--(\ref{rfc}) transform into self-similar Ricci soliton
equations (\ref{rsolit1}) and (\ref{rsolit2}) which for the off--diagonal
ansatz can be written
\begin{eqnarray}
\ _{\flat }\psi ^{\bullet \bullet }+\ _{\flat }\psi ^{\prime \prime } &=&2\
\ \ \ _{\flat }^{\thicksim }\overline{\Upsilon }\mbox{ and }  \label{rs1} \\
\ \ _{\flat }\Psi ^{\ast }\ _{\flat }h_{4}^{\ast } &=&2\ _{\flat }h_{3}\
_{\flat }h_{4}\ _{\flat }\overline{\Upsilon }\ _{\flat }\Psi .  \label{rs2}
\end{eqnarray}%
The equation (\ref{rs1}) is just the 2-d Poisson equation which can be
solved in general form for any given source $\ _{\flat }^{\thicksim }%
\overline{\Upsilon }(x^{k}).$

Let us show how we can integrate the system system (\ref{genfunct}) and (\ref%
{rs2}) for arbitrary source $\ _{\flat }\Upsilon (x^{k},y^{3}).$ Here we
elaborate a new approach which is different from that considered in \cite%
{sv2014,svvvey,tgovsv}. In that work, it was applied the property that such
systems are invariant under re-definition of generating function, $\Psi
\longleftrightarrow \widetilde{\Psi },$ and the effective source, $(\Upsilon
+\ ^{\phi }\Lambda +2\varsigma ^{2})\longleftrightarrow (\Lambda _{0}+\
^{\phi }\Lambda +2\varsigma ^{2})=const,\Lambda _{0}\neq 0.$\footnote{%
Such nonlinear transforms are given by formulas
\begin{equation*}
(\Lambda _{0}+\ ^{\phi }\Lambda +2\varsigma ^{2})(\Psi ^{2})^{\ast
}=|\Upsilon +\ ^{\phi }\Lambda +2\varsigma ^{2}|(\widetilde{\Psi }%
^{2})^{\ast },\mbox{  or  }(\Lambda _{0}+\ ^{\phi }\Lambda +2\varsigma
^{2})\Psi ^{2}=\widetilde{\Psi }^{2}|\Upsilon +\ ^{\phi }\Lambda +2\varsigma
^{2}|-\int dy^{3}\widetilde{\Psi }^{2}|\Upsilon |^{\ast }.
\end{equation*}%
They can be used for re--definition of generation and source functions and
constructing new classes of solutions.} For generating off--diagonal locally
anisotropic cosmological solutions depending on $y^{3},$ we have to consider
generating functions for which $\Psi ^{\ast }\neq 0.$ \ We obtain such a
system nonlinear PDEs%
\begin{eqnarray}
\ \ _{\flat }\Psi ^{\ast }\ _{\flat }h_{4}^{\ast } &=&2\ _{\flat }h_{3}\
_{\flat }h_{4}\ _{\flat }\overline{\Upsilon }\ _{\flat }\Psi ,  \label{rsa1}
\\
\sqrt{|\ _{\flat }h_{3}\ _{\flat }h_{4}|}\ _{\flat }\Psi &=&\ _{\flat
}h_{4}^{\ast }  \label{rsa1a} \\
\ _{\flat }\Psi ^{\ast }\ _{\flat }w_{i}-\partial _{i}\ _{\flat }\Psi &=&0,\
\label{rsa2} \\
\ _{\flat }n_{i}^{\ast \ast }+\left( \ln |\ _{\flat }h_{4}|^{3/2}/|\ _{\flat
}h_{3}|\right) ^{\ast }\ _{\flat }n_{i}^{\ast } &=&0.\   \label{rsa3}
\end{eqnarray}%
This system for nonholonomic Ricci solitons (\ref{rs1}), (\ref{rs2}) and (%
\ref{rf3a})--(\ref{rfc}) can be solved in very general forms by prescribing $%
~\ _{\flat }\widetilde{\Upsilon },\ _{\flat }\overline{\Upsilon }$ and $\
_{\flat }\Psi $ and integrating the equations "step by step" for a fixed
parameter $\tau _{0}.$ Introducing the function
\begin{equation}
q^{2}:=\epsilon _{3}\epsilon _{4}\ _{\flat }h_{3}\ _{\flat }h_{4},
\label{qf2}
\end{equation}%
for $\epsilon _{3,4}=\pm 1$ depending on signature of the metrics, we
consider that the system (\ref{rsa1}) and (\ref{rsa1a}) can be expressed
respectively as
\begin{equation}
\ _{\flat }\Psi ^{\ast }\ _{\flat }h_{4}^{\ast }=2\epsilon _{3}\epsilon
_{4}q^{2}\ _{\flat }\overline{\Upsilon }\ _{\flat }\Psi \mbox{ and }\
_{\flat }h_{4}^{\ast }=q\ _{\flat }\Psi .  \label{rsa1b}
\end{equation}%
Introducing $\ _{\flat }h_{4}^{\ast }$ form the second equation into the
first one, we find%
\begin{equation}
q=\frac{\epsilon _{3}\epsilon _{4}}{2}\frac{\ _{\flat }\Psi ^{\ast }}{\
_{\flat }\overline{\Upsilon }}.  \label{qf1}
\end{equation}%
We can use this value in the second equation of (\ref{rsa1b}) and find
\begin{equation}
\ _{\flat }h_{4}=h_{4}^{[0]}(x^{k})+\frac{\epsilon _{3}\epsilon _{4}}{4}\int
dy^{3}\frac{(\ _{\flat }\Psi ^{2})^{\ast }}{\ _{\flat }\overline{\Upsilon }},
\label{h4}
\end{equation}%
where $h_{4}^{[0]}(x^{k})$ is an integration function. We compute $h_{3}$
considering (\ref{qf1}), (\ref{qf2}) and formula (\ref{h4}),%
\begin{equation}
\ _{\flat }h_{3}=\frac{\epsilon _{3}\epsilon _{4}}{4h_{4}}(\frac{\ _{\flat
}\Psi ^{\ast }}{\ _{\flat }\overline{\Upsilon }})^{2}=\frac{\epsilon
_{3}\epsilon _{4}}{4}\frac{(\ _{\flat }\Psi ^{\ast })^{2}}{\ _{\flat }%
\overline{\Upsilon }^{2}}\left( h_{4}^{[0]}+\frac{\epsilon _{3}\epsilon _{4}%
}{4}\int dy^{3}\frac{(\ _{\flat }\Psi ^{2})^{\ast }}{\ _{\flat }\overline{%
\Upsilon }}\right) ^{-1}.  \label{h3}
\end{equation}

For a given $\ _{\flat }\Psi ,$ we can solve the linear algebraic equations (%
\ref{rsa2}) and express
\begin{equation*}
\ _{\flat }w_{i}=\partial _{i}\ _{\flat }\Psi /\ _{\flat }\Psi ^{\ast }.
\end{equation*}%
The second part of N--connection coefficients are found by integrating two
times on $y^{3}$ in (\ref{rsa3}) expressed as
\begin{equation*}
\ _{\flat }n_{i}^{\ast \ast }=\partial _{3}(\ _{\flat }n_{i}^{\ast })=-\
_{\flat }n_{i}^{\ast }\partial _{3}(\ln |\ _{\flat }h_{4}|^{3/2}/|\ _{\flat
}h_{3}|)
\end{equation*}%
for the coefficient $\gamma $ in (\ref{coefgenf}). The first integration
results in $\ _{\flat }n_{i}^{\ast }=\ _{2}n_{i}(x^{k})|\ _{\flat }h_{3}|/|\
_{\flat }h_{4}|^{3/2},$ for certain integration functions $\
_{2}n_{i}(x^{k}).$ Integrating second time on $y^{3},$ including the
signature signs and certain coefficients in integration functions and using
formulas (\ref{h3}) and (\ref{h4}), we obtain
\begin{eqnarray*}
\ _{\flat }n_{k}(\tau _{0}) &=&\ _{1}n_{k}+\ _{2}n_{k}\int dy^{3}\ \frac{\
_{\flat }h_{3}}{|\ _{\flat }h_{4}|^{3/2}}=\ _{1}n_{k}+\ _{2}\widetilde{n}%
_{k}\int dy^{3}\ \frac{(\ _{\flat }\Psi ^{\ast })^{2}}{|\ _{\flat
}h_{4}|^{5/2}\ _{\flat }\overline{\Upsilon }^{2}} \\
&=&\ _{1}n_{k}+\ _{2}n_{k}\int dy^{3}\frac{(\ _{\flat }\Psi ^{\ast })^{2}}{\
_{\flat }\overline{\Upsilon }^{2}}\left\vert h_{4}^{[0]}+\frac{\epsilon
_{3}\epsilon _{4}}{4}\int dy^{3}\frac{(\ _{\flat }\Psi ^{2})^{\ast }}{\
_{\flat }\overline{\Upsilon }}\right\vert ^{-5/2},
\end{eqnarray*}%
containing also a second set of integration functions $\ _{1}n_{k}(x^{i})$
and redefined $\ _{2}\widetilde{n}_{k}(x^{i}).$

Putting together all above formulas and writing in explicit form the
effective source (\ref{effs}), we obtain the formulas for the coefficients
of a d--metric and a N--connection determining a Ricci soliton solution for
the system (\ref{rsolit1}) and (\ref{rsolit2}),%
\begin{eqnarray}
\ _{\flat }g_{i}(\tau _{0}) &=&g_{i}[\ _{\flat }\psi ,~\ _{\flat }\widetilde{%
\Upsilon },~\ \ _{\flat }^{\phi }\widetilde{\Lambda },\varsigma ^{2}]\simeq
e^{\ _{\flat }\psi (\tau _{0},x^{k})}%
\mbox{ as
a solution of 2-d Poisson equations (\ref{rs1})};  \notag \\
\ _{\flat }h_{3}(\tau _{0}) &=&\frac{\epsilon _{3}\epsilon _{4}}{4}\frac{(\
_{\flat }\Psi ^{\ast })^{2}}{\ _{\flat }\overline{\Upsilon }^{2}}\left(
h_{4}^{[0]}+\frac{\epsilon _{3}\epsilon _{4}}{4}\int dy^{3}\frac{(\ _{\flat
}\Psi ^{2})^{\ast }}{\ _{\flat }\overline{\Upsilon }}\right) ^{-1};
\label{solut1t} \\
\ _{\flat }h_{4}(\tau _{0}) &=&h_{4}^{[0]}(x^{k})+\frac{\epsilon
_{3}\epsilon _{4}}{4}\int dy^{3}\frac{(\ _{\flat }\Psi ^{2})^{\ast }}{\
_{\flat }\overline{\Upsilon }};  \notag \\
\ _{\flat }w_{i}(\tau _{0}) &=&\partial _{i}\ _{\flat }\Psi /\ _{\flat }\Psi
^{\ast };  \notag \\
\ _{\flat }n_{k}(\tau _{0}) &=&\ _{1}n_{k}+\ _{2}n_{k}\int dy^{3}\frac{(\
_{\flat }\Psi ^{\ast })^{2}}{\ _{\flat }\overline{\Upsilon }}\left\vert
h_{4}^{[0]}+\frac{\epsilon _{3}\epsilon _{4}}{4}\int dy^{3}\frac{(\ _{\flat
}\Psi ^{2})^{\ast }}{\ _{\flat }\overline{\Upsilon }}\right\vert ^{-5/2};
\notag \\
\ _{\flat }\omega (\tau _{0}) &=&\omega \lbrack \ _{\flat }\Psi ,\ _{\flat }%
\overline{\Upsilon }]\mbox{ is any solution of 1st order system
(\ref{rfc})}.  \notag
\end{eqnarray}%
In these formulas, we state that the coefficients $\ _{\flat }h_{a}$ depend
functionally on $\ _{\flat }\Psi $ and $\ _{\flat }\Upsilon ,$ which (in
their turn) may depend on the flow evolution parameter $\tau $ which is
fixed to a value $\tau _{0}.$

We can solve the equations (\ref{rfc}) for a nontrivial $\ _{\flat }\omega
^{2}=|\ _{\flat }h_{3}|^{-1}.$

Using coefficients (\ref{solut1t}), we define such a class of quadratic
elements for off--diagonal stationary Ricci solitons with nonholonomically
induced torsion (tRs),
\begin{eqnarray}
&&ds_{tRs}^{2}=\ _{\flat }g_{\alpha \beta }(x^{k},y^{3})du^{\alpha
}du^{\beta }=e^{\ _{\flat }\psi }[(dx^{1})^{2}+(dx^{2})^{2}]+\ _{\flat
}\omega ^{2}\frac{\epsilon _{3}\epsilon _{4}}{4\ _{\flat }h_{4}}(\frac{\
_{\flat }\Psi ^{\ast }}{\ _{\flat }\overline{\Upsilon }})^{2}\ [dy^{3}+\frac{%
\partial _{i}\ _{\flat }\Psi }{\ _{\flat }\Psi ^{\ast }}dx^{i}]^{2}  \notag
\\
&&+\ _{\flat }\omega ^{2}\ _{\flat }h_{4}[\ _{\flat }\Psi ,\ _{\flat }%
\overline{\Upsilon }][dt+(\ _{1}n_{k}+\ _{2}\widetilde{n}_{k}\int dy^{3}%
\frac{(\ _{\flat }\Psi ^{\ast })^{2}}{\ _{\flat }\overline{\Upsilon }^{2}|\
_{\flat }h_{4}|^{5/2}})dx^{k}]^{2}.  \label{riccisolt}
\end{eqnarray}%
This class of metrics defines also exact solutions for the canonical
d--connection $\widehat{\mathbf{D}}$ in $R^{2}$ gravity with effective
scalar field encoded into a nonholonomically polarized vacuum.

\subsubsection{Geometric evolution with factorized dependence on flow
parameter of d--metric and N--connection coefficients}

Such classes of solutions are defined by generated functions, $\psi (\tau
,x^{k})$ and $\Psi (\tau ,x^{k},y^{3}),$ and effective sources, $\widetilde{%
\Upsilon }(\tau ,x^{k})$ and $\Upsilon (\tau ,x^{k},y^{3}),$ depending in
factorized form on flow parameter $\tau .$ For simplicity, we shall analyze
stationary solutions with $\omega =1.$ We can integrate such equations in
explicit form if we consider subclasses of solutions with separation of
variables, when {\small
\begin{eqnarray}
&&\psi (\tau ,x^{k}) =\ _{\bot }\psi (\tau )+\ \ _{\flat }\psi (x^{k}),\Psi
=\ \ _{\bot }\Psi (\tau )\ \ _{\flat }\Psi (x^{k},y^{3}),  \notag \\
&&\mbox{ for }h_{3} =\ \ _{\bot }h_{3}(\tau )\ \ _{\flat
}h_{3}(x^{k},y^{3}),h_{4}=\ \ _{\bot }h_{4}(\tau )\ \ _{\flat
}h_{4}(x^{k},y^{3})  \label{factoriz} \\
&&\mbox{ and }~\widetilde{\Upsilon }(\tau ,x^{k}) = \ _{\bot }\widetilde{%
\Upsilon }(\tau )+\ \ _{\flat }\widetilde{\Upsilon }(x^{k}),\Upsilon (\tau
,x^{k},y^{3})=\ _{\bot }\Upsilon (\tau )+\ \ _{\flat }\Upsilon (x^{k},y^{3})%
\mbox{ and }  \notag \\
&&\ ^{\phi }\widetilde{\Lambda } \rightarrow \ ^{\phi }\widetilde{\Lambda }%
+\ ^{\phi }\widetilde{\check{\Lambda}}(\tau ), \ ^{\phi }\Lambda \rightarrow
\ ^{\phi }\Lambda +\ ^{\phi }\check{\Lambda}(\tau ), \varsigma
^{2}\rightarrow \varsigma ^{2}+\check{\varsigma}(\tau ), \ \widetilde{%
\Lambda }(\tau ) = \ _{\bot }\widetilde{\Upsilon }(\tau )+~\ ^{\phi }%
\widetilde{\check{\Lambda}}(\tau )+2\check{\varsigma}(\tau ),  \notag \\
&& \ ^{\thicksim }\overline{\Upsilon } = \ _{\bot }^{\thicksim }\overline{%
\Upsilon }(\tau )+\ \ \ _{\flat }^{\thicksim }\overline{\Upsilon }(x^{k}),\
\overline{\Upsilon }=\ \ _{\bot }\overline{\Upsilon }(\tau )+\ \ _{\flat }%
\overline{\Upsilon }(x^{k},y^{3}),  \notag \\
&& \ _{\bot }\overline{\Upsilon }(\tau ) = \ _{\bot }\Upsilon (\tau )+\
^{\phi }\widetilde{\check{\Lambda}}(\tau )+2\check{\varsigma}(\tau )%
\mbox{
and }\ \ _{\flat }\overline{\Upsilon }=\ \Upsilon +\ ^{\phi }\Lambda
+2\varsigma ^{2},  \notag
\end{eqnarray}%
} see (\ref{scfsourc}). For simplicity, we shall consider in this section a
constant value $\ _{\flat }\overline{\Upsilon }(x^{k},y^{3})=$ $\ \overline{%
\Upsilon }_{[0]}\ =\ \Upsilon _{\lbrack 0]}+\ ^{\phi }\Lambda +2\varsigma
^{2}=const\neq 0,$ i.e. for $\ _{\flat }\Upsilon =\ \Upsilon _{\lbrack
0]}=const,$ which is enough to study various classes of geometric flow
evolution models. If $\ _{\flat }\Upsilon (x^{k},y^{3})$ is not constant, it
is a more difficult task to construct exact solutions in explicit form (see
such examples in subsection \ref{cgfdmncon}).

We can solve in explicit form the equations (\ref{rf1})--(\ref{rf4a})
considering (for simplicity) $\omega =1$ and corresponding factorizations of
the generating functions and sources. Via separation of variables, we obtain
the system of equations
\begin{eqnarray}
\ _{\flat }\psi ^{\bullet \bullet }+\ _{\flat }\psi ^{\prime \prime } &=&2\
_{\flat }\overline{\Upsilon },\ \partial _{\tau }\ \ _{\bot }\psi (\tau )=2\
\widetilde{\Lambda }(\tau );  \label{rff1} \\
\ \ \ _{\flat }\Psi ^{\ast }\ \ _{\flat }h_{4}^{\ast } &=&2\ \ _{\bot
}h_{3}\ \ _{\flat }h_{3}\ \ _{\flat }h_{4}(\ \ _{\bot }\overline{\Upsilon }%
(\tau )+\overline{\Upsilon }_{[0]}-\partial _{\tau }\ln |\ \ _{\bot
}h_{4}|)\ \ _{\flat }\Psi ,\   \label{rff2} \\
\partial _{\tau }\ln |\ \ _{\bot }h_{3}| &=&\partial _{\tau }\ln |\ \ _{\bot
}h_{4}|=\partial _{\tau }\ln |\breve{N}^{2}|\ ,  \label{rff2a} \\
w_{i}(\tau ,x^{k},y^{3}) &=&\partial _{i}\ _{\flat }\Psi /\ _{\flat }\Psi
^{\ast };  \label{rff3a} \\
n_{k}(\tau ,x^{k},y^{3}) &=&\ _{1}n_{k}+\ _{2}n_{k}\int dy^{3}\ \frac{h_{3}}{%
|h_{4}|^{3/2}}  \label{rff4a}
\end{eqnarray}%
This system can be integrated in explicit form "step by step" as follows:

The first equation in (\ref{rff1}) is just the 2-d Poisson equation for $\ \
_{\flat }\psi (x^{k})$ corresponding to the first line in the solution for
Ricci solitons (\ref{solut1t}). The second equation with dependence on flow
parameter$\ $can be solved and expressed as
\begin{equation*}
e^{\ _{\bot }\psi }=A_{0}e^{2\int d\tau \ \widetilde{\Lambda }(\tau
)},A_{0}=const,
\end{equation*}%
where the integration constant can be taken $A_{0}=1.$ Such a solution has
physical meaning if $\int d\tau \ \widetilde{\Lambda }(\tau )\leq 0$ for an
interval $0\leq \tau \leq \tau _{0}$ which correlates possible variation of
constants induced by effective scalar fields and effective cosmological
constant and other possible matter sources.

We have to solve together both equations (\ref{rff2}) with separation of
spacetime coordinates and flow parameter. To model the evolution of certain
Ricci soliton configurations it is necessary to satisfy the conditions $|\ \
_{\bot }h_{3}|=|\ _{\bot }h_{4}|$ and
\begin{equation*}
\ \ _{\bot }h_{3}[1+\frac{1}{\overline{\Upsilon }_{[0]}}\left( \ \ _{\bot }%
\overline{\Upsilon }(\tau )-\frac{\partial _{\tau }\ \ _{\bot }h_{3}}{\ \
_{\bot }h_{3}}\right) ]\ =1.
\end{equation*}%
The solution of this equation is
\begin{eqnarray}
\ \ _{\bot }h_{3}(\tau ) &=&1+S_{0}e^{\lambda _{1}\tau }+S_{1}e^{\lambda
_{1}\tau }\int d\tau e^{-\lambda _{1}\tau }[\ \ _{\bot }\overline{\Upsilon }%
(\tau )]  \label{tauh3} \\
&&1+\ _{\bot }\varepsilon (\tau )  \label{tauh3s}
\end{eqnarray}%
for certain integration constants $S_{0}$ and $S_{1}$ and $\lambda _{1}:=(\
\overline{\Upsilon }_{[0]})$ and
\begin{equation*}
\ _{\bot }\varepsilon (\tau ):=S_{0}e^{\lambda _{1}\tau }+S_{1}e^{\lambda
_{1}\tau }\int d\tau e^{-\lambda _{1}\tau }[\ \ _{\bot }\overline{\Upsilon }%
(\tau )].
\end{equation*}
Such configurations have physical importance if there is an interval $0\leq
\tau \leq \tau _{0}$ $\ ^{0}h_{3}\rightarrow 1$ for increasing $\tau _{0}.$
For certain deformations of stationary solutions in MGTs, \ the function $\
_{\bot }\varepsilon (\tau ),|\ _{\bot }\varepsilon (\tau )|\ll 1,$ has to be
found from experimental data. We can express
\begin{equation*}
h_{3}=|\ \ _{\bot }h_{3}(\tau )|\ \ _{\flat }h_{3}(x^{k},y^{3})\mbox{ and }%
h_{4}=|\ \ _{\bot }h_{3}(\tau )|\ \ _{\flat }h_{4}(x^{k},y^{3})
\end{equation*}%
where $\ \ _{\flat }h_{a}$ are taken as $\ _{\flat }h_{a}(\tau _{0})$ from (%
\ref{solut1t}) but with
\begin{equation*}
\ \ _{\flat }h_{3}=\frac{\epsilon _{3}\epsilon _{4}}{4\ \ \ _{\flat }h_{4}}(%
\frac{\ \ _{\flat }\Psi ^{\ast }}{\overline{\Upsilon }_{[0]}})^{2}%
\mbox{ and
}\ \ \ _{\flat }h_{4}=h_{4}^{[0]}(x^{k})+\frac{\epsilon _{3}\epsilon _{4}}{4%
\overline{\Upsilon }_{[0]}}(\ _{\flat }\Psi )^{2}.
\end{equation*}

Putting together above formulas, we find the d--metric coefficients,
\begin{eqnarray}
g_{1}(\tau ,x^{k}) &=&g_{2}=e^{\ _{\bot }\psi }e^{\ \ \ _{\flat }\psi
(x^{k})}\mbox{ for }e^{\ _{\bot }\psi }=A_{0}e^{2\int d\tau \widetilde{%
\Lambda }(\tau )},A_{0}=const;  \notag \\
\ h_{3}(\tau ,x^{k},y^{3}) &=&|\ _{\bot }h_{3}(\tau )|\frac{\epsilon
_{3}\epsilon _{4}}{4\ \ \ _{\flat }h_{4}}\left( \frac{\ \ _{\flat }\Psi
^{\ast }}{\overline{\Upsilon }_{[0]}}\right) ^{2}\mbox{ for }\ _{\bot
}h_{3}(\tau )\mbox{ taken as in (\ref{tauh3})};  \notag \\
h_{4}(\tau ,x^{k},y^{3}) &=&|\ _{\bot }h_{3}(\tau )|\ \left[
h_{4}^{[0]}(x^{k})+\frac{\epsilon _{3}\epsilon _{4}}{4\overline{\Upsilon }%
_{[0]}}(\ _{\flat }\Psi )^{2}\right] ;  \label{geomflcoef}
\end{eqnarray}
and N--connection coefficients,%
\begin{eqnarray}
w_{i}(x^{k},y^{3}) &=&\partial _{i}\ \ _{\flat }\Psi /\ \ _{\flat }\Psi
^{\ast };  \notag \\
n_{k}(\tau ,x^{i},y^{3}) &=&\ _{1}n_{k}(\tau ,x^{i})+\ _{2}n_{k}(\tau
,x^{i})\int dy^{3}\ \frac{h_{3}}{|h_{4}|^{3/2}}  \notag \\
&=&\ _{1}n_{k}(\tau ,x^{i})+\ _{2}\widetilde{n}_{k}(\tau ,x^{i})\int dy^{3}\
\frac{(\ \ _{\flat }\Psi )^{2}}{|\ \ \ _{\flat }h_{4}|^{5/2}}  \notag \\
&=&\ _{1}n_{k}(\tau ,x^{i})+\ _{2}\widetilde{n}_{k}(\tau ,x^{i})\int
dy^{3}(\ \ _{\flat }\Psi )^{2}\left\vert h_{4}^{[0]}(x^{k})+\frac{\epsilon
_{3}\epsilon _{4}}{4\overline{\Upsilon }_{[0]}}(\ _{\flat }\Psi
)^{2}\right\vert ^{-5/2},  \notag
\end{eqnarray}%
for certain re-defined integration and generation functions.

In such formulas, the generation functions and sources, the integration
functions and constants depend on evolution parameter which determine
additional anisotropic polarizations of physical values and running of
physical constants. The off--diagonal terms $w_{i}(x^{k},y^{3})$ do not
depend on evolution parameter. If we take $_{2}n_{k}=0$ and $\ _{1}n_{k}=\
_{1}n_{k}(x^{k}),$ the N--connection coefficients do not depend on geometric
evolution parameter being determined by a prescribed Ricci soliton
configuration.

The coefficients (\ref{geomflcoef}) determine generic off--diagonal
quadratic elements for solutions of relativistic geometric flows inducing
anisotropic polarizations and running of constants and of Ricci solitons,
\begin{eqnarray}
ds_{tRs}^{2} &=&g_{\alpha \beta }(\tau ,x^{k},y^{3})du^{\alpha }du^{\beta
}=e^{2\int d\tau \widetilde{\Lambda }(\tau )}e^{\ ^{1}\psi
(x^{k})}[(dx^{1})^{2}+(dx^{2})^{2}]+  \label{runningconst} \\
&&\{1+\ _{\bot }\varepsilon (\tau )\}\{\frac{\epsilon _{3}\epsilon _{4}}{4\
\ \ _{\flat }h_{4}}\left( \frac{\ \ _{\flat }\Psi ^{\ast }}{\overline{%
\Upsilon }_{[0]}}\right) ^{2}\ \left[ dy^{3}+\frac{\partial _{i}\ \ _{\flat
}\Psi }{\ \ _{\flat }\Psi ^{\ast }}dx^{i}\right] ^{2}+\   \notag \\
&&\ \ \ _{\flat }h_{4}(x^{k},y^{3})\left[ \ dt+(\ _{1}n_{k}(\tau ,x^{i})+\
_{2}n_{k}(\tau ,x^{i})\int dy^{3}\ \frac{h_{3}}{|h_{4}|^{3/2}})dx^{k}\right]
^{2}\},  \notag
\end{eqnarray}%
In these formulas, the flow evolution is determined by certain parameters
and nonholonomic constraints in $R^{2}$ gravity \ and small corrections with
dependence of type $\ _{\bot }\varepsilon (\tau )$ (\ref{tauh3s}).\ For this
class of solutions, the off--diagonal coefficients are determined by a Ricci
solitonic background which became dependent on the evolution parameter $\tau
$ via the vertical part of d--metric and N--connection coefficients.
Nevertheless, we can fix such integration fuctions when $w_{i}=\ \ _{\flat
}w_{i}(x^{k},y^{3})$ and $\ _{1}n_{k}=\ _{1}n_{k}(x^{i})$ and $\ _{2}n_{k}=0$
with N--connection and anholonomy coefficients not subjected to geometric
flows. Such a nonholonomic geometric flow evolution is for the canonical
d--connection $\widehat{\mathbf{D}}$ in $R^{2}$ gravity with effective
scalar field encoded into a nonholonomically polarized vacuum.

In explicit form, we generate exact solutions the geometric flow/ Ricci
soliton equations for certain prescribed values of $\Psi $ and $\Upsilon ,$
corresponding "prime" constants lile $\Lambda ,\ ^{\phi }\Lambda $ and $%
\varsigma ^{2}$ and following certain assumptions on
initial/boundary/asymptotic conditions, physical arguments on symmetries of
solutions, compatibility with observational data etc. \ Variations of
constants $\ ^{\phi }\Lambda (\tau ),\varsigma (\tau )$ etc should be taken
from certain observational data which are provided, for instance, in \cite%
{flambaum}.

\subsubsection{Geometric flows of effective sources and d--metric and
N--connection coefficients}

\label{cgfdmncon} For certain conditions, we can find exact solutions of the
geometric flow equations when the d--metric and N--connection coefficients
and of generating functions depend in a general form form on evolution
parameter $\tau .$ In the simplest way, we have to impose necessary type
constraints on the generating functions and then to compute the
corresponding horizontal and vertical effective sources.

\bigskip Using a necessary effective source $~\widetilde{\Upsilon }(\tau
,x^{k})$ constrained to satisfy the conditions
\begin{equation*}
~\widetilde{\Upsilon }+~\ ^{\phi }\widetilde{\Lambda }_{0}+2\varsigma ^{2}-%
\frac{1}{2}\partial _{\tau }\psi =\widetilde{\Lambda }_{0}(\tau ),
\end{equation*}%
for an effective $\widetilde{\Lambda }_{0}(\tau ),$ we find $\psi (\tau
,x^{k})$ for (\ref{rf1}) as a solution of parametric 2-d Poisson equation,

\begin{equation*}
\psi ^{\bullet \bullet }+\psi ^{\prime \prime }=2\widetilde{\Lambda }%
_{0}(\tau )~.
\end{equation*}

We can generate a class of solutions of geometric flow equations (\ref{rf2}%
)--(\ref{rfsc}) for arbitrary $h_{4}(\tau ,x^{k},y^{3}),h_{4}^{\ast }\neq 0$
taken as a generating function if we consider an effective source $\Upsilon
(\tau ,x^{k},y^{3})$ determined by the condition
\begin{equation}
\Upsilon +\ ^{\phi }\Lambda +2\varsigma ^{2}-\partial _{\tau }\ln |\omega
^{2}h_{4}|=\Lambda _{0}\neq 0,  \label{effsourc}
\end{equation}%
where $\Lambda _{0}$ is an effective cosmological constant. For such a
condition, the system of equations (\ref{genfunct}) and (\ref{rf2})
transforms into
\begin{equation*}
\sqrt{|h_{3}|}=\frac{h_{4}^{\ast }}{\Psi \sqrt{|h_{4}|}}\mbox{ and }h_{3}=%
\frac{\Psi ^{\ast }}{\Psi }\frac{h_{4}^{\ast }}{2h_{4}}\Lambda _{0},
\end{equation*}%
for two un--known functions $h_{3}(\tau ,x^{k},y^{3})$ and $\Psi (\tau
,x^{k},y^{3}).$ Taking the square of the first equation with $h_{a}=\epsilon
_{a}|h_{a}|,\epsilon _{a}=\pm 1,$ we compute
\begin{eqnarray}
\Psi ^{2} &=&B(\tau ,x^{k})+\frac{4\epsilon _{3}\epsilon _{4}}{\Lambda _{0}}%
h_{4}\mbox{ and }  \label{auxf} \\
h_{3} &=&\epsilon _{3}\epsilon _{4}\frac{(h_{4}^{\ast })^{2}}{h_{4}[B(\tau
,x^{k})+\frac{4\epsilon _{3}\epsilon _{4}}{\Lambda _{0}}h_{4}]}  \label{h3a}
\end{eqnarray}%
for an integration function $B(\tau ,x^{k}).$

We can solve the equation (\ref{rf2a}) if we take $h_{3}=h_{4}$ considering
both such values determined by the same generating function $h_{4}(\tau
,x^{k},y^{3}).$ In general, there are similar solutions with $h_{3}\neq
h_{4} $ (being involved the formula (\ref{h3a})) but it is a difficult task
to solve the mentioned equation for arbitrary $\omega .$

The algebraic equation (\ref{rf3a}) are solved in general form using the
formula (\ref{auxf}),%
\begin{equation}
w_{i}(\tau ,x^{k},y^{3})=\frac{\partial _{i}\Psi }{\Psi ^{\ast }}=\frac{%
\partial _{i}\Psi ^{2}}{\partial _{3}(\Psi ^{2})}=(h_{4}^{\ast
})^{-1}\partial _{i}[\frac{\Lambda _{0}}{4\epsilon _{3}\epsilon _{4}}B(\tau
,x^{k})+h_{4})].  \label{wa}
\end{equation}%
We find the complete set of N--connection coefficients by integrating two
times on $y^{3}$ in (\ref{rf4a}) using the condition $h_{3}=h_{4},$
\begin{equation}
n_{k}(\tau ,x^{i},y^{3})=\ _{1}n_{k}(\tau ,x^{i})+\ _{2}\widetilde{n}%
_{k}(\tau ,x^{i})\int dy^{3}\ (\sqrt{|h_{4}|})^{-1}.  \label{na}
\end{equation}

In general, we can use any $\omega (\tau ,x^{k},y^{3})$ as a solution of the
equation (\ref{rfsc})
\begin{equation*}
\partial _{k}\omega +w_{k}(\tau ,x^{i},y^{3})\omega ^{\ast }+n_{k}(\tau
,x^{i},y^{3})\partial _{4}\omega =0,
\end{equation*}%
for coefficients determined by $h_{4}$ and respective integration functions,
see (\ref{wa}) an (\ref{na}). In particular, we can take $\omega =1$ and
generate solutions for geometric evolution of stationary configurations. As
solutions of the equations (\ref{rfsc}), we can consider distributions of a
scalar field subjected to the conditions (\ref{scfl}) and (\ref{scfsourc})
resulting in modifications with an effective cosmological constant.

Above formulas determine a quadratic element
\begin{eqnarray}
&&ds_{tRs}^{2}=g_{\alpha \beta }(\tau ,x^{k},y^{3})du^{\alpha }du^{\beta
}=e^{\psi (\tau ,x^{k})}[(dx^{1})^{2}+(dx^{2})^{2}]+\omega ^{2}(\tau
,x^{i},y^{3})h_{4}(\tau ,x^{i},y^{3})  \notag \\
&& \{\ [dy^{3}+\frac{\partial _{i}(\frac{\Lambda _{0}}{4\epsilon
_{3}\epsilon _{4}}B(\tau ,x^{k})+h_{4})}{h_{4}^{\ast }}dx^{i}]^{2} +[dt+(\
_{1}n_{k}+\ _{2}\widetilde{n}_{k}\int dy^{3}\sqrt{|h_{4}|}%
)^{-1})dx^{k}]^{2}\}.  \label{ggeomfl}
\end{eqnarray}%
This class of solutions is a general one with evolution of N--connection
coefficients and flows of the nonholonomically induced torsion. Such
geometric flows may transform one class of Ricci solitons into another one,
i.e. a MGT into another MGT, or into a solution in GR (if the final torsion
is constrained to be zero). Mutual transforms of classes of (off-) diagonal
solutions in GR can be described as some particular examples of such
geometric flow evolution models.

\subsection{Extracting Levi--Civita configurations}

The solutions for Ricci solitons (\ref{riccisolt}) and their factorized
geometric evolution (\ref{runningconst}) and non-factorized geometric flow
evolution solutions are defined for the canonical d--connection $\widehat{%
\mathbf{D}}.$ There are nontrivial coefficients of nonholonomically induced
torsion which can be computed by introducing the coefficients (\ref{coefft})
(with fixed flow parameter in the metric anstaz) into (\ref{candcon}) and (%
\ref{dtors}). We have to consider certain restricted classes of
parameterizations and nonholonomic constraints on the d--metric and
N--connection coefficients in order to satisfy the zero torsion conditions (%
\ref{lccond}) and extract Levi--Civita, LC, configurations. For ansatz (\ref%
{offdans}) with dependence on flow parameter $\tau ,$ such conditions are
equivalent to the system equations (see details in \cite%
{sv2014,svvvey,tgovsv})
\begin{eqnarray}
w_{i}^{\ast } &=&(\partial _{i}-w_{i}\partial _{3})\ln \sqrt{|h_{4}[\tau ]|}%
,(\partial _{i}-w_{i}\partial _{3})\ln \sqrt{|h_{3}[\tau ]|}=0,
\label{lccond1} \\
\partial _{i}w_{j} &=&\partial _{j}w_{i},n_{i}^{\ast }[\tau ]=0,\partial
_{i}n_{j}[\tau ]=\partial _{j}n_{i}[\tau ],  \notag
\end{eqnarray}%
where we denoted in brief, for instance, $h_{4}[\tau ]=h_{4}(\tau
,x^{i},y^{3}).$

Let us consider, for simplicity, certain classes of solutions with
factorized parameterizations of d--metrics like (\ref{factoriz}) which
allows to model geometric evolution of self--similar fixed
LC--configurations for Ricci solitons.

Any functional $\widetilde{\Psi }[\Psi \lbrack \tau ]]$ satisfies the
conditions $\ $%
\begin{equation*}
\mathbf{e}_{i}\widetilde{\Psi }[\tau ]=(\partial _{i}-w_{i}\partial _{3})%
\widetilde{\Psi }[\tau ]=\frac{\partial \widetilde{\Psi }}{\partial \Psi }%
(\partial _{i}-w_{i}\partial _{3})\Psi \lbrack \tau ]\equiv 0,
\end{equation*}%
which follow from (\ref{rf4a}). We can chose, for instance, $\widetilde{\Psi
}=\ln \sqrt{|\ h_{4}[\tau ]|}$ when $\mathbf{e}_{i}\ln \sqrt{|\ h_{4}[\tau ]|%
}=0.$ If we work with classes of generating functions $\Psi =\check{\Psi}%
(\tau ,x^{k},y^{3})$ for which there are satisfied the integrability
conditions
\begin{equation}
(\partial _{i}\check{\Psi}[\tau ])^{\ast }=\partial _{i}(\check{\Psi}^{\ast
}[\tau ]),  \label{cond1}
\end{equation}%
we obtain $w_{i}^{\ast }[\tau ]=\mathbf{e}_{i}\ln |\check{\Psi}^{\ast }[\tau
]|.$ For a given functional dependence $h_{3}[\Psi \lbrack \tau ],\Upsilon
,\ ^{\phi }\Lambda (\tau ),\varsigma (\tau )]$ and using $\mathbf{e}_{i}%
\check{\Psi}=0,$ we can express
\begin{equation*}
\mathbf{e}_{i}\ln \sqrt{|\ h_{3}[\tau ]|}=\mathbf{e}_{i}[\ln |\check{\Psi}%
^{\ast }[\tau ]|-\ln \sqrt{|\ \Upsilon \lbrack \tau ]|}].
\end{equation*}%
In result, $w_{i}^{\ast }=\mathbf{e}_{i}\ln \sqrt{|\ h_{3}[\tau ]|}$ if $%
\mathbf{e}_{i}\ln \sqrt{|\ \Upsilon \lbrack \tau ]|}=0.$ This is possible
for any $\Upsilon =const,$ or any effective source expressed as a functional
$\ \Upsilon (x^{i},y^{3})=\ \Upsilon \lbrack \check{\Psi},^{\phi }\Lambda
(\tau ),\varsigma (\tau )]$ with parametric coordinate dependencies.

The conditions that $\partial _{i}w_{j}=\partial _{j}w_{i}$ can be expressed
in conventional form via any function $\check{A}=\check{A}(\tau
,x^{k},y^{3}) $ for which
\begin{equation}
w_{i}=\check{w}_{i}=\partial _{i}\check{\Psi}/\check{\Psi}^{\ast }=\partial
_{i}\check{A}.  \label{cond2}
\end{equation}%
If a functional $\check{\Psi}$ is prescribed, we have to solve a system of
first order PDEs which allows to find a function $\check{A}[\check{\Psi}].$
For the second set of N--coefficients, we chose $\ _{1}n_{j}(\tau
,x^{k})=\partial _{j}n(\tau ,x^{k})$ for a function $n(\tau ,x^{k}).$ As a
matter of principle, we can consider running on a geometric flow parameter,
like $n(\tau ,x^{k})$ considering a more generalized class of integration
functions.

We can generate off--diagonal torsionless solutions of the Ricci soliton
equations and generalized Einstein equations for $R^{2}$--gravity, with
possible polarizations of fundamental constants determined by geometric
flows if we chose certain subclasses of generating functions and effective
sources in (\ref{riccisolt}) and (\ref{runningconst}), when
\begin{equation}
\check{\Upsilon}=\Upsilon (\tau ,x^{i},y^{3})=\ \Upsilon \lbrack \check{\Psi}%
],w_{i}=\partial _{i}\check{A}[\check{\Psi}[\tau ]],n_{i}=\partial _{i}n,
\label{cond3}
\end{equation}%
and the generating function $\Psi =\check{\Psi}$ and "associated" $\check{A}$
for $N_{i}^{4}$--coefficients are subjected to the conditions (\ref{cond1})
and (\ref{cond2}).

If the generating/effective functions and sources are subjected to the
LC--conditions (\ref{cond1})--(\ref{cond3}), we obtain such quadratic linear
elements:

For Ricci solitons with zero nonholonomic torsion (LCRs) [a particular case
of solutions (\ref{riccisolt})],
\begin{eqnarray}
ds_{LCRs}^{2} &=&\ _{\flat }g_{\alpha \beta }(x^{k},y^{3})du^{\alpha
}du^{\beta }=e^{\ _{\flat }\psi }[(dx^{1})^{2}+(dx^{2})^{2}]+
\label{riccisolitlc} \\
&&\ _{\flat }\omega ^{2}\frac{\epsilon _{3}\epsilon _{4}}{4\ _{\flat }h_{4}}(%
\frac{\ _{\flat }\check{\Psi}^{\ast }}{\ _{\flat }\overline{\Upsilon }}%
)^{2}\ [dy^{3}+\partial _{i}\ \check{A}[\ _{\flat }\check{\Psi}%
])dx^{i}dx^{i}]^{2}+\ _{\flat }\omega ^{2}\ _{\flat }h_{4}[\ _{\flat }\check{%
\Psi},\ _{\flat }\overline{\Upsilon }][dt+\partial _{k}n(x^{k})dx^{k}]^{2},
\notag
\end{eqnarray}%
where $\ _{\flat }\Psi \rightarrow \ _{\flat }\check{\Psi}.$

In the case of geometric flows with zero nonholonomic torsion (i.e.
torsionless geometric evolution of type (\ref{runningconst})),
\begin{eqnarray}
&&ds_{LC}^{2}=g_{\alpha \beta }(\tau ,x^{k},y^{3})du^{\alpha }du^{\beta
}=e^{2\int d\tau \widetilde{\Lambda }(\tau )}e^{\ ^{1}\psi
(x^{k})}[(dx^{1})^{2}+(dx^{2})^{2}]+\{1+\ _{\bot }\varepsilon (\tau )\}
\label{ricciflsollc} \\
&&\{\frac{\epsilon _{3}\epsilon _{4}}{4(\overline{\Upsilon }_{[0]})\ \ \
_{\flat }h_{4}}\left( \ \ _{\flat }\check{\Psi}^{\ast }\right) ^{2}\ \ \left[
dy^{3}+(\partial _{i}\ \check{A}[\ \ _{\flat }\check{\Psi}])dx^{i}\right]
^{2}+\ _{\flat }h_{4}(x^{k},y^{3})\left[ \ dt+\partial _{k}n(\tau
,x^{k})dx^{k}\right] ^{2}\},  \notag
\end{eqnarray}%
where the LC--conditions (\ref{cond1})--(\ref{cond3}) hold for $\ \ _{\flat
}\Psi =\ \ _{\flat }\check{\Psi}(\tau ,x^{k},y^{3})$ and $\ \overline{%
\Upsilon }_{[0]}=const.$ The coefficients of these generic off--diagonal
metrics also generate exact solutions for geometric flows and Ricci solitons
with effective matter source but with zero torsion. We note that the metrics
are generic off--diagonal if the anholonomic coefficients $\ W_{\alpha \beta
}^{\gamma }$ (\ref{anhcoef}) are not zero.

In a similar form, we can model LC geometric evolution of metrics of type (%
\ref{ggeomfl}) [in brief, LCgev], with $\Psi \rightarrow \check{\Psi}$ and $%
\partial _{i}\ \check{A}[\check{\Psi}]=\partial _{i}(\frac{\Lambda _{0}}{%
4\epsilon _{3}\epsilon _{4}}B(\tau ,x^{k})+h_{4})/h_{4}^{\ast },$ $\ $when
\begin{eqnarray}
ds_{LCgev}^{2}&=&g_{\alpha \beta }(\tau ,x^{k},y^{3})du^{\alpha }du^{\beta
}=e^{\psi (\tau ,x^{k})}[(dx^{1})^{2}+(dx^{2})^{2}] +  \label{ggemfllc} \\
&&\omega ^{2}(\tau ,x^{i},y^{3})h_{4}(\tau ,x^{i},y^{3}) \{\
[dy^{3}+\partial _{i}\ \check{A}[\check{\Psi}(\tau
,x^{i},y^{3})]dx^{i}]^{2}+[dt+\partial _{k}n(\tau ,x^{k})dx^{k}\}.  \notag
\end{eqnarray}

We can generate subsets of solutions of (\ref{ricciflsollc}) with
N--coefficients which do not depend on flow parameter $\tau $ but only the
d--metric coefficients $g_{i}$ and $h_{a}$ are functions on $\tau $ and
spacetime coordinates preserving the Killing symmetry on $\partial
_{4}=\partial /\partial t.$ If $\tilde{\Psi}\rightarrow \ ^{1}\widetilde{%
\Psi },$ and $n=n(x^{k})$ in the two last formulas, we can consider that for
$S_{0}=S_{1}=0$ and $\widetilde{\Lambda }(\tau )=0,$ the solutions for
geometric evolution transform into a Ricci soliton configuration (\ref%
{riccisolitlc}). The LC--configurations with $\tau $--dependence describe a
self--consistent geometric evolution of LC Ricci solitons for any interval $%
0\leq \tau \leq \tau _{0}$ when the exponential on $\tau $ terms are not
singular.

The class of generic off--diagonal metrics (\ref{ggemfllc}) define LC
configurations of geometric evolution of exact solutions in $R^{2}$ and/or
GR theory. The N--connection structure for such solutions possess a
nontrivial dependence on parameter $\tau .$ Such torsionless configurations
may mix under geometric evolution different types of Ricci solitons and
transform mutually solutions from a MGT into a another MGTs, or in GR.

Variations of values $\ ^{0}\widetilde{\Upsilon }(\tau ),~\ ^{\phi }%
\widetilde{\Lambda }(\tau ),\ ^{1}\Upsilon ,$ $\ ^{\phi }\Lambda (\tau
),\varsigma (\tau )$ etc. have to be taken from observational data \cite%
{flambaum} (the Dirac's idea on variation of physical constants is
re--considered for modified theories of gravity). We conclude that geometric
flow solutions can explain possible locally anisotropic polarizations and
running of d--metric and N--connection coefficients and of fundamental
physical constants.

\subsection{Small parametric deformations of off--diagonal solutions for
geometric flows and Ricci solitons}

\label{ssedef}

It is quite difficult to provide any physical interpretation of general
classes of solutions for geometric flows and Ricci solitons constructed in
previous subsections. Nevertheless, such theoretical and phenomenological
problems can be solved in a more simple form if we consider sub--sets of
solutions generated as deformations on a small parameter. It is supposed
that in certain limits and/or for special classes of (non) holonomic
constraints transform into well defined and/or known classes of physically
important solutions. We emphasize that even for small parameters, the
corresponding systems of nonlinear PDEs are generic nonlinear ones with
decoupling properties. Mathematically, the solutions can be constructed as
exact ones for certain sets of prescribed parameters and generating
functions. Small generic off--diagonal deformations of some known (or to
certain almost known solutions) are considered in this work only with the
aim to understand the physical meaning of some classes of geometric
evolution/ Ricci soliton solutions with small polarization/ running of
constants and nonlinear off-diagonal gravitational interactions in MGTs.

Let us consider a "prime " pseudo--Riemannian metric $\mathbf{\mathring{g}}=[%
\mathring{g}_{i},\mathring{h}_{a},\mathring{N}_{b}^{j}],$ when
\begin{eqnarray}
ds^{2} &=&\mathring{g}_{i}(x^{k})(dx^{i})^{2}+\mathring{h}%
_{a}(x^{k},y^{3})(dy^{a})^{2}(\mathbf{\mathring{e}}^{a})^{2},  \label{pm} \\
\mathbf{\mathring{e}}^{3} &=&dy^{3}+\mathring{w}_{i}(x^{k},y^{3})dx^{i},%
\mathbf{\mathring{e}}^{4}=dt+\mathring{n}_{i}(x^{k},y^{3})dx^{i}.  \notag
\end{eqnarray}%
Such a metric is diagonalizable if there is a coordinate transform $%
u^{\alpha \prime }=u^{\alpha \prime }(u^{\alpha })$%
\begin{equation*}
ds^{2}=\mathring{g}_{i^{\prime }}(x^{k\prime })(dx^{i^{\prime }})^{2}+%
\mathring{h}_{a\prime }(x^{k\prime })(dy^{a\prime })^{2},
\end{equation*}%
with $\mathring{w}_{i}=\mathring{n}_{i}=0.$ In order to construct exact
solutions with non--singular coordinate conditions it may be important to
work with "formal" off--diagonal parameterizations when the coefficients $%
\mathring{w}_{i}(x^{k},y^{3})$ and/or $\mathring{n}_{i}(x^{k},y^{3})$ are
not zero but the anholonomy coefficients $\mathring{W}_{\beta \gamma
}^{\alpha }(u^{\mu })=0,$ see (\ref{anhcoef}). We suppose that some data $(%
\mathring{g}_{i},\mathring{h}_{a})$ may define a diagonal exact solution in
MGT or in GR (for instance, a black hole, BH, configuration). Our goal is to
study certain small generic off--diagonal parametric deformations into
certain target metrics
\begin{eqnarray}
ds^{2} &=&\eta _{i}(x^{k})\mathring{g}_{i}(x^{k})(dx^{i})^{2}+\eta
_{a}(x^{k},y^{3})\mathring{g}_{a}(x^{k},y^{3})(\mathbf{e}^{a})^{2},
\label{targm} \\
\mathbf{e}^{3} &=&dy^{3}+\ ^{w}\eta _{i}\mathring{w}_{i}(x^{k},y^{3})dx^{i},%
\mathbf{e}^{4}=dt+\ ^{n}\eta _{i}\mathring{n}_{i}(x^{k},y^{3})dx^{i},  \notag
\end{eqnarray}%
where the coefficients $(g_{\alpha }=\eta _{\alpha }\mathring{g}_{\alpha
},^{w}\eta _{i}\mathring{w}_{i},\ ^{n}\eta _{i}n_{i})$ define, for instance,
a Ricci soliton configuration determined by a class of solutions (\ref%
{solut1t}). For certain well--defined conditions, we can express
\begin{eqnarray}
\eta _{i} &=&1+\varepsilon \chi _{i}(x^{k},y^{3}),\eta _{a}=1+\varepsilon
\chi _{a}(x^{k},y^{3})\mbox{ and }  \label{smpolariz} \\
\ ^{w}\eta _{i} &=&1+\varepsilon \ ^{w}\chi _{i}(x^{k},y^{3}),\ ^{n}\eta
_{i}=1+\varepsilon \ \ ^{n}\eta _{i}(x^{k},y^{3}),  \notag
\end{eqnarray}%
for a small parameter $0\leq \varepsilon \ll 1,$ when (\ref{targm})
transforms into (\ref{pm}) for $\varepsilon \rightarrow 0$ and $%
w_{i}=n_{i}=0.$ In general, there are not smooth limits from such
nonholonomic deformations which can be satisfied for arbitrary generation
and integration functions, integration constants and general (effective)
sources. The goal of this subsection is to analyze such conditions when $%
\varepsilon $-deformations with nontrivial N--connection coefficients can be
related to new classes of solutions of geometric flow and/or Ricci soliton
equations.

\subsubsection{$\protect\varepsilon $--deformations for stationary Ricci
solitons}

Let us denote nonholonomic $\varepsilon $--deformations of certain prime
d--metric (\ref{pm}) into a target one (\ref{targm}) with polarizations (\ref%
{smpolariz}) in the form $\mathbf{\mathring{g}}\rightarrow \ ^{\varepsilon }%
\mathbf{g=(}\ ^{\varepsilon }g_{i},\ ^{\varepsilon }h_{a},\ ^{\varepsilon
}N_{b}^{j}).$ The goal of this subsection is to compute the formulas for $%
\varepsilon $--deformations of prime d--metrics resulting in solutions of
type (\ref{riccisolt}), or (\ref{riccisolitlc}), for $\ _{\flat }\omega =1.$

Deformations of $h$-components are characterized by
\begin{equation*}
\ ^{\varepsilon }g_{i}=\mathring{g}_{i}(1+\varepsilon \chi _{i})=e^{\psi
(x^{k})}
\end{equation*}%
being a solution of (\ref{rs1}). For $\ _{\flat }\psi =\ \ _{\flat }^{0}\psi
(x^{k})+\varepsilon \ \ _{\flat }^{1}\psi (x^{k})$ and $_{\flat }^{\thicksim
}\overline{\Upsilon }=\ _{\flat }^{\tilde{0}}\overline{\Upsilon }(x^{k})+\
_{\flat }^{\tilde{1}}\overline{\Upsilon }(x^{k})$, we compute the
deformation polarization functions%
\begin{equation*}
\chi _{i}=e^{\ _{\flat }^{0}\psi }\ _{\flat }^{1}\psi /\mathring{g}_{i}\ \
_{\flat }^{\tilde{0}}\overline{\Upsilon }.
\end{equation*}%
In this formula, the generating and source functions are solutions of
\begin{equation*}
\ _{\flat }^{0}\psi ^{\bullet \bullet }+\ _{\flat }^{0}\psi ^{\prime \prime
}=\ _{\flat }^{\tilde{0}}\overline{\Upsilon }\mbox{ and }\ _{\flat }^{1}\psi
^{\bullet \bullet }+\ _{\flat }^{1}\psi ^{\prime \prime }=\ \ _{\flat }^{%
\tilde{1}}\overline{\Upsilon }.
\end{equation*}

We compute $\varepsilon $--deformations of $v$--components,
\begin{eqnarray}
\ _{\flat }^{\varepsilon }h_{3} &=&\frac{\epsilon _{3}\epsilon _{4}}{4}\frac{%
(\ _{\flat }\Psi ^{\ast })^{2}}{\ _{\flat }\overline{\Upsilon }^{2}}\left(
h_{4}^{[0]}+\frac{\epsilon _{3}\epsilon _{4}}{4}\int dy^{3}\frac{(\ _{\flat
}\Psi ^{2})^{\ast }}{\ _{\flat }\overline{\Upsilon }}\right)
^{-1}=(1+\varepsilon \ \chi _{3})\mathring{g}_{3};  \label{h3b} \\
\ _{\flat }^{\varepsilon }h_{4} &=&h_{4}^{[0]}(x^{k})+\frac{\epsilon
_{3}\epsilon _{4}}{4}\int dy^{3}\frac{(\ _{\flat }\Psi ^{2})^{\ast }}{\
_{\flat }\overline{\Upsilon }}=(1+\varepsilon \chi _{4})\mathring{g}_{4}.
\label{h4b}
\end{eqnarray}%
Parameterizing the generation function
\begin{equation}
\ _{\flat }\Psi =\ _{\flat }^{\varepsilon }\Psi =\mathring{\Psi}%
(x^{k},y^{3})[1+\varepsilon \chi (x^{k},y^{3})],  \label{aux5}
\end{equation}%
we introduce this value in (\ref{h4b}). We obtain
\begin{equation}
\chi _{4}=\frac{\epsilon _{3}\epsilon _{4}}{4\mathring{g}_{4}}\int dy^{3}%
\frac{(\mathring{\Psi}^{2}\chi )^{\ast }}{\ _{\flat }\overline{\Upsilon }}%
\mbox{ and }\int dy^{3}\frac{(\mathring{\Psi}^{2})^{\ast }}{\ _{\flat }%
\overline{\Upsilon }}=4\epsilon _{3}\epsilon _{4}(\mathring{g}%
_{4}-h_{4}^{[0]}).  \label{cond2a}
\end{equation}%
Such formulas show that we can compute $\chi _{4}$ for any deformation $\chi
$ from a 2-hypersurface $y^{3}=y^{3}(x^{k})$ defined in non-explicit form
from $\mathring{\Psi}=\mathring{\Psi}(x^{k},y^{3})$ when the integration
function $h_{4}^{[0]}(x^{k}),$ the prime value $\mathring{g}_{4}(x^{k})$ and
the fraction $(\mathring{\Psi}^{2})^{\ast }/\ _{\flat }\overline{\Upsilon }$
satisfy the condition (\ref{cond2a}).

We can find the formula for hypersurface $\mathring{\Psi}(x^{k},y^{3})$ by
finding the value of $\overline{\Upsilon }.$ Introducing (\ref{aux5}) into (%
\ref{h3b}), we get%
\begin{equation*}
\chi _{3}=2(\chi +\frac{\mathring{\Psi}}{\mathring{\Psi}^{\ast }}\chi ^{\ast
})-\chi _{4}=2(\chi +\frac{\mathring{\Psi}}{\mathring{\Psi}^{\ast }}\chi
^{\ast })-\frac{\epsilon _{3}\epsilon _{4}}{4\mathring{g}_{4}}\int dy^{3}%
\frac{(\mathring{\Psi}^{2}\chi )^{\ast }}{\ _{\flat }\overline{\Upsilon }}
\end{equation*}%
which allows to compute $\chi _{3}$ for any data $\left( \mathring{\Psi},%
\mathring{g}_{4},\chi \right) .$ $\ $The formula for a compatible source is
\begin{equation*}
\ _{\flat }\overline{\Upsilon }=\pm \mathring{\Psi}^{\ast }/2\sqrt{|%
\mathring{g}_{3}h_{4}^{[0]}|},
\end{equation*}%
which transforms (\ref{cond2a}) into a 2-d hypersurface formula $%
y^{3}=y^{3}(x^{k})$ defined in non-explicit form from
\begin{equation}
\int dy^{3}\mathring{\Psi}=\pm (h_{4}^{[0]}-\mathring{g}_{4})/\sqrt{|%
\mathring{g}_{3}h_{4}^{[0]}|}.  \label{cond2b}
\end{equation}

The $\varepsilon $--deformations of d--metric and N--connection coefficients
$\ _{\flat }w_{i}(\tau _{0})=\partial _{i}\ _{\flat }\Psi /\ _{\flat }\Psi
^{\ast }$ for nontrivial $\ \mathring{w}_{i}(\tau _{0})=\partial _{i}\
\mathring{\Psi}/\ \mathring{\Psi}^{\ast }$ are found following formulas (\ref%
{aux5}) and (\ref{smpolariz}),%
\begin{equation*}
\ ^{w}\chi _{i}=\frac{\partial _{i}(\chi \ \mathring{\Psi})}{\partial _{i}\
\mathring{\Psi}}-\frac{(\chi \ \mathring{\Psi})^{\ast }}{\mathring{\Psi}%
^{\ast }},
\end{equation*}%
where there is not summation on index $i.$ In a similar way, one computes
the deformation on the $n$--coefficients (we omit such details).

Summarizing above formulas, we obtain such coefficients for $\varepsilon $%
--deformations of a prime metric (\ref{pm}) into a target Ricci soliton
stationary metric:
\begin{eqnarray}
\ ^{\varepsilon }g_{i}(\tau _{0}) &=&\mathring{g}_{i}[1+\varepsilon \chi
_{i}]=[1+\varepsilon e^{\ _{\flat }^{0}\psi }\ _{\flat }^{1}\psi /\
\mathring{g}_{i}\ _{\flat }^{\tilde{0}}\overline{\Upsilon }]\mathring{g}_{i}%
\mbox{ as
a solution of 2-d Poisson equations (\ref{rs1})};  \notag \\
\ _{\flat }^{\varepsilon }h_{3}(\tau _{0}) &=&[1+\varepsilon \ \chi _{3}]%
\mathring{g}_{3}=\left[ 1+\varepsilon \ \left( 2(\chi +\frac{\mathring{\Psi}%
}{\mathring{\Psi}^{\ast }}\chi ^{\ast })-\frac{\epsilon _{3}\epsilon _{4}}{4%
\mathring{g}_{4}}\int dy^{3}\frac{(\mathring{\Psi}^{2}\chi )^{\ast }}{\
_{\flat }\overline{\Upsilon }}\right) \right] \mathring{g}_{3};  \notag \\
\ _{\flat }^{\varepsilon }h_{4}(\tau _{0}) &=&[1+\varepsilon \ \chi _{4}]%
\mathring{g}_{4}=\left[ 1+\varepsilon \frac{\epsilon _{3}\epsilon _{4}}{4%
\mathring{g}_{4}}\int dy^{3}\frac{(\mathring{\Psi}^{2}\chi )^{\ast }}{\
_{\flat }\overline{\Upsilon }}\right] \mathring{g}_{4};  \label{ersdef} \\
\ _{\flat }^{\varepsilon }w_{i}(\tau _{0}) &=&[1+\varepsilon \ ^{w}\chi _{i}]%
\mathring{w}_{i}=\left[ 1+\varepsilon (\frac{\partial _{i}(\chi \ \mathring{%
\Psi})}{\partial _{i}\ \mathring{\Psi}}-\frac{(\chi \ \mathring{\Psi})^{\ast
}}{\mathring{\Psi}^{\ast }})\right] \mathring{w}_{i};  \notag \\
\ _{\flat }^{\varepsilon }n_{i}(\tau _{0}) &=&[1+\varepsilon \ ^{n}\chi _{i}]%
\mathring{n}_{i}=\left[ 1+\varepsilon \ \widetilde{n}_{i}\int dy^{3}\ \frac{1%
}{_{\flat }\overline{\Upsilon }^{2}}\left( \chi +\frac{\mathring{\Psi}}{%
\mathring{\Psi}^{\ast }}\chi ^{\ast }-\frac{5}{8}\frac{\epsilon _{3}\epsilon
_{4}}{\mathring{g}_{4}}\frac{(\mathring{\Psi}^{2}\chi )^{\ast }}{\ _{\flat }%
\overline{\Upsilon }}\right) \right] \mathring{n}_{i},  \notag
\end{eqnarray}%
where $\ \widetilde{n}_{i}(x^{k})$ is a re-defined integration function
including contributions from the prime metric. The corresponding quadratic
element
\begin{eqnarray}
ds_{\varepsilon tRs}^{2} &=&\ _{\flat }^{\varepsilon }g_{\alpha \beta
}(x^{k},y^{3})du^{\alpha }du^{\beta }=\ ^{\varepsilon }g_{i}\left(
x^{k}\right) [(dx^{1})^{2}+(dx^{2})^{2}]+  \label{riccisoltdef} \\
&&\ _{\flat }^{\varepsilon }h_{3}(x^{k},y^{3})\ [dy^{3}+\ _{\flat
}^{\varepsilon }w_{i}(x^{k},y^{3})dx^{i}]^{2}+\ _{\flat }^{\varepsilon
}h_{4}(x^{k},y^{3})[dt+\ _{\flat }^{\varepsilon }n_{k}\
(x^{k},y^{3})dx^{k}]^{2}.  \notag
\end{eqnarray}%
We can subject additional constraints in order to extract LC--configurations
as we considered in (\ref{riccisolitlc}).

\subsubsection{Geometric flow evolution of $\protect\varepsilon $--deformed
stationary Ricci solitons}

Introducing the data (\ref{riccisoltdef}) into (\ref{runningconst}), we find
the quadratic element
\begin{eqnarray}
ds_{tRs}^{2} &=&g_{\alpha \beta }(\tau ,x^{k},y^{3})du^{\alpha }du^{\beta
}=e^{2\int d\tau \widetilde{\Lambda }(\tau )}\ \ ^{\varepsilon }g_{i}\left(
x^{k}\right) [(dx^{1})^{2}+(dx^{2})^{2}]+[1+\ _{\bot }\varepsilon (\tau )]
\label{runningconstdef} \\
&&\{\frac{\epsilon _{3}\epsilon _{4}}{4\ \ _{\flat }^{\varepsilon
}h_{4}(x^{k},y^{3})}\left( \frac{\ \ _{\flat }^{\varepsilon }\Psi ^{\ast }}{%
\overline{\Upsilon }_{[0]}}\right) ^{2}\ \left[ dy^{3}+\ _{\flat
}^{\varepsilon }w_{i}dx^{i}\right] ^{2}+\ \ \ \ _{\flat }^{\varepsilon
}h_{4}(x^{k},y^{3})\left[ \ dt+\ _{\flat }^{\varepsilon }n_{k}dx^{k}\right]
^{2}\}.  \notag
\end{eqnarray}%
For simplicity, we do not linearize on $\varepsilon $ in $(\ _{\flat
}^{\varepsilon }\Psi ^{\ast })^{2}/\ _{\flat }^{\varepsilon }h_{4}$, which
is determined by any generating function $\chi (x^{k},y^{3})$ and
corresponding integration functions.

\section{Geometric Evolution of Black Ellipsoids for Ricci Solitons and $%
R^{2}$ Gravity}

\label{s4} The goal of this section is to construct explicit examples of
stationary geometric flow and Ricci soliton exact solutions of type (\ref%
{riccisolt}), (\ref{runningconst}), (\ref{riccisolitlc}) and (\ref%
{ricciflsollc}) which for certain classes of nonholonomic constraints and
well--defined limits transform into black hole solutions for $R^{2}$ gravity
\cite{kehagias}. We construct also a model of geometric evolution of $R^{2}$
black holes into 3-d KdP solitonic (from Kadomtsev--Petviashvili, references
and geometric methods in \cite{kdp,belinski,sv2001,vsolit1,vsolit2,tgovsv})
configurations for non--factorizable solutions.

\subsection{Ricci solitonic black ellipsoids and limits to black hole
solutions in $R^{2}$ and GR theories}

\label{ssrsbel}

The techniques of $\varepsilon $--deformations outlined in section \ref%
{ssedef} is applyied for off--diagonal generalizations of "prime" black hole
solutions.

\subsubsection{Prime and target metrics}

Let us consider a "prime" metric
\begin{eqnarray}
ds^{2} &=&\mathring{g}_{\alpha ^{\prime }\beta ^{\prime }}(x^{k\prime
})du^{\alpha ^{\prime }}du^{\beta ^{\prime }}=(1-\frac{M}{r}+\frac{K}{r^{2}}%
)^{-1}dr^{2}+r^{2}d\theta ^{2}+r^{2}\sin \theta d\varphi ^{2}-(1-\frac{M}{r}+%
\frac{K}{r^{2}})dt^{2}  \notag \\
&=&\mathring{g}_{1^{\prime }}(dx^{1^{\prime }})^{2}+\mathring{g}_{2^{\prime
}}(x^{1^{^{\prime }}})(dx^{2^{\prime }})^{2}+\mathring{h}_{3^{\prime
}}(x^{1^{\prime }},x^{2^{\prime }})(dy^{3^{\prime }})^{2}+\mathring{h}%
_{4^{\prime }}(x^{1^{\prime }})(dy^{4^{\prime }})^{2},  \label{pm1}
\end{eqnarray}%
for some constants $M$ and $K,$ where%
\begin{eqnarray*}
x^{1^{^{\prime }}}(r) &=&\int dr(1-\frac{M}{r}+\frac{K}{r^{2}}%
)^{-1/2},x^{2^{\prime }}=\theta ,y^{3^{\prime }}=\varphi ,y^{4^{\prime }}=t;
\\
\mathring{g}_{1^{\prime }} &=&1,\mathring{g}_{2^{\prime }}(x^{1^{\prime
}})=r^{2}(x^{1^{\prime }}),\mathring{h}_{3^{\prime }}=r^{2}(x^{1^{\prime
}})\sin (x^{2^{\prime }}),\mathring{h}_{4^{\prime }}=-(1-\frac{M}{%
r(x^{1^{\prime }})}+\frac{K}{r^{2}(x^{1^{\prime }})})
\end{eqnarray*}%
are defined on certain carts on an open region $U\subset V,$ where $%
x^{1^{\prime }}(r)$ allows to find $r(x^{1^{\prime }})$ in a unique form.
This metric was studied as a spherical symmetric vacuum solution in $R^{2}$
gravity \cite{kehagias} (in our approach, of the Ricci soliton equations (%
\ref{riccisol4d})). Such a solution does not exist if $R=0$ (for
LC-configurations) because it is not allowed by transforms (\ref{overlmetric}%
).

We consider a coordinate transform $u^{\alpha ^{\prime }}=u^{\alpha ^{\prime
}}(u^{\alpha })$ with $\varphi =\varphi (y^{3},x^{k})$ and $%
t=t(y^{4},x^{k}). $ In such cases,
\begin{equation*}
d\varphi =\frac{\partial \varphi }{\partial y^{3}}[dy^{3}+(\partial
_{3}\varphi )^{-1}(\partial _{k}\varphi )dx^{k}]\mbox{ and }dt=\frac{%
\partial t}{\partial y^{4}}[dy^{4}+(\partial _{4}t)^{-1}(\partial
_{k}t)dx^{k}]
\end{equation*}%
for $\partial _{i}\varphi =\partial \varphi /\partial x^{i}$ and $\partial
_{a}\varphi =\partial \varphi /\partial y^{a}.$ Choosing
\begin{equation*}
\mathring{w}_{i}=\partial _{i}\ \mathring{\Psi}/\ \mathring{\Psi}^{\ast
}=(\partial _{3}\varphi )^{-1}(\partial _{i}\varphi )\mbox{ and }\mathring{n}%
_{i}=\partial _{i}n(x^{k})=(\partial _{4}t)^{-1}(\partial _{i}t),
\end{equation*}%
for any $\mathring{\Psi}$ (\ref{cond2b}), we express (\ref{pm1}) as
\begin{eqnarray}
ds^{2}=\mathring{g}_{1^{\prime }}(dx^{1^{\prime }})^{2}+\mathring{g}%
_{2^{\prime }}(x^{1^{^{\prime }}})(dx^{2^{\prime }})^{2}+\mathring{g}%
_{3}[dy^{3}+\mathring{w}_{i}(x^{k})dx^{i}]^{2}+\mathring{g}%
_{4}(x^{k}(x^{k^{\prime }}))[dy^{4}+\mathring{n}_{i}(x^{k})dx^{i}]^{2},
\label{pm1b} \\
\mbox{ for } \mathring{g}_{3}(x^{k}(x^{k^{\prime }}))=(\partial _{3}\varphi
)^{2}r^{2}(x^{1^{\prime }})\sin (x^{2^{\prime }})\mbox{ and }\mathring{g}%
_{4}(x^{k}(x^{k^{\prime }}))=-(\partial _{4}t)^{2}(1-\frac{M}{r}+\frac{K}{%
r^{2}}).  \label{pm1bv}
\end{eqnarray}%
This is a "formal" off--diagonal metric of type (\ref{pm}) with nontrivial
values $\mathring{h}_{a}^{\ast },\mathring{w}_{i}$ and $\mathring{n}_{i},$
but $\mathring{W}_{\beta \gamma }^{\alpha }(u^{\mu })=0,$ see (\ref{anhcoef}%
). Using such an ansatz, we can apply the AFDM with $\varepsilon $%
--deformation of geometric/ physical objects and physical parameters as we
described in subsection \ref{ssedef}.

Our goal is to show how the metric (\ref{pm1}) and/or (\ref{pm1b}) can be
off--diagonally deformed into certain classes of "target" new solutions of
type (\ref{riccisolitlc}) and (\ref{ricciflsollc}) with ellipsoidal
configurations which have a well--defined physical interpretation. In a
similar way, we can consider solutions with nonholonomically induced torsion
if $\mathring{W}_{\beta \gamma }^{\alpha }(u^{\mu })\neq 0.$ \ The condition
$R\neq 0$ together with $\widehat{R}\neq 0,$ see (\ref{sdcurv}), can be
preserved for such solutions. Here we note that geometric flows with $%
\widehat{R}\neq 0$ may allow evolution of solutions via $R=0$ by
transforming one class of Ricci soliton solutions for $R^{2}$ gravity into
another one. We consider target metrics (\ref{targm}) with $\varepsilon $%
--deformations (\ref{smpolariz}) resulting in solutions of geometric flow or
Ricci soliton equations,
\begin{eqnarray}
ds^{2} &=&\left[ 1+\varepsilon \chi _{i}(\tau ,x^{k},y^{3})\right] \mathring{%
g}_{i}(x^{k})(dx^{i})^{2}+[1+\varepsilon \chi _{a}(\tau ,x^{k},y^{3})]%
\mathring{g}_{a}(x^{k},y^{3})(\mathbf{e}^{a})^{2},  \label{target1aa} \\
\mathbf{e}^{3} &=&dy^{3}+[1+\varepsilon \ ^{w}\chi _{i}(\tau ,x^{k},y^{3})]%
\mathring{w}_{i}(x^{k},y^{3})dx^{i},\mathbf{e}^{4}=dy^{4}+[1+\varepsilon \ \
^{n}\chi _{i}(\tau ,x^{k},y^{3})]\mathring{n}_{i}(x^{k},y^{3})dx^{i}.  \notag
\end{eqnarray}%
For such target metrics, we can fix $\tau =\tau _{0}$ in order to generate
new Ricci soliton configurations or to consider factorizations of type (\ref%
{factoriz}). The generating function $\Psi =\check{\Psi}$ and $N_{i}^{4}$%
--coefficients are subjected to the conditions (\ref{cond1}) and (\ref{cond2}%
) for a function $\check{A}$ determined as a solution of $\partial _{i}%
\check{\Psi}=(\partial _{i}\check{A})\check{\Psi}^{\ast }.$ In this
quadratic element, there are also considered the so--called polarization
functions $\eta _{\alpha }(\tau ,x^{k},y^{a})\simeq 1+\varepsilon \chi
_{\alpha }(\tau ,x^{k},y^{a})$ which can be used for computing small
parametric geometric flow and/or generic off--diagonal deformation effects
from a prime metric. If for a class of solutions there are smooth limits $%
\varepsilon \rightarrow 0$ and $N_{i}^{a}\rightarrow 0,$ we obtain that $%
g_{\alpha \beta }$ (\ref{target1aa}) $\longrightarrow \mathring{g}_{\alpha
\beta }$ (\ref{pm1b}). For general nonlinear generic off--diagonal geometric
evolution and/or gravitational interactions such limits with $\varepsilon
\rightarrow 0$ do not exist. Nevertheless, it is important to study
subclasses of solutions with smooth configurations for $\varepsilon
\rightarrow 0$ and $N_{i}^{a}\rightarrow 0$ because it is more "easy" to
provide certain physical interpretation for such metrics.

Finally, we note that we can impose additional constraints on (\ref%
{target1aa}) in order to model, for instance, geometric evolution of Ricci
solitons with a $\tau =\tau _{0}$ fixed N--connection structure when $\
^{w}\chi _{i}=\ ^{w}\chi _{i}(x^{k},y^{3})$ and $\ ^{n}\chi _{i}=\ ^{n}\chi
_{i}(x^{k},y^{3}).$ We chose $\ _{\flat }^{\varepsilon }\Psi =\mathring{\Psi}%
(x^{k},y^{3})[1+\varepsilon \chi (x^{k},y^{3})]$ as in (\ref{aux5}) and
consider
\begin{eqnarray}
&&ds^{2}=g_{\alpha \beta }(\tau ,x^{k},y^{3})du^{\alpha }du^{\beta
}=[1+2\int d\tau \widetilde{\Lambda }(\tau )]\left[ 1+\varepsilon \chi
_{i}(\tau ,x^{k},y^{3})\right] \mathring{g}%
_{i}(x^{k})[(dx^{1})^{2}+(dx^{2})^{2}]+  \notag \\
&&[1+\ _{\bot }\varepsilon (\tau )]\{[1+\varepsilon \chi _{3}(x^{k},y^{3})]%
\mathring{g}_{3}(x^{k},y^{3})\left[ dy^{3}+\left( 1+\varepsilon \ ^{w}\chi
_{i}(x^{k},y^{3})\right) \mathring{w}_{i}(x^{k},y^{3})dx^{i}\right] ^{2}+
\notag \\
&&\ \ \ [1+\varepsilon \chi _{4}(x^{k},y^{3})]\mathring{g}_{4}(x^{k},y^{3})%
\left[ dy^{4}+[1+\varepsilon \ ^{n}\chi _{i}(x^{k},y^{3})]\mathring{w}%
_{i}(x^{k},y^{3})dx^{i}\right] ^{2}\}.  \label{target1b}
\end{eqnarray}%
We consider $\tau $--depending terms $2\int d\tau \widetilde{\Lambda }(\tau
) $ and $\ _{\bot }\varepsilon (\tau )$ to be of order $\varepsilon $\
introducing additional dependencies of physical constants and polarization
functions on flow parameter.

\subsubsection{Black ellipsoids in $R^{2}$ gravity as Ricci solitons}

Let us model an $\varepsilon $--deformation of (\ref{pm1}) into an
ellipsoidal Ricci soliton configuration when
\begin{eqnarray}
h_{4^{\prime }} &=&-(1-\frac{M}{r}+\frac{K}{r^{2}})\left[ 1-\varepsilon
\frac{M}{r}\times \frac{\cos (\omega _{0}\varphi +\varphi _{0})}{1-\frac{M}{r%
}+\frac{K}{r^{2}}}\right]  \label{h4epolar} \\
&=&\mathring{h}_{4^{\prime }}(x^{1^{\prime }})\left[ 1+\varepsilon \frac{M}{r%
}(\mathring{h}_{4^{\prime }})^{-1}\cos (\omega _{0}\varphi +\varphi _{0})%
\right] \simeq -\left[ 1-\frac{\ \widetilde{M}(\varphi )}{r}+\frac{K}{r^{2}}%
\right]  \label{h4horiz}
\end{eqnarray}%
for some constant values $\omega _{0}\varphi +\varphi _{0}$ and
anisotropically polarized mass
\begin{equation}
\ \widetilde{M}(\varphi )=M[1+\varepsilon \cos (\omega _{0}\varphi +\varphi
_{0})].  \label{anispolmass}
\end{equation}%
We obtain a zero value of $h_{4^{\prime }},$ i.e. the effective horizons
for (\ref{h4horiz}), if  $r_{\pm }(\varphi )=\frac{\widetilde{M}}{2}\pm
\frac{\widetilde{M}}{2}\sqrt{1-\frac{4K}{\widetilde{M}^{2}}}$. In the linear
approximations on $\varepsilon $ and $K,$ we write%
\begin{equation*}
r_{+}\simeq \frac{M}{1-\varepsilon \cos (\omega _{0}\varphi +\varphi _{0})},
\end{equation*}%
which is just the parametric equation of an ellipse with radial parameter $%
\mathring{r}_{+}=M$ and eccentricity $\varepsilon .$

Following formulas (\ref{ersdef}) for $\ _{\flat }^{\varepsilon }h_{4}(\tau
_{0})\simeq h_{4^{\prime }}$ from (\ref{h4epolar}), we can identify up to
coordinate transforms
\begin{equation*}
-\frac{1}{4}\int dy^{3}\frac{(\mathring{\Psi}^{2}\chi )^{\ast }}{\ _{\flat }%
\overline{\Upsilon }}=\frac{M}{r}\cos (\omega _{0}\varphi +\varphi _{0}).
\end{equation*}%
For $\ _{\flat }\overline{\Upsilon }=const,$ we find the polarization
function for ellipsoidal configurations
\begin{equation*}
\chi =\ ^{e}\chi =4\frac{M}{r}\ _{\flat }\overline{\Upsilon }\ \mathring{\Psi%
}^{-2}\cos (\omega _{0}\varphi +\varphi _{0}).
\end{equation*}%
Introducing $\ \ ^{e}\chi $ into formulas for d--metric coefficients (\ref%
{ersdef}), we compute for ellipsoid deformations of (\ref{pm1}),
\begin{eqnarray}
\ ^{e}g_{i}(\tau _{0}) &=&\mathring{g}_{i}[1+\varepsilon \chi
_{i}]=[1+\varepsilon e^{\ _{\flat }^{0}\psi }\ _{\flat }^{1}\psi /\
\mathring{g}_{i}\ _{\flat }^{\tilde{0}}\overline{\Upsilon }]\mathring{g}_{i}%
\mbox{ as
a solution of 2-d Poisson equations (\ref{rs1})};  \notag \\
\ _{\flat }^{e}h_{3}(\tau _{0}) &=&[1+\varepsilon \ \ ^{e}\chi _{3}]%
\mathring{g}_{3}=\left[ 1+\varepsilon \ \left( 2(\ ^{e}\chi +\frac{\mathring{%
\Psi}}{\mathring{\Psi}^{\ast }}\ ^{e}\chi ^{\ast })+\frac{1}{4\mathring{g}%
_{4}}\frac{\mathring{\Psi}^{2}\ ^{e}\chi }{\ _{\flat }\overline{\Upsilon }}%
\right) \right] \mathring{g}_{3};  \notag \\
\ _{\flat }^{e}h_{4}(\tau _{0}) &=&[1+\varepsilon \ \ ^{e}\chi _{4}]%
\mathring{g}_{4}=\left[ 1-\varepsilon \frac{1}{4\mathring{g}_{4}}\frac{(%
\mathring{\Psi}^{2}\ ^{e}\chi )}{\ _{\flat }\overline{\Upsilon }}\right]
\mathring{g}_{4};  \label{ellipsoidcoef} \\
\ _{\flat }^{e}w_{i}(\tau _{0}) &=&[1+\varepsilon \ ^{w}\chi _{i}]\mathring{w%
}_{i}=\left[ 1+\varepsilon (\frac{\partial _{i}(\ ^{e}\chi \ \mathring{\Psi})%
}{\partial _{i}\ \mathring{\Psi}}-\frac{(\ ^{e}\chi \ \mathring{\Psi})^{\ast
}}{\mathring{\Psi}^{\ast }})\right] \mathring{w}_{i};  \notag \\
\ _{\flat }^{e}n_{i}(\tau _{0}) &=&[1+\varepsilon \ ^{n}\chi _{i}]\mathring{n%
}_{i}=\left[ 1+\varepsilon \ \widetilde{n}_{i}\int dy^{3}\left( \ ^{e}\chi +%
\frac{\mathring{\Psi}}{\mathring{\Psi}^{\ast }}\ ^{e}\chi ^{\ast }-\frac{5}{8%
}\frac{\epsilon _{3}\epsilon _{4}}{\mathring{g}_{4}}\frac{(\mathring{\Psi}%
^{2}\ ^{e}\chi )^{\ast }}{\ _{\flat }\overline{\Upsilon }}\right) \right]
\mathring{n}_{i},  \notag
\end{eqnarray}%
where $\ \widetilde{n}_{i}(x^{k})$ is a re-defined integration function
including contributions from the prime metric. Re--defining the coordinates,
the corresponding quadratic element can be written in the form
\begin{eqnarray}
ds_{\varepsilon tRs}^{2} &=&\ _{\flat }^{e}g_{\alpha \beta }(x^{k},\varphi
)du^{\alpha }du^{\beta }=\ ^{e}g_{i}\left( x^{k}\right)
[(dx^{1})^{2}+(dx^{2})^{2}]+  \label{dmbellips} \\
&&\ _{\flat }^{e}h_{3}(x^{k},\varphi )\ [d\varphi +\ _{\flat
}^{e}w_{i}(x^{k},\varphi )dx^{i}]^{2}+\ _{\flat }^{e}h_{4}(x^{k},\varphi
)[dt+\ _{\flat }^{e}n_{k}\ (x^{k},\varphi )dx^{k}]^{2},  \notag
\end{eqnarray}%
where the coefficients are given by formulas (\ref{ellipsoidcoef}). We can
impose additional constraints in order to extract stationary
LC--configurations as we considered in (\ref{riccisolitlc}). Such solutions
with rotoid deformations were studied in \cite{tgovsv,svvvey} for certain
MGTs and higher dimension, see references therein. For small values of $%
\varepsilon $ and well defined asymptotic conditions, the metrics of type (%
\ref{dmbellips}) define black ellipsoid configurations which are stable. In
this section, such solutions were derived as generic off--diagonal Ricci
solitons.

Finally, we emphasize that black ellipsoids exist also in $R^{2}$ gravity as
it is encoded into effective source $\ _{\flat }\overline{\Upsilon }$ (\ref%
{effs}) determined by constants of such a theory, or for GR with effective
scalar field. The limit $\varepsilon \rightarrow 0$ is not allowed because
in such cases $R\rightarrow 0,$ as it was found in \cite{kehagias}. We
conclude that Ricci solitons with spherical symmetry do not exist in $R^{2}$
gravity but deformations to rotoid configurations are allowed in such a
theory.

\subsubsection{Geometric evolution of black ellipsoid Ricci solitons}

Geometric flows of ellipsoidal Ricci solitonic configurations with
factorized $\tau $--evolution can be described by metrics of type (\ref%
{target1b}). We have to use a set of coefficients (\ref{ellipsoidcoef}) for
a fixed self--similar configuration. The corresponding quadratic line
element is
\begin{eqnarray*}
&&ds^{2}=g_{\alpha \beta }(\tau ,x^{k},y^{3})du^{\alpha }du^{\beta
}=[1+2\int d\tau \widetilde{\Lambda }(\tau )]\ ^{e}g_{i}\left( x^{k}\right)
[(dx^{1})^{2}+(dx^{2})^{2}]+ \\
&&[1+\ _{\bot }\varepsilon (\tau )]\{\ _{\flat }^{e}h_{3}(x^{k},\varphi )
\left[ d\varphi +\ _{\flat }^{e}w_{i}(x^{k},\varphi )dx^{i}\right] ^{2}+\
_{\flat }^{e}h_{4}(x^{k},\varphi )[dt+\ _{\flat }^{e}n_{k}\ (x^{k},\varphi
)dx^{k}]^{2}\}.\
\end{eqnarray*}%
The explicit computation of the conditions of vanishing of the time-time
coefficient, $[1+\ _{\bot }\varepsilon (\tau )]$ $\ _{\flat
}^{e}h_{4}(x^{k},\varphi )=0,$ emphasizes two physical effects:

\begin{enumerate}
\item There are a running on $\tau $ mass (induced by geometric flows)  $\
_{\bot }M(\tau )=M(1+\ _{\bot }\varepsilon (\tau ))$ with an effective
locally anistropic mass $\ ^{e}M=M(1+\varepsilon \cos (\omega _{0}\varphi
+\varphi _{0})+\ _{\bot }\varepsilon (\tau ))$ containing contributions both
from evolution of geometric objects and generic off--diagonal deformations.

\item We can compute horizon deformations and modifications determined by
additional $\tau $-- and $\varphi $--depending ellipsoid deformations  $%
r_{+}(\tau ,\varphi ,\varepsilon )\simeq {M_{\bot }(\tau )}/  {1-\varepsilon
\cos (\omega _{0}\varphi +\varphi _{0})}$.
\end{enumerate}

In general, N--adapted geometric flow evolution results in running and
anisotropic polarization of physical constants, horizon deformations and
locally anisotropic polarizations of d--metric and N--connection
coefficients.

\subsection{Ricci flows and solitons for asymptotically de Sitter solutions}

Asymptotically de Sitter solutions with spherical symmetry for $R^{2}$
gravity were studied in \cite{kehagias}. Generic off--diagonal
ellipsoid--solitonic deformations of similar Kerr Sen black holes were
constructed in \cite{vkerrads}. Combining the results and methods of the
mentioned works, we can construct exact solutions for ellipsoidal de Sitter
Ricci solitons in MGTs and geometric flow evolution of such theories and
corresponding classes of solutions.

The metric
\begin{equation}
d\overline{s}^{2}=\frac{3\lambda }{2\varsigma ^{2}}\left\{ (1-\frac{M}{r}%
-\lambda r^{2})^{-1}dr^{2}+r^{2}d\theta ^{2}+r^{2}\sin \theta d\varphi
^{2}-(1-\frac{M}{r}-\lambda r^{2})dt^{2}\right\}  \label{ads}
\end{equation}%
for
\begin{equation*}
e^{\sqrt{1/3}\phi }=\frac{3\lambda }{2\varsigma ^{2}}=\frac{1}{8\varsigma
^{2}}R\text{\mbox{ and }}\overline{g}_{\mu \nu }=e^{\sqrt{1/3}\phi }g_{\mu
\nu }=\frac{R}{8\varsigma ^{2}}g_{\mu \nu },R\neq 0,
\end{equation*}%
define an exact solution with spherical symmetry in $R^{2}$ gravity,%
\begin{equation*}
\overline{R}_{\mu \nu }=2\varsigma ^{2}\overline{g}_{\mu \nu }.
\end{equation*}%
The asymptotically de Sitter solutions with $\lambda >0$ and $R\neq 0$
correspond to equations Re--defining the coordinates,
\begin{eqnarray*}
\widetilde{x}^{1^{^{\prime }}}(r) &=&\sqrt{\left\vert \frac{3\lambda }{2}%
\right\vert }\frac{1}{\varsigma }\int dr(1-\frac{M}{r}-\lambda r^{2})^{-1/2},%
\widetilde{x}^{2^{\prime }}=\theta ,y^{3^{\prime }}=\varphi ,y^{4^{\prime
}}=t; \\
\underline{\mathring{g}}_{1^{\prime }} &=&1,\underline{\mathring{g}}%
_{2^{\prime }}(\widetilde{x}^{1^{\prime }})=r^{2}(\widetilde{x}^{1^{\prime
}}),\mathring{h}_{3^{\prime }}=r^{2}(\widetilde{x}^{1^{\prime }})\sin
(x^{2^{\prime }}),\mathring{h}_{4^{\prime }}=-(1-\frac{M}{r(\widetilde{x}%
^{1^{\prime }})}+\lambda r^{2}(\widetilde{x}^{1^{\prime }})),
\end{eqnarray*}%
the metric (\ref{ads}) is written as a "prime" metric
\begin{equation*}
ds^{2}=\underline{\mathring{g}}_{\alpha ^{\prime }\beta ^{\prime }}(%
\widetilde{x}^{k\prime })du^{\alpha ^{\prime }}du^{\beta ^{\prime }}=%
\underline{\mathring{g}}_{1^{\prime }}(d\widetilde{x}^{1^{\prime }})^{2}+%
\underline{\mathring{g}}_{2^{\prime }}(\widetilde{x}^{1^{^{\prime }}})(d%
\widetilde{x}^{2^{\prime }})^{2}+\underline{\mathring{h}}_{3^{\prime }}(%
\widetilde{x}^{1^{\prime }},\widetilde{x}^{2^{\prime }})(dy^{3^{\prime
}})^{2}+\underline{\mathring{h}}_{4^{\prime }}(\widetilde{x}^{1^{\prime
}})(dy^{4^{\prime }})^{2},
\end{equation*}%
for some constants $M$ and $\lambda $ and $u^{\alpha }=(\widetilde{x}%
^{k\prime },y^{a}).$ To construct a "formal" off--diagonal metric of type (%
\ref{pm}) with nontrivial values $\mathring{h}_{a}^{\ast },\mathring{w}_{i}$
and $\mathring{n}_{i},$ but $\mathring{W}_{\beta \gamma }^{\alpha }(%
\widetilde{u}^{\mu })=0,$ see (\ref{anhcoef}), we consider a coordinate
transform $u^{\alpha ^{\prime }}=u^{\alpha ^{\prime }}(u^{\alpha })$ with $%
\varphi =\varphi (y^{3},\widetilde{x}^{k})$ and $t=t(y^{4},\widetilde{x}%
^{k}).$ For such transforms,
\begin{equation*}
d\varphi =\frac{\partial \varphi }{\partial y^{3}}[dy^{3}+(\partial
_{3}\varphi )^{-1}(\widetilde{\partial }_{k}\varphi )\widetilde{d}x^{k}]%
\mbox{ and }dt=\frac{\partial t}{\partial y^{4}}[dy^{4}+(\partial
_{4}t)^{-1}(\widetilde{\partial }_{k}t)\widetilde{d}x^{k}]
\end{equation*}%
for $\widetilde{\partial }_{i}\varphi =\partial \varphi /\partial \widetilde{%
x}^{i}$ and $\partial _{a}\varphi =\partial \varphi /\partial y^{a}.$
Choosing
\begin{equation*}
\underline{\mathring{w}}_{i}=\widetilde{\partial }_{i}\ \mathring{\Psi}/\
\mathring{\Psi}^{\ast }=(\partial _{3}\varphi )^{-1}(\widetilde{\partial }%
_{i}\varphi )\mbox{ and }\underline{\mathring{n}}_{i}=\widetilde{\partial }%
_{i}n(x^{k})=(\partial _{4}t)^{-1}(\widetilde{\partial }_{i}t),
\end{equation*}%
for any $\mathring{\Psi}$ (\ref{cond2b}), we express (\ref{ads}) as
\begin{eqnarray}
ds^{2}=\underline{\mathring{g}}_{1^{\prime }}(d\widetilde{x}^{1^{\prime
}})^{2}+\underline{\mathring{g}}_{2^{\prime }}(\widetilde{x}^{1^{^{\prime
}}})(d\widetilde{x}^{2^{\prime }})^{2}+\underline{\mathring{g}}_{3}[dy^{3}+%
\underline{\mathring{w}}_{i}(\widetilde{x}^{k})dx^{i}]^{2}+\underline{%
\mathring{g}}_{4}(x^{k}(\widetilde{x}^{k^{\prime }}))[dy^{4}+\underline{%
\mathring{n}}_{i}(\widetilde{x}^{k})d\widetilde{x}^{i}]^{2},  \label{pm2a} \\
\mbox{ for } \underline{\mathring{g}}_{3}(\widetilde{x}^{k}(\widetilde{x}%
^{k^{\prime }}))=(\partial _{3}\varphi )^{2}r^{2}(\widetilde{x}^{1^{\prime
}})\sin (\widetilde{x}^{2^{\prime }})\mbox{ and }\underline{\mathring{g}}%
_{4}(\widetilde{x}^{k}(\widetilde{x}^{k^{\prime }}))=-(\partial _{4}t)^{2}(1-%
\frac{M}{r}+\lambda r^{2}).  \label{pm2avert}
\end{eqnarray}%
The prime d--metric (\ref{pm2a}) allows us to apply the AFDM and construct $%
\varepsilon $--deformation of geometric/ physical objects and physical
parameters as we considered in details in subsections \ref{ssedef} and \ref%
{ssrsbel}.

\subsubsection{Asymptotically de Sitter black ellipsoids in $R^{2}$ gravity
as Ricci solitons}

For $\underline{\mathring{g}}_{4}=\underline{\mathring{h}}_{4}(\widetilde{x}%
^{1^{\prime }})=(1-\frac{M}{r}+\lambda r^{2})$ and $(\partial _{4}t)^{2}=1$
and anisotropically polarized mass $\ \widetilde{M}(\varphi
)=M[1+\varepsilon \cos (\omega _{0}\varphi +\varphi _{0})],$ we obtain
\begin{eqnarray*}
\ ^{s}h_{4} &=&-(1-\frac{M}{r}+\lambda r^{2})[1-\varepsilon \frac{M}{r}\frac{%
\cos (\omega _{0}\varphi +\varphi _{0})}{1-\frac{M}{r}+\lambda r^{2}}] \\
&=&\underline{\mathring{h}}_{4}(\widetilde{x}^{1^{\prime }})\left[
1-\varepsilon \frac{M}{r}(\underline{\mathring{h}}_{4})^{-1}\cos (\omega
_{0}\varphi +\varphi _{0})\right] \simeq -\left[ 1-\frac{\ \widetilde{M}%
(\varphi )}{r}+\lambda r^{2}\right]
\end{eqnarray*}%
The parametric equation of an ellipse with radial parameter $\mathring{r}%
_{+}=M$ and eccentricity $\varepsilon ,$
\begin{equation*}
r_{+}\simeq \frac{M}{1-\varepsilon \cos (\omega _{0}\varphi +\varphi _{0})},
\end{equation*}%
can be determined in a simple way for $\lambda =0.$ We have to \ find
solutions of a third order algebraic equation in order to determine possible
horizons for nontrivial $\lambda .$

We construct ellipsoidal deformations of d--metric (\ref{pm2a}) if
\begin{equation*}
\chi =\ ^{s}\chi =8\frac{M}{r}\varsigma ^{2}\ \mathring{\Psi}^{-2}\cos
(\omega _{0}\varphi +\varphi _{0}),
\end{equation*}%
when the former value $\ _{\flat }\overline{\Upsilon }$ is substituted into $%
2\varsigma ^{2}.$ \ Following the same method as in section \ref{ssrsbel}
but for $\ \ ^{s}\chi $ used for d--metric coefficients (\ref{ersdef}), we
compute
\begin{eqnarray}
\ ^{s}g_{i}(\tau _{0}) &=&\underline{\mathring{g}}_{i}[1+\varepsilon \chi
_{i}]=[1+\varepsilon e^{\ _{\flat }^{0}\psi }\ _{\flat }^{1}\psi /\
\underline{\mathring{g}}_{i}\ 2\varsigma ^{2}]\underline{\mathring{g}}_{i}%
\mbox{ solution of 2-d Poisson equations (\ref{rs1})};  \notag \\
\ ^{s}h_{3}(\tau _{0}) &=&[1+\varepsilon \ ^{s}\chi _{3}]\mathring{g}_{3}=%
\left[ 1+\varepsilon \ \left( 2(\ ^{s}\chi +\frac{\mathring{\Psi}}{\mathring{%
\Psi}^{\ast }}\ ^{s}\chi ^{\ast })+\frac{1}{8\varsigma ^{2}\underline{%
\mathring{g}}_{4}}\mathring{\Psi}^{2}\ ^{s}\chi \right) \right] \underline{%
\mathring{g}}_{3};  \notag \\
\ \ ^{s}h_{4}(\tau _{0}) &=&[1+\varepsilon \ \ ^{s}\chi _{4}]\mathring{g}%
_{4}=\left[ 1-\varepsilon \frac{1}{8\varsigma ^{2}\underline{\mathring{g}}%
_{4}}\mathring{\Psi}^{2}\ ^{s}\chi \right] \underline{\mathring{g}}_{4};
\label{scoeff} \\
\ \ ^{s}w_{i}(\tau _{0}) &=&[1+\varepsilon \ ^{w}\chi _{i}]\mathring{w}_{i}=%
\left[ 1+\varepsilon (\frac{\partial _{i}(\ ^{s}\chi \ \mathring{\Psi})}{%
\partial _{i}\ \mathring{\Psi}}-\frac{(\ ^{s}\chi \ \mathring{\Psi})^{\ast }%
}{\mathring{\Psi}^{\ast }})\right] \underline{\mathring{w}}_{i};  \notag \\
\ \ ^{s}n_{i}(\tau _{0}) &=&[1+\varepsilon \ ^{n}\chi _{i}]\mathring{n}_{i}=%
\left[ 1+\varepsilon \ \widetilde{n}_{i}\int dy^{3}\left( \ ^{e}\chi +\frac{%
\mathring{\Psi}}{\mathring{\Psi}^{\ast }}\ ^{e}\chi ^{\ast }+\frac{5}{%
16\varsigma ^{2}}\frac{1}{\underline{\mathring{g}}_{4}}(\mathring{\Psi}^{2}\
^{s}\chi )^{\ast }\right) \right] \underline{\mathring{n}}_{i},  \notag
\end{eqnarray}%
where $\ \widetilde{n}_{i}(x^{k})$ is a re-defined integration function
including contributions from the prime metric (\ref{pm2a}). \ The generating
functions $\ ^{s}\chi $ and $\ _{\flat }^{0}\psi $ can be determined for an
ellipsoid configuration induced by the effective cosmological constant \ $%
\varsigma ^{2}$ in $R^{2}$ gravity.

The solutions for stationary generic off--diagonal Ricci solitons (\ref%
{scoeff}) encode also the data for a black hole metric $[\underline{%
\mathring{g}}_{i},\underline{\mathring{g}}_{a},\underline{\mathring{w}}_{i},%
\underline{\mathring{n}}_{i}]$ with a prime generating function $\mathring{%
\Psi}$ fixed by a 2-d hypersurface (\ref{cond2b}). This reflects the fact
that we parameterize the ellipsoid small deformations in N--adapted form.

\subsubsection{Geometric evolution of asymptotically de Sitter black
ellipsoid Ricci solitons}

The corresponding quadratic element with $\varepsilon $--deformations and
factorized $\tau $--evolution are computed as in (\ref{target1b}),
\begin{eqnarray*}
ds^{2} &=&\ \ ^{s}g_{\alpha \beta }(\tau ,\widetilde{x}^{k},\varphi
)du^{\alpha }du^{\beta }=[1+2\int d\tau \widetilde{\Lambda }(\tau )]\
^{s}g_{i}\left( \widetilde{x}^{k}\right) [(d\widetilde{x}^{i})^{2}]+[1+\
_{\bot }\varepsilon (\tau )] \\
&&\{\ ^{s}h_{3}(\widetilde{x}^{k},\varphi )\ [d\varphi +\ ^{s}w_{i}(%
\widetilde{x}^{k},\varphi )dx^{i}]^{2}+\ ^{s}h_{4}(\widetilde{x}^{k},\varphi
)[dt+\ ^{s}n_{k}\ (\widetilde{x}^{k},\varphi )d\widetilde{x}^{k}]^{2}\},
\end{eqnarray*}%
is determined by the coefficients $\ ^{s}g_{\alpha \beta }$ (\ref{scoeff}).
The evolution of such self--similar ellipsoidal configurations is
characterized by locally anisotropic polarizations and running of physical
constants. For instance, the effective mass modifications are parameterized
in a form similar as for black ellipsoids considered in previous subsection
when $\ ^{e}M=M(1+\varepsilon \cos (\omega _{0}\varphi +\varphi _{0})+\
_{\bot }\varepsilon (\tau ).$ Such values have to be defined from
experimental data.

LC--configurations can be extracted by additional nonholonomic constraints
as we described in previous sections. This is also an issue for experimental
verifications of MGTs and possible limits to GR and equivalent modelling.

\subsection{ Geometric evolution as 3-d KdV configurations}

In a different context, the geometric evolution of certain black hole/
ellipsoid and/or Ricci soliton configurations can be characterized by
solitonic wave solutions which provide examples of generic nonlinear
evolution models.

Let us consider the class of metrics (\ref{ggeomfl}) when, for simplicity, $%
\omega =1.$ We generate families of 3--d solitonic wave equation of
Kadomtev--Petviashvili (KP) type, see details in \cite%
{kdp,belinski,vsolit1,vsolit2,vkerrads}, if it is taken as a generating
function any $h_{4}(\tau ,x^{1},y^{3})=h(\tau ,x^{1},y^{3})$ being a
solution of\footnote{%
in a similar form, we can consider solution of any 3-d solitonic equations,
for instance, of generalized sine--Gordon ones}
\begin{equation}
\pm \partial _{11}^{2}h+(\partial _{\tau }h+hh^{\ast }+\epsilon h^{\ast \ast
\ast })^{\ast }=0.  \label{kdveq}
\end{equation}%
The so--called dispersionless limit is characterized by $\epsilon
\rightarrow 0$ and corresponding Burgers' equation $\partial _{\tau
}h+hh^{\ast }=0.$ Integrating above equation on $y^{3},$ we obtain
\begin{equation*}
\partial _{\tau }h_{4}=-h_{4}h_{4}^{\ast }-\epsilon h_{4}^{\ast \ast \ast
}\mp \int dy^{3}\partial _{11}^{2}h_{4}.
\end{equation*}%
Substituting this value in \ (\ref{effsourc}), we construct an effective
solitonic source%
\begin{equation}
\Upsilon =\Lambda _{0}-\ ^{\phi }\Lambda -2\varsigma ^{2}-hh^{\ast
}-\epsilon h^{\ast \ast \ast }\mp \int dy^{3}\partial _{11}^{2}h.
\label{sourckdv}
\end{equation}

Having a solution $h_{4}(\tau ,x^{1},y^{3}),$ we compute
\begin{equation*}
\Psi ^{2}=B(\tau ,x^{1})-\frac{4}{\Lambda _{0}}h_{4}\mbox{ and }h_{3}=-\frac{%
(h_{4}^{\ast })^{2}}{h_{4}[B(\tau ,x^{1})-\frac{4}{\Lambda _{0}}h_{4}]}
\end{equation*}%
for an integration function $B(\tau ,x^{1}).$ For simplicity, we can take $%
h_{3}=h_{4}=h$ and solve (\ref{rf2a}).

The next step is to use the algebraic equation (\ref{rf3a}) and find a
solution of type (\ref{auxf}),%
\begin{equation*}
w_{1}(\tau ,x^{1},y^{3})=\frac{\partial _{i}\Psi }{\Psi ^{\ast }}=\frac{%
\partial _{i}\Psi ^{2}}{\partial _{3}(\Psi ^{2})}=(h^{\ast })^{-1}\partial
_{i}[-\frac{\Lambda _{0}}{4}B(\tau ,x^{1})+h)],w_{2}=0,
\end{equation*}%
when $\Psi =\Psi (\tau ,x^{1},y^{3}).$ Integrating two times on $y^{3}$ in (%
\ref{rf4a}) and using the condition $h_{3}=h_{4},$ we obtain
\begin{equation*}
n_{k}(\tau ,x^{1},y^{3})=\ _{1}n_{k}(\tau ,x^{1})+\ _{2}\widetilde{n}%
_{k}(\tau ,x^{1})\int dy^{3}\ (\sqrt{|h|})^{-1}.
\end{equation*}

Summarizing the results in this subsection, we constructed a 3-d KdP
solitonic quadratic element
\begin{eqnarray}
&&ds_{KdP}^{2}=g_{\alpha \beta }(\tau ,x^{1},y^{3})du^{\alpha }du^{\beta
}=e^{\psi (\tau ,x^{k})}[(dx^{1})^{2}+(dx^{2})^{2}]+h(\tau ,x^{1},y^{3})
\label{solitevol} \\
&&\{\ [dy^{3}+\frac{\partial _{i}(-\frac{\Lambda _{0}}{4}B(\tau ,x^{1})+h)}{%
h^{\ast }}dx^{1}]^{2}+[dt+(\ _{1}n_{k}(x^{1})+\ _{2}\widetilde{n}%
_{k}(x^{1})\int dy^{3}\sqrt{|h|})^{-1})dx^{1}]^{2}\}.  \notag
\end{eqnarray}%
This class of solutions possesses two Killing vectors, $\partial _{2}$ and $%
\partial _{4}.$ Nevertheless, this defines a model with a quite general
evolution of N--connection coefficients and flows of the nonholonomically
induced torsion. Such stationary on time metrics are generic off--diagonal
and can be characterized by solitonic symmetries and derived solitonic
hierarchies, see details in Refs. \cite{vsolit1,vsolit2,vkerrads}.

\section{W--thermodynamics for Black Ellipsoids and Solitonic Flows in $R^2$
Gravity}

\label{s5} In this work, we constructed generic off--diagonal stationary
solutions of geometric flow and Ricci soliton equations modeling nonlinear
evolution and interactions in MGTs and GR. Using the W--entropy (\ref%
{perelm3w}), we can elaborate a statistical thermodynamics model
characterizing both the spacetime geometric evolution and fixed parameter
3--d configurations embedded in 4--d relativistic spacetimes.

Any d--metric can be parameterized in the form (\ref{coefft}). For 3--d
thermodynamical values, we obtain
\begin{eqnarray}
\ _{\shortmid }\widehat{\mathcal{E}}\ &=&\tilde{\tau}^{2}\int_{\widehat{\Xi }%
_{t}}\tilde{\mu}\sqrt{|q_{1}q_{2}q_{3}|}d\grave{x}^{3}\left( \_{\shortmid }%
\widehat{R}+|\ _{\shortmid }\widehat{\mathbf{D}}\tilde{f}|^{2}- \frac{3}{%
\tilde{\tau}}\right) ,  \label{thval3d} \\
\ _{\shortmid }\widehat{S} &=&\int_{\widehat{\Xi }_{t}}\tilde{\mu}\sqrt{%
|q_{1}q_{2}q_{3}|}d\grave{x}^{3}\left[ \tilde{\tau}\left( \ \mathbf{\ }%
_{\shortmid }\widehat{R}+|\ _{\shortmid }\widehat{\mathbf{D}}\tilde{f}%
|^{2}\right) +\tilde{f}-6\right] ,  \notag \\
\ _{\shortmid }\widehat{\sigma } &=&-2\ \tilde{\tau}^{4}\int_{\widehat{\Xi }%
_{t}}\tilde{\mu}\sqrt{|q_{1}q_{2}q_{3}|}d\grave{x}^{3}[|\ _{\shortmid }%
\widehat{\mathbf{R}}_{\grave{\imath}\grave{j}}+\ _{\shortmid }\widehat{%
\mathbf{D}}_{\grave{\imath}}\ _{\shortmid }\widehat{\mathbf{D}}_{\grave{j}}%
\tilde{f}-\frac{1}{2\tilde{\tau}}q_{\grave{\imath}\grave{j}}|^{2}],  \notag
\end{eqnarray}%
up to any parametric function $\tilde{\tau}(\tau )$ in $\tilde{\mu}=\left(
4\pi \tilde{\tau}\right) ^{-3}e^{-\tilde{f}}$ with any $\tilde{\tau}(\tau )$
for $\partial \tilde{\tau}/\partial \tau =-1$ and $\tau >0$. Taking
respective 3-d coefficients of a d-metric (\ref{riccisolt}), or (\ref%
{runningconst}), or (\ref{ggemfllc}) [or any solution of type ellipsoidal
deformed black hole solutions (\ref{ellipsoidcoef}), or (\ref{scoeff}), or a
KdP evolution model (\ref{solitevol})], and prescribing a closed 3-d
hypersurface $\widehat{\Xi }_{0}$, we can compute such values for any
effective source (\ref{effs}).

The vertical conformal factor $\omega (\tau ,x^{k},y^{3},t)$ in (\ref%
{thval3d}) depends (in general, for non-stationary solutions) on a time like
coordinate $t.$ In such cases, we have to consider relativistic evolution
models and integrate additionally on a time interval in order to compute for
$\ _{\shortmid }\widehat{\mathcal{E}}\ ,$ $\ _{\shortmid }\widehat{S},$ and $%
\ _{\shortmid }\widehat{\sigma }$ some values of type~(\ref{thermodv}). Such
constructions are elaborated for relativistic hydrodynamical geometric
models in \cite{vvrfthbh}. \ In this work, for simplicity we shall consider
stationary solutions with $\omega =1.$ We have to fix an explicit N--adapted
system of reference and scaling function $\tilde{f}$ \ in order to find
certain explicit values for corresponding average energy, entropy and
fluctuations for evolution on a time like parameter $t$ of any family of
closed hypersurfaces. For explicit examples, we can decide if certain
solutions with effective Lorentz-Ricci soliton source and/or with
contributions from additional MGT sources may be more convenient
thermodynamically than other configurations.

\subsection{Perelman's energy and entropy for stationary Ricci solitons and
their factorized geometric evolution}

Stating a configuration with $\tilde{f}=0$ and $_{\shortmid }\widehat{%
\mathbf{D}}\tilde{f}=0,$ we compute the values $\ _{\shortmid }\widehat{%
\mathcal{E}}$ and $\ _{\shortmid }\widehat{S}$ from (\ref{thval3d}) (for
simplicity, we omit more cumbersome computations for $\ _{\shortmid }%
\widehat{\sigma }$) for a d--metric \ of type (\ref{runningconst}). From
effective Einstein equations $\widehat{\mathbf{R}}_{\alpha \beta }=\mathbf{%
\Upsilon }_{\alpha \beta }$ with effective N--adapted source
\begin{eqnarray*}
\mathbf{\Upsilon }_{\alpha \beta } &=&diag[\ ^{\thicksim }\overline{\Upsilon
}(\tau ,x^{k}):=~\widetilde{\Upsilon }(\tau ,x^{k})+~\ ^{\phi }\widetilde{%
\Lambda }(\tau )+2\varsigma ^{2}(\tau ) \\
\mbox{ and }\overline{\Upsilon }(\tau ,x^{k},y^{3}):= &&\Upsilon (\tau
,x^{k},y^{3})+\ ^{\phi }\Lambda (\tau )+2\varsigma ^{2}(\tau )],
\end{eqnarray*}%
we find $\mathbf{\ }_{\shortmid }\widehat{R}=\ ^{\thicksim }\overline{%
\Upsilon }+\frac{1}{2}\overline{\Upsilon }.$ We have (see relevant formulas (%
\ref{solut1t}))
\begin{eqnarray*}
q_{1} &=&q_{2}=e^{2\int d\tau \widetilde{\Lambda }(\tau )}e^{\ ^{1}\psi
(x^{k})},q_{3}=-\{1+\ _{\bot }\varepsilon (\tau )\}\frac{1}{4\ \ \ _{\flat
}h_{4}}\left( \frac{\ \ _{\flat }\Psi ^{\ast }}{\overline{\Upsilon }_{[0]}}%
\right) ^{2}, \\
\ \mbox{ for }_{\flat }h_{4} &=&h_{4}^{[0]}(x^{k})-\frac{1}{4}\int dy^{3}%
\frac{(\ _{\flat }\Psi ^{2})^{\ast }}{\ _{\flat }\overline{\Upsilon }},
\end{eqnarray*}%
when for Ricci soliton, Rs, evolution
\begin{equation*}
\ ^{Rs}Q(\tau ,x^{k},y^{3}):=\sqrt{|q_{1}q_{2}q_{3}|}=\left( 1+2\int d\tau
\widetilde{\Lambda }(\tau )+\frac{1}{2}\ _{\bot }\varepsilon (\tau )\right)
\frac{e^{\ ^{1}\psi (x^{k})}\ _{\flat }\Psi ^{\ast }}{2\overline{\Upsilon }%
_{[0]}\sqrt{\ |\ _{\flat }h_{4}|}}
\end{equation*}%
is considered for small values $|2\int d\tau \widetilde{\Lambda }(\tau )|,|\
_{\bot }\varepsilon (\tau )|\ll 1.$ Introducing such data in in respective
formulas in (\ref{thval3d}) for redefined flow parameter, we obtain
\begin{eqnarray*}
\ _{\shortmid }\widehat{\mathcal{E}}\ \ &=&\tau ^{2}\int_{\widehat{\Xi }_{0}}%
\frac{dx^{1}dx^{2}dy^{3}}{\left( 4\pi \tau \right) ^{3}}\ ^{Rs}Q(\tau
,x^{k},y^{3})\left[ \ ^{\thicksim }\overline{\Upsilon }(\tau ,x^{k})+\frac{1%
}{2}\overline{\Upsilon }(\tau ,x^{k},y^{3})-\frac{3}{\tau }\right] , \\
\ _{\shortmid }\widehat{S} &=&\int_{\widehat{\Xi }_{0}}\frac{%
dx^{1}dx^{2}dy^{3}}{\left( 4\pi \tau \right) ^{3}}\ ^{Rs}Q(\tau ,x^{k},y^{3})%
\left[ \tau \left( \ \ ^{\thicksim }\overline{\Upsilon }(\tau ,x^{k})+\frac{1%
}{2}\overline{\Upsilon }(\tau ,x^{k},y^{3})\right) -6\right] .
\end{eqnarray*}%
In explicit form, such values can be computed if we prescribe corresponding
generating and integration functions, integration constants and fix a closed
3-d hypersurface. For a fixed $\tau _{0},$ these formulas can be used for
determining gravitational thermodynamic values of Ricci solitons.

\subsection{Non-factorized thermodynamic configurations for N--adapted
effective sources}

Similar formulas can be considered for the class of solutions of geometric
evolution equations (\ref{ggemfllc}) with
\begin{eqnarray*}
q_{1} &=&q_{2}=e^{\psi (\tau ,x^{k})},q_{3}=h_{4}(\tau ,x^{i},y^{3}), \\
&&\mbox{ for arbitrary generating function}h_{4}(\tau ,x^{i},y^{3}),
\end{eqnarray*}%
when related to source $\Upsilon +\ ^{\phi }\Lambda +2\varsigma
^{2}-\partial _{\tau }\ln |h_{4}|=\Lambda _{0}\neq 0,$ see (\ref{effsourc})
for $\omega =1.$ We obtain%
\begin{eqnarray*}
Q(\tau ,x^{k},y^{3}) &:=&\sqrt{|q_{1}q_{2}q_{3}|}=e^{\psi (\tau ,x^{k})}%
\sqrt{|h_{4}|}\mbox{ and } \\
\mathbf{\ }_{\shortmid }\widehat{R} &=&\ ^{\thicksim }\overline{\Upsilon }+%
\frac{1}{2}\Upsilon =\ ^{\thicksim }\overline{\Upsilon }(\tau ,x^{k})+\frac{1%
}{2}[\Lambda _{0}+\partial _{\tau }\ln |h_{4}(\tau ,x^{i},y^{3})|+\ ^{\phi
}\Lambda (\tau )+2\varsigma ^{2}(\tau )]
\end{eqnarray*}%
The thermodynamic values are
\begin{eqnarray*}
\ _{\shortmid }\widehat{\mathcal{E}} &=&\tau ^{2}\int_{\widehat{\Xi }_{0}}%
\frac{dx^{1}dx^{2}dy^{3}}{\left( 4\pi \tau \right) ^{3}}e^{\psi (\tau
,x^{k})}\sqrt{|h_{4}|} \\
&& \left\{ \ ^{\thicksim }\overline{\Upsilon }(\tau ,x^{k})+\frac{1}{2}%
[\Lambda _{0}+\partial _{\tau }\ln |h_{4}(\tau ,x^{i},y^{3})|+\ ^{\phi
}\Lambda (\tau )+2\varsigma ^{2}(\tau )]\ -\frac{3}{\tau }\right\} , \\
\ _{\shortmid }\widehat{S} &=&\int_{\widehat{\Xi }_{0}}\frac{%
dx^{1}dx^{2}dy^{3}}{\left( 4\pi \tau \right) ^{3}}\ e^{\psi (\tau ,x^{k})}%
\sqrt{|h_{4}|} \\
&& \left\{ \tau \left( \ ^{\thicksim }\overline{\Upsilon }(\tau ,x^{k})+%
\frac{1}{2}[\Lambda _{0}+\partial _{\tau }\ln |h_{4}(\tau ,x^{i},y^{3})|+\
^{\phi }\Lambda (\tau )+2\varsigma ^{2}(\tau )]\right) -6\right\} .
\end{eqnarray*}%
We can compute such values in explicit form for any generating functions $%
h_{4}(\tau ,x^{i},y^{3})$ and $\psi (\tau ,x^{k})$ and above mentioned
sources.

\subsection{W--energy and W--entropy for black ellipsoids and solitons in $%
R^{2}$ gravity}

\subsubsection{Thermodynamic values for asymptotic de Sitter black ellipsoids%
}

We use the d--metric coefficients (\ref{scoeff}) constructed as $\varepsilon
$--deformations of the prime black hole solution (\ref{ads}) for generating
function $\ ^{s}\chi =8\frac{M}{r\mathring{\Psi}^{2}(r,\theta ,\varphi )}%
\varsigma ^{2}\ \cos (\omega _{0}\varphi +\varphi _{0}),$ when $\mathbf{\ }%
_{\shortmid }\widehat{R}=6\varsigma ^{2}.$ Parameterizing
\begin{equation*}
q_{i}=[1+\varepsilon e^{\ _{\flat }^{0}\psi }\ _{\flat }^{1}\psi /\
\underline{\mathring{g}}_{i}\ 2\varsigma ^{2}]\mathring{g}_{i},q_{3}=\left[
1+\varepsilon \ \left( 2(\ ^{s}\chi +\frac{\mathring{\Psi}}{\mathring{\Psi}%
^{\ast }}\ ^{s}\chi ^{\ast })+\frac{1}{8\varsigma ^{2}\underline{\mathring{g}%
}_{4}}\mathring{\Psi}^{2}\ ^{s}\chi \right) \right] \underline{\mathring{g}}%
_{3},
\end{equation*}%
we find%
\begin{eqnarray*}
\ ^{ds}Q(x^{k},y^{3}):=\sqrt{|q_{1}q_{2}q_{3}|} &= &\sqrt{|\underline{%
\mathring{g}}_{1}\underline{\mathring{g}}_{2}\underline{\mathring{g}}_{3}|}%
[1+\varepsilon (e^{\ _{\flat }^{0}\psi }\ _{\flat }^{1}\psi /\ \underline{%
\mathring{g}}_{1}\ 4\varsigma ^{2}+e^{\ _{\flat }^{0}\psi }\ _{\flat
}^{1}\psi /\ \underline{\mathring{g}}_{2}\ 4\varsigma ^{2})+ \\
&&\varepsilon (\ ^{s}\chi +\frac{\mathring{\Psi}}{\mathring{\Psi}^{\ast }}\
^{s}\chi ^{\ast }\frac{1}{16\varsigma ^{2}\underline{\mathring{g}}_{4}}%
\mathring{\Psi}^{2}\ ^{s}\chi )],
\end{eqnarray*}%
for $\sqrt{|\underline{\mathring{g}}_{1}\underline{\mathring{g}}_{2}%
\underline{\mathring{g}}_{3}|}=r^{2}(\widetilde{x}^{1^{\prime }})\sin
\widetilde{x}^{2^{\prime }}(\theta ).$ Introducing such values in (\ref%
{thval3d}), we get %\begin{eqnarray*}
\begin{equation*}
\ _{\shortmid }\widehat{\mathcal{E}}\ =\tau _{0}^{2}\int_{\widehat{\Xi }_{0}}%
\frac{dx^{1}dx^{2}dy^{3}}{\left( 4\pi \tau _{0}\right) ^{3}}\
^{ds}Q(x^{k},y^{3})\left[ 6\varsigma ^{2}\ -\frac{3}{\tau _{0}}\right] ,\ \
_{\shortmid }\widehat{S} =\int_{\widehat{\Xi }_{0}}\frac{dx^{1}dx^{2}dy^{3}}{%
\left( 4\pi \tau _{0}\right) ^{3}}\ ^{Rs}Q(x^{k},y^{3})\left[ 6\varsigma
^{2}\tau _{0}-6\right] .
\end{equation*}
%\end{eqnarray*}%
For $\varepsilon \rightarrow 0,$ $\ ^{ds}Q\rightarrow \sqrt{|\underline{%
\mathring{g}}_{1}\underline{\mathring{g}}_{2}\underline{\mathring{g}}_{3}|}.$
We can chose such $\tau _{0}$ and $\mathring{\Psi}^{2}(r,\theta ,\varphi )$
which would allow to relate such values to those of Hawking-Bekenstein black
hole thermodynamics. Nevertheless, it should be emphasized that Perelman's
thermodynamics for 3-d hypersurfaces is different from the standard black
hole thermodynamics determined by 2-d surface geometries.

\subsubsection{Thermodynamic values for 3-d soliton KdV evolution}

A stationary geometric flow evolution thermodynamics can be associated also
to 3-d soliton KdV flows of type (\ref{solitevol}), $q_{1}=q_{2}=e^{\psi
(\tau ,x^{k})},q_{3}=h(\tau ,x^{1},y^{3}),\ $for $h$ being a solution of KdV
equation (\ref{kdveq}). The related source is
\begin{equation*}
\Upsilon =\Lambda _{0}-\ ^{\phi }\Lambda -2\varsigma ^{2}-hh^{\ast
}-\epsilon h^{\ast \ast \ast }\mp \int dy^{3}\partial _{11}^{2}h.
\end{equation*}%
In result, we compute%
\begin{eqnarray*}
^{KdV}Q(\tau ,x^{k},y^{3}) &:=&\sqrt{|q_{1}q_{2}q_{3}|}=e^{\psi (\tau
,x^{k})}\sqrt{|h|}\mbox{ and } \\
\mathbf{\ }_{\shortmid }\widehat{R} &=&\ ^{\thicksim }\overline{\Upsilon }+%
\frac{1}{2}\Upsilon =\frac{3}{2}\Lambda _{0}-\ ^{\phi }\Lambda -2\varsigma
^{2}-hh^{\ast }-\epsilon h^{\ast \ast \ast }\mp \int dy^{3}\partial
_{11}^{2}h.
\end{eqnarray*}%
The thermodynamic values are
\begin{eqnarray*}
\ _{\shortmid }\widehat{\mathcal{E}}\ \ &=&\tau ^{2}\int_{\widehat{\Xi }_{0}}%
\frac{dx^{1}dx^{2}dy^{3}}{\left( 4\pi \tau \right) ^{3}}e^{\psi (\tau
,x^{k})}\sqrt{|h|}\left\{ \frac{3}{2}\Lambda _{0}-\ ^{\phi }\Lambda
-2\varsigma ^{2}-hh^{\ast }-\epsilon h^{\ast \ast \ast }\mp \int
dy^{3}\partial _{11}^{2}h-\frac{3}{\tau }\right\} , \\
\ _{\shortmid }\widehat{S} &=&\int_{\widehat{\Xi }_{0}}\frac{%
dx^{1}dx^{2}dy^{3}}{\left( 4\pi \tau \right) ^{3}}\ e^{\psi (\tau ,x^{k})}%
\sqrt{|h|}\left\{ \tau \left( \frac{3}{2}\Lambda _{0}-\ ^{\phi }\Lambda
-2\varsigma ^{2}-hh^{\ast }-\epsilon h^{\ast \ast \ast }\mp \int
dy^{3}\partial _{11}^{2}h\right) -6\right\} .
\end{eqnarray*}%
It is obvious that certain parametric 3-d solitonic waves can be not
admissible as physical solutions if they result in negative effective
thermodynamics energy and/or entropy.

\section{Discussion and Conclusions}

\label{s6} We studied a model of relativistic geometric flow theory which
for self--similar stationary configurations defines Ricci solitons modelling
modified $R^{2}$ gravity theories. Although the Lorentz signature changes
substantially the physical character of geometric evolution which in such
cases is not governed by a nonlinear diffusion operator with modified Laplacian (but by nonlinear
generalizations of d 'Alambert operator), such models seem to be more realistic
and important for research in modified gravity theories, MGTs, and
understanding generic off--diagonal interactions in GR. Our key idea was to
define such nonholonomic variables when the generalized geometric evolution
(and Ricci soliton) equations decouple in very general forms. In certain
sense, the bulk of  MGTs can be modelled geometrically by a corresponding
nonholonomic Ricci soliton configuration.

Applying the anholonomic frame deformation method, AFDM, very general
classes of exact solutions of generalized R. Hamilton and modified Ricci
soliton equations can be constructed. Such solutions are, in general, with
nontrivial torsion structure and depend on all spacetime coordinates via
corresponding classes of generating and integration functions, generalized
effective sources, integration parameters etc. Metrics are generic
off--diagonal and the nonlinear and linear connections  can be
nonholonomically constrained in order to extract Levi-Civita, LC,
configurations. This geometric method of generating exact solutions allow to
integrate in very general forms different nonlinear systems of PDEs for
geometric flow evolution and MGTs, string and brane models with nonholonomic
/ noncommutative / supersymmetric variables, see reviews of results in \cite%
{vnrflnc,sv2014,svvvey,tgovsv,vsolit2,vkerrads,vtamsuper}.

Mathematically, one has not been elaborated yet necessary methods of
geometric analysis for Lorentzian manifolds with pseudo-Euclidean signature
and for non-Riemannian manifolds (for instance, with nontrivial torsion
structure and/or endowed with additional distributions of Lagrange densities
for gravitational and matter fields defining MGTs and nonholonomic GR
models). In result, it is not possible at present to elaborate a
mathematical rigorous theory of relativistic/ supersymmetric / nonholonomic
geometric flows like it was possible for for Riemannian manifolds.
Nevertheless, we can study a number of applications and possible physically
important effects for various types of relativistic and MGTs modifications
using exact solutions generated following the AFDM. For certain nonholonomic
configurations, we can solve the Cauchy problem, or satisfy certain
boundary/ asymptotic conditions, analyse the necessary criteria for
gravitational (nonlinear) diffusion, consider noncommutative interactions,
topological changing etc.

Positively, we can apply methods of standard Ricci flow theory for 3+1
splitting. Such constructions were considered, for instance, in the
super-renormalizable versions of Ho\v{r}ava-Lifshitz gravity, with
Ricci--Cotton flows, focusing on Bianchi cosmological models, see \cite%
{bakaslust}, for study low dimensional Ricci flow equations etc. \cite%
{carstea,vrfijmpa,vacvis}. The AFDM allows to construct generic
off--diagonal solutions in MGTs of arbitrary dimension \cite{tgovsv}. This
geometric method can be developed for finding solutions of geometric flow
equations by considering additional dependencies on evolution parameter.
Even, in general, the parametric dependence and relativistic evolution of
generalized Ricci flow models may change the type of corresponding nonlinear
PDE (for instance, locally parabolic systems can be transformed into certain
hyperbolic ones etc.) we can investigate and understand main properties of
such nonlinear systems working with nonholonomic variables which allows to
find exact solutions.

One should be emphasized here that the AFDM works effectively, and the
resulting solutions admit certain realistic physical interpretation, if we
consider auxiliary linear connections with nonholonomically induced torsion
all determined by certain off--diagonal deformations of physically important
solutions (like black holes, wormholes, locally anisotropic cosmological
models etc.\cite{sv2001,sv2014,svvvey,tgovsv}). This way we work with very
general ansatz for metrics and connections when the corresponding geometric
evolution / gravitational field modified equations can be integrated in
certain general forms. The bulk of exact solutions constructed by other
authors were obtained for much "simple" ansatz with diagonalizable metrics
when coefficients depend on one spacelike/time like coordinates and the
corresponding effective Einstein equations transforms into a nonlinear
system of ordinary differential equations, ODEs. Even such an approach with
ansatz of high symmetry offers certain possibilities to construct exact and
very important astrophysical and cosmological solutions for some special
classes of systems of nonlinear PDEs, it is very restrictive comparing to
the AFDM. Transforming a system of PDEs into a a system of ODEs for special
ansatz, we cut from the very beginning the possibility to find exact
solutions with generic off--diagonal metrics depending on 3-4 and extra
dimension variables. For researchers on
physical mathematics, there is a very important question: Shall we really
modify the GR theory or preserve the physical paradigm by considering
generic nonlinear off--diagonal solutions which for certain conditions mimic
MGTs effects and provide a theoretical explanation of observable data in
modern acceleration cosmology?

MGTs can be treated alternatively as some nonholonomic Ricci soliton
configurations of relativistic geometric flow models. Various classes of
exact solutions for corresponding evolution / self--similar equations can be
related to important physically solutions, and their off--diagonal
deformations, via certain locally anisotropic polarization functions and
variation of constants. This may provide a theoretical background for recent
experimental and phenomenological work on variation of constants \cite%
{flambaum}. In another turn, observational data in modern cosmology and
related research on MGTs and dark energy and dark matter physics may serve
as certain crucial indication how a realistic geometric flow theories can be
developed in relativistic and physically motivated forms. In result, we
addressed the issue how the $R^2$ gravity (which is of grate interest for
physicists beginning original cosmological papers \cite{starob,much}) can be
involved into a realistic geometric flow scenarios and realized as a
nonholonomic Ricci flow model.

As a toy model for testing our constructions on physically motivated
geometric flow and Ricci soliton models we chosen the black hole solutions
for $R^2$ gravity \cite{kehagias}. Generalizations of such classes of
solutions can be obtained by applying the AFDM to modified R. Hamilton and
Ricci soliton equations written in nonholonomic variables. For small
parametric deformations, we can construct stationary black ellipsoid
configurations when the "eccentricity" is related to possible locally
anisotropic polarization and/or running of physical constants. It should be
noted that black ellipsoids have spheroidal topology and, in consequence,
such objects are not prohibited by black hole uniqueness theorems in GR.
They positively exist in $R^2$ gravity and other modifications, see \cite%
{svvvey,tgovsv,vsolit2,vkerrads,vtamsuper}. Vacuum black hole solutions of
Kerr type are not admitted in certain $R^2$ models for the Levi-Civita
connection, but such solutions can be obtained for a nontrivial cosmological
constant, nonholonomic deformations of connection structures, off--diagonal
modifications of metrics, contributions from geometric flows etc.

The AFDM allows us to integrate systems of nonlinear PDEs (for modified
geometric flow and gravity theories, in particular, in $R^2$) in very
general forms without small parametric limits to well known classes of exact
solutions with very special symmetries. It is not clear what physical
importance may have such general classes of solutions. We provided some
examples for the cases when nontrivial vacuum configurations and
polarizations of effective cosmological constants in $R^2$ gravity are
determined by 3-d solitonic waves, for instance, of KdV type \cite%
{kdp,belinski,vsolit1,vsolit2,vkerrads}. Such new types of solutions have a
well defined physical interpretation as nonlinear solitonic waves for
gravitational and matter field interactions.

Our approach to geometric flows and MGTs is based on generalizations of
Perelman's functionals reformulated in nonholonomic variables. Such
functionals for the LC--connection and 3-d Riemannian metrics played a
crucial role in the proof of the Poincar\'{e} conjecture. The so--called
W--functional is a Lyapunov type functional which play the role of effective
entropy which was used for formulating an analogous statistical
thermodynamics characterizing Ricci flows. Geometrically, it is possible to
generalize the constructions for various types of gravity theories, for
generalized connections and new physical objects but the Lorentz signature
does not allow to treat directly the W--functional as an entropy one. We
have to consider additional nonholonomic 3+1 and 2+2 decompositions and, in
general, to elaborate models of locally anisotropic relativistic geometric
flow by analogy to relativistic hydrodynamics and relativistic kinetics
theories, as we discuss in \cite{vvrfthbh}. For stationary configurations in
different MGTs realized as nonholonomic Ricci solitons, the Perelman's
functionals can be determined almost in a standard way on 3-d spacelike
hypersurfaces. This is very important because nonholonomic versions of
W--functionals provide a thermodynamic interpretation to various classes of
generalize off--diagonal solutions in such theories (like black ellipsoids /
holes, wormholes etc.). The standard Hawking-Bekenshtein black hole
thermodynamics is based on 2-d hypersurface gravity which is not applicable
for more general classes of solutions in MGTs. One of the goals of this work
was to show in explicit form how to compute Perelman's thermodynamical
energy and entropy for black ellipsoid and KdV solitons in $R^2$ gravity.

Finally, another interesting problem is the application of MGTs (in
particular, of $R^2$ gravity) in order to test physically viable
supersymmetric generalizations of geometric flows and supergravity models.
We have a self--consistent variant of noncommutative geometric flow theory
in the A. Connes approach, see \cite{vnrflnc} with generalized Perelman's
functionals, nonholonomic Dirac operators and spectral triples. Such
noncommutative Ricci flow models can be elaborated for other approaches to
noncommutative geometry. There is a number of formulations of modified
supergravity and superstring theories which do not allow to elaborate an
unified model of supergeometric flows. Mathematically, the problem is also
less clear because different groups of mathematicians work with different
definitions of supermanifolds \cite{vnpfinsl}. In order to study possible
indications from modern gravity and cosmology how a supersymmetric
modification of geometric flow theory could be physically motivated, we plan
to apply and develop the results of this work and paper \cite{kounnas3} in
\cite{vsuperfl} (a research on supersymmetric Ricci flows and $\mathcal{R}^2$
inflation from scale invariant supergravity).

\vskip2pt \textbf{Acknowledgments:} S. V. research is partially supported by
IDEI, PN-II-ID-PCE-2011-3-0256 and DAAD. He is grateful for DAAD hosting to  D. L\"{u}st and O. Lechtenfeld.

\appendix

\setcounter{equation}{0} \renewcommand{\theequation}
{A.\arabic{equation}} \setcounter{subsection}{0}
\renewcommand{\thesubsection}
{A.\arabic{subsection}}

\section{Some Formulas for N-adapted 2+2 splitting}

\label{sa}For convenience, we summarize in this Appendix some important
N--adapted coefficient formulas (see details and proofs in Refs. \cite%
{sv2001,sv2014,svvvey,tgovsv}).

The N--adapted coefficients of the canonical d--connection $\widehat{\mathbf{%
D}}=\{$ $\widehat{\mathbf{\Gamma }}_{\ \alpha \beta }^{\gamma }=(\widehat{L}%
_{jk}^{i},\widehat{L}_{bk}^{a},\widehat{C}_{jc}^{i},\widehat{C}_{bc}^{a})\}$
are
\begin{eqnarray}
\widehat{L}_{jk}^{i} &=&\frac{1}{2}g^{ir}\left( \mathbf{e}_{k}g_{jr}+\mathbf{%
e}_{j}g_{kr}-\mathbf{e}_{r}g_{jk}\right),\ \widehat{L}_{bk}^{a} =
e_{b}(N_{k}^{a})+\frac{1}{2}h^{ac}\left( e_{k}h_{bc}-h_{dc}\
e_{b}N_{k}^{d}-h_{db}\ e_{c}N_{k}^{d}\right) ,  \notag \\
\widehat{C}_{jc}^{i} &=&\frac{1}{2}g^{ik}e_{c}g_{jk},\ \widehat{C}_{bc}^{a}=%
\frac{1}{2}h^{ad}\left( e_{c}h_{bd}+e_{c}h_{cd}-e_{d}h_{bc}\right) .
\label{candcon}
\end{eqnarray}%
The nonholonomically induced torsion $\widehat{\mathcal{T}}$ $=\{\widehat{%
\mathbf{T}}_{\ \alpha \beta }^{\gamma }\}$ of $\ $(\ref{candcon}) satisfy
the conditions $\widehat{T}_{\ jk}^{i}=0$ and $\widehat{T}_{\ bc}^{a}=0,$
but with nontrivial h--v-- coefficients
\begin{equation}
\widehat{T}_{\ jk}^{i} = \widehat{L}_{jk}^{i}-\widehat{L}_{kj}^{i},\widehat{T%
}_{\ ja}^{i}=\widehat{C}_{jb}^{i},\widehat{T}_{\ ji}^{a}=-\Omega _{\
ji}^{a},\ \widehat{T}_{aj}^{c} = \widehat{L}_{aj}^{c}-e_{a}(N_{j}^{c}),%
\widehat{T}_{\ bc}^{a}=\ \widehat{C}_{bc}^{a}-\ \widehat{C}_{cb}^{a}.
\label{dtors}
\end{equation}

We can consider N--splitting with zero noholonomically induced d--torsion,
when $\widehat{\mathbf{T}}_{\ \alpha \beta }^{\gamma }=0,$ i.e.
\begin{equation}
\widehat{C}_{jb}^{i}=0,\Omega _{\ ji}^{a}=0\mbox{ and }\widehat{L}%
_{aj}^{c}=e_{a}(N_{j}^{c}).  \label{lccond}
\end{equation}%
These conditions follow from formulas (\ref{candcon}) and (\ref{dtors}). If
the Levi--Civita conditions, LC--conditions, (\ref{lccond}) are satisfied,
we obtain that in N--adapted frames (\ref{nader}) and (\ref{nadif}) $%
\widehat{\mathbf{Z}}_{\ \alpha \beta }^{\gamma }=0$ and $\widehat{\mathbf{%
\Gamma }}_{\ \alpha \beta }^{\gamma }=\Gamma _{\ \alpha \beta }^{\gamma }.$
Here we note that the definition and the frame/coordinate transformation
laws of a d--connection are different from that of a "usual" linear
connection (for instance, $\widehat{\mathbf{D}}\neq \nabla ),$ we can impose
additional conditions on coefficients $(\mathbf{g}_{\alpha \beta
},N_{j}^{c}) $ which allow us to generate LC--configurations.

The curvature $\widehat{\mathcal{R}}=\{\widehat{\mathbf{R}}_{\ \beta \gamma
\delta }^{\alpha }\}$ of the canonical d--connection $\widehat{\mathbf{D}}$
is characterized by six groups of N--adapted coefficients,
\begin{eqnarray}
\widehat{R}_{\ hjk}^{i} &=&e_{k}\widehat{L}_{\ hj}^{i}-e_{j}\widehat{L}_{\
hk}^{i}+\widehat{L}_{\ hj}^{m}\widehat{L}_{\ mk}^{i}-\widehat{L}_{\ hk}^{m}%
\widehat{L}_{\ mj}^{i}-\widehat{C}_{\ ha}^{i}\Omega _{\ kj}^{a},  \notag \\
\widehat{R}_{\ bjk}^{a} &=&e_{k}\widehat{L}_{\ bj}^{a}-e_{j}\widehat{L}_{\
bk}^{a}+\widehat{L}_{\ bj}^{c}\widehat{L}_{\ ck}^{a}-\widehat{L}_{\ bk}^{c}%
\widehat{L}_{\ cj}^{a}-\widehat{C}_{\ bc}^{a}\Omega _{\ kj}^{c},
\label{dcurv} \\
\widehat{R}_{\ jka}^{i} &=&e_{a}\widehat{L}_{\ jk}^{i}-\widehat{D}_{k}%
\widehat{C}_{\ ja}^{i}+\widehat{C}_{\ jb}^{i}\widehat{T}_{\ ka}^{b},\widehat{%
R}_{\ bka}^{c}=e_{a}\widehat{L}_{\ bk}^{c}-D_{k}\widehat{C}_{\ ba}^{c}+%
\widehat{C}_{\ bd}^{c}\widehat{T}_{\ ka}^{c},  \notag \\
\widehat{R}_{\ jbc}^{i} &=&e_{c}\widehat{C}_{\ jb}^{i}-e_{b}\widehat{C}_{\
jc}^{i}+\widehat{C}_{\ jb}^{h}\widehat{C}_{\ hc}^{i}-\widehat{C}_{\ jc}^{h}%
\widehat{C}_{\ hb}^{i},\widehat{R}_{\ bcd}^{a}=e_{d}\widehat{C}_{\
bc}^{a}-e_{c}\widehat{C}_{\ bd}^{a}+\widehat{C}_{\ bc}^{e}\widehat{C}_{\
ed}^{a}-\widehat{C}_{\ bd}^{e}\widehat{C}_{\ ec}^{a}.  \notag
\end{eqnarray}

The Ricci d--tensor $\widehat{\mathbf{R}}_{\alpha \beta }:=\widehat{\mathbf{R%
}}_{\ \alpha \beta \gamma }^{\gamma }$ of $\widehat{\mathbf{D}}$ is defined
by standard formulas and characterized by four groups of N--adapted
coefficients
\begin{equation}
\widehat{\mathbf{R}}_{\alpha \beta }=\{\widehat{R}_{ij}:=\widehat{R}_{\
ijk}^{k},\ \widehat{R}_{ia}:=-\widehat{R}_{\ ika}^{k},\ \widehat{R}_{ai}:=%
\widehat{R}_{\ aib}^{b},\ \widehat{R}_{ab}:=\widehat{R}_{\ abc}^{c}\}.
\label{driccic}
\end{equation}%
The corresponding scalar curvature $\widehat{R}$ of $\ \widehat{\mathbf{D}}$
is also a usual one when by definition
\begin{equation}
\widehat{R}:=\mathbf{g}^{\alpha \beta }\widehat{\mathbf{R}}_{\alpha \beta
}=g^{ij}\widehat{\mathbf{R}}_{ij}+g^{ab}\widehat{\mathbf{R}}_{ab}.
\label{sdcurv}
\end{equation}%
Now, we \ can define and compute the Einstein tensor $\widehat{\mathbf{E}}%
_{\alpha \beta }$ of $\widehat{\mathbf{D}},$
\begin{equation}
\widehat{\mathbf{E}}_{\alpha \beta }:=\widehat{\mathbf{R}}_{\alpha \beta }-%
\frac{1}{2}\mathbf{g}_{\alpha \beta }\widehat{R}.  \label{enstdt}
\end{equation}%
This d--tensor is different from that for the Levi--Civita connection $%
\nabla ,$ but related via distortion relation depending only on $\mathbf{g}%
_{\alpha \beta }$ and $N_{a}^{i}$ to the Einstein tensor $E_{\alpha \beta }$
computed for data $(\mathbf{g}_{\alpha \beta },\nabla ).$ Using formulas (%
\ref{distr}), we can compute distortions of connections, torsions and
curvatures, Ricci and Einstein tensors and, respective, scalars.

The N--adapted coefficients $\widehat{\mathbf{\Gamma }}_{\ \alpha \beta
}^{\gamma }$ of $\ \widehat{\mathbf{D}}$ are equal to the coefficients $%
\Gamma _{\ \alpha \beta }^{\gamma }$ of $\ \nabla ,$ both sets computed with
respect to N--adapted frames (\ref{nader}) and (\ref{nadif}), if and only if
there are satisfied the conditions $\widehat{L}_{aj}^{c}=e_{a}(N_{j}^{c}),%
\widehat{C}_{jb}^{i}=0$ and $\Omega _{\ ji}^{a}=0.$ In such a case, all
N--adapted coefficients of the torsion $\widehat{\mathbf{T}}_{\ \alpha \beta
}^{\gamma }$ \ (\ref{dtors}) and the distortion d--tensor $\widehat{\mathbf{Z%
}}_{\ \alpha \beta }^{\gamma }$ are zero.


\begin{thebibliography}{99}
\bibitem{hamilt1} R. S. Hamilton, Three-manifolds with postive Ricci
curvature, \emph{\ J. Diff. Geom } \textbf{\ 17 } (1982) 255-306

\bibitem{hamilt2} R. S. Hamilton, The Ricci flow on surfaces, in:
Mathematics and General Relativity, \emph{\ Contemp. Math.} \textbf{\ 71},
p. 237-262, Amer. Math. Soc., Providence, 1988

\bibitem{hamilt3} R. S. Hamilton, in: Surveys in Differential Geometry, vol.
2 (International Press, 1995), pp. 7-136

\bibitem{perelman1} G. Perelman, The entropy formula for the Ricci flow and
its geometric applications, arXiv: math. DG/0211159

\bibitem{perelman2} G. Perelman, Ricci flow with surgery on
three--manifolds, arXiv: math.DG/0303109

\bibitem{perelman3} G. Perelman, Finite extintion time for the solutions to
the Ricci flow on certain three-manifolds, arXiv: math.DG/0307245

\bibitem{friedan1} D. Friedan, Nonlinear models in $2+\varepsilon $
dimensions, PhD Thesis (Berkely) LBL-11517, UMI-81-13038, Aug 1980. 212pp

\bibitem{friedan2} D. Friedan, Nonlinear models in $2+\varepsilon $
dimensions, \emph{\ Phys. Rev. Lett. } \textbf{45} (1980) 1057-1060

\bibitem{friedan3} D. Friedan, Nonlinear models in $2+\varepsilon $
dimensions, \emph{\ Ann. of Physics } \textbf{163} (1985) 318-419

\bibitem{vnhrf} S. Vacaru, Nonholonomic Ricci flows: II. Evolution equations
and dynamics, \emph{\ J. Math. Phys. } \textbf{\ 49 } (2008) 043504

\bibitem{vnrflnc} S. Vacaru, Spectral functionals, nonholonomic Dirac
operators, and noncommutative Ricci flows, \emph{\ J. Math. Phys. } \textbf{%
\ 50 } (2009) 073503

\bibitem{vvrfthbh} O. Vacaru and S. Vacaru, On relativistic generalization
of Perelman's W-entropy and statistical thermodynamic description of
gravitational fields, arXiv: 1312.2580v3

\bibitem{bakaslust} I. Bakas, F. Fourliot, D. L\"{u}st and M. Petropoulos,
Geometric flows in Ho\v{r}ava--Lifshitz gravity, \emph{JHEP} \textbf{4}
(2010) 131

\bibitem{carfora} M. Carfora, The Wasserstein geometry of non-linear $\sigma
$ modles and the Hamilton-Perelman Ricci flow, arXiv: 1405.0827

\bibitem{sv2001} S. Vacaru, Anholonomic soliton-dilaton and black hole
solutions in general relativity, \emph{JHEP} \textbf{04} (2001) 009

\bibitem{sv2014} S. Vacaru, Exact solutions in modified massive gravity and
off-diagonal wormhole deformations, \emph{EPJC} \textbf{74} (2014) 2781

\bibitem{svvvey} S. Vacaru, E. Veliev, E. Yazici, A geometric method of
constructing exact solutions in modified f(R,T) gravity with Yang-Mills and
Higgs Interactions, \emph{IJGMMP} \textbf{11} (2014) 1450088

\bibitem{tgovsv} T. Gheorghiu, O. Vacaru, S. Vacaru, Off-diagonal
deformations of Kerr black holes in Einstein and modified massive gravity
and higher dimensions, \emph{EPJC} \textbf{74} (2014) 3152

\bibitem{kehagias} A. Kehagias, C. Kounnas, D. L\"{u}st and A. Riotto, Black
hole solutions in $R^{2}$ gravity, \emph{JHEP} \textbf{5} (2015) 143

\bibitem{polyakov} A. M. Polyakov, Interactions of Goldstone particles in
two dimensions. Applications to ferromagnets and massive Yang--Mills fields,
\emph{\ Phys. Lett. B} \textbf{\ 59 } (1975) 79-81

\bibitem{starob} A. A. Starobinsky, A new type of isotropic cosmological
models without singularity, \emph{Phys. Lett.} \textbf{B 91} (1980) 99

\bibitem{much} V. F. Mukhanov and B. V. Chibisov, Quantum fluctuation and
nonsingurlar universes, \emph{JETP Lett.} \textbf{33} (1981) 532 [\emph{%
Pisma Zh. Eksp. Teor. Fiz.} \textbf{33} (1981) 549, in Russian]

\bibitem{guth} A. H. Guth, The inflationary universe: a possible solution to
the horizon and flatness problems, \emph{Phys. Rev.} \textbf{D 23} (1981) 347

\bibitem{linde} A. D. Linde, A new inflationary universe scenario: a
possible soluton of the horizon, flatness, homogeneity, isotropy and
promordial monopole problems, \emph{Phys. Lett.} \textbf{B 108} (1982) 389

\bibitem{albrecht} A. Albrecht and P. J. Steinhardt, Cosmology for grand
unified theories with ratiatively induces symmetry breaking, \emph{Phys.
Rev. Lett.} \textbf{48} (1982) 1220

\bibitem{eghnsr} E. Guendelman, H. Nishino and S. Rajpoot, Lorentz-covariant
four-vector formalism for two-measure theory, \emph{Phys. Rev. D} \textbf{87}
(2013) 027702

\bibitem{srsv2} S. Rajpoot and S. Vacaru, Cosmological attractors and
anisotropies in two measure theories, effective EYMH systems, and
off-diagonal inflation models, under elaboration.

\bibitem{hl1} P. Ho\v{r}ava, Membranes at quantum criticality, \emph{JHEP}
\textbf{03} (2009) 020

\bibitem{hl2} P. Ho\v{r}ava, Quantum gravity at a Lifshitz point, \emph{%
Phys.\ Rev.} \textbf{D 79} (2009) 084008

\bibitem{flambaum} A. Windberger, J. R. Creso Lopez-Urrutia, H. Bekker et
all, Identification of the predicted 5s-4f level crossing optical lines with
application to metrology and searchers for the variation of fundamental
constants, \emph{Phys.\ Rev. Lett.} \textbf{114} (2015) 150801

\bibitem{kdp} B. B. Kadomtsev and V. I. Petviashvili, On stability of
solitary waves in weakly dispersive media, \emph{Sov. Phys. Dokl.} \textbf{15%
} (1970) 539-541 [Russian translation: \emph{Doklady Akademii Nauk} \emph{%
SSSR} \textbf{192} (1970) 753-756]

\bibitem{belinski} V.\ Belinski and E. Verdaguer, Gravitational Solitons
(Cambriged Universtity Press, 2001)

\bibitem{vsolit1} S. Vacaru, Generic Off-Diagonal Solutions and Solitonic
Hierarchies in Einstein and Modified Gravity, \emph{Mod. Phys. Lett.} A
\textbf{30} (2015) 1550090

\bibitem{vsolit2} S. Vacaru and D. Singleton, Warped solitonic deformations
and propagation of black holes in 5D vacuum gravity, \emph{Class. Quant.
Grav. }\textbf{19} (2002) 3583-3602

\bibitem{vkerrads} S. Vacaru, Hidden symmetries for ellipsoid-solitonic
deformations of Kerr-Sen black holes and quantum anomalies, \emph{Eur. Phys.
J.} C \textbf{73} (2013) 2287

\bibitem{vtamsuper} T. Gheorghiu, O. Vacaru and S. Vacaru, Modified
Dynamical Supergravity Breaking and Off-Diagonal Super-Higgs Effects, \emph{%
Class. Quant. Grav.} \textbf{\ 32 } (2015) 065004

\bibitem{vnpfinsl} S. Vacaru, Superstrings in higher order extensions of
Finsler superspaces, \emph{Nucl. Phys.} \textbf{B 434} (1997) 590 -656

\bibitem{kounnas3} C. Kounnas, D. L\"{u}st and N. Toumbas, $\mathcal{R}^2$
inflation from scale invariant supergravity and anomaly free superstrings
with fluxes, \emph{Fortsch. Phys. } \textbf{63} (2015) 12-35

\bibitem{vsuperfl} V. Ruchin and S. Vacaru, On supersymmetric Ricci flows and
$\mathcal{R}^2$ inflation from scale invariant supergravity [under elaboration]

\bibitem{carstea} S. A. Carstea and M. Visinescu, Special solutions for
Ricci flow equation in 2D using the linearization approach, \emph{Mod. Phys.
Lett. }\textbf{\ A20 } (2005) 2993-3002

\bibitem{vrfijmpa} S. Vacaru, Ricci flows and solitonic pp-waves, \emph{Int.
J. Mod. Phys.} \textbf{\ A21 } (2006) 4899-4912

\bibitem{vacvis}  S. Vacaru and M. Visinescu, Nonholonomic Ricci flows
and running cosmological constant: I. 4D Taub-NUT metrics, \emph{Int. J.
Mod. Phys.} \textbf{\ A22 } (2007) 1135-1159
\end{thebibliography}
\end{document}